\documentclass[article]{aa}  

\usepackage{graphicx}
\usepackage{txfonts}
\usepackage{lscape}
\usepackage{booktabs}
\usepackage{threeparttablex}
\usepackage{wasysym}
\usepackage{amssymb}
\usepackage{float}
\usepackage{diagbox}
\usepackage{capt-of} 
\usepackage[bookmarks=false,colorlinks=true,linkcolor=cyan,citecolor=blue,filecolor=black,urlcolor=cyan]{hyperref}
\newcommand*{\distance}{\ensuremath{\mathrm{D}}}
\newcommand*{\sfr}{\ensuremath{\mathrm{SFR}}}
\newcommand*{\stellarmass}{\ensuremath{\mathrm{M}_{\star}}}
\newcommand*{\lx}{\ensuremath{\mathrm{L}_{X}}}
\newcommand*{\solarmass}{\ensuremath{\mathrm{M}_{\odot}}}
\newcommand*{\xmm}{\ensuremath{\textit{XMM}}}
\newcommand*{\chandra}{\ensuremath{\textit{Chandra}}}
\newcommand*{\erosita}{\ensuremath{\rm{eROSITA}}}
\newcommand*{\fluxunits}
{\ensuremath{\mathrm{erg\,cm^{-2}\,s^{-1}}}}
\newcommand*{\coldens}{\ensuremath{\mathrm{N}_{H}}}

\newcommand*{\energyunits}{\ensuremath{\mathrm{keV}}}

\begin{document}

   \title{The first all-sky survey of star-forming galaxies with eROSITA}

   \subtitle{Scaling relations and a population of X-ray luminous starbursts}

   \author{E. Kyritsis\inst{1,2}
          \and
          A. Zezas\inst{1,2}
          \and
          F. Haberl\inst{3}
          \and
          P. Weber\inst{4}
          \and
          A. Basu-Zych\inst{6,7,8}
          \and
          N. Vulic\inst{5,6,7,8}
          \and
          C. Maitra\inst{3}
          \and
          S. H\"ammerich\inst{4}
          \and
          J. Wilms\inst{4}
          \and
          M. Sasaki\inst{4}
          \and
          A. Hornschemeier\inst{6,9}
          \and
          A. Ptak\inst{6}
          \and
          A. Merloni\inst{3}
          \and 
          J. Comparat\inst{3}
          }
          \authorrunning{Kyritsis et al.}

   \institute{Physics Department, \& Institute of Theoretical and Computational Physics,   University of Crete, GR 71003, Heraklion, Greece\\
   \email{ekyritsis@physics.uoc.gr}
   \and
    Institute of Astrophysics, Foundation for Research and Technology-Hellas, GR 71110 Heraklion, Greece
    \and
    Max-Planck-Institut für extraterrestrische Physik (MPE), Gießenbachstraße 1, D-85748 Garching bei München, Germany
    \and
    Remeis Observatory and ECAP, Universität Erlangen-Nürnberg, Sternwartstraße 7, 96049 Bamberg, Germany
    \and
    Eureka Scientific, Inc., 2452 Delmer St., Suite 100, Oakland, CA 94602-3017, USA 
    \and
    NASA Goddard Space Flight Center, Code 662, Greenbelt, MD 20771, USA
    \and
    Center for Space Science and Technology, University of Maryland Baltimore County, 1000 Hilltop Circle, Baltimore, MD 21250, USA
    \and
    Center for Research and Exploration in Space Science and Technology, NASA/GSFC, Greenbelt, MD 20771, USA
    \and
    Department of Physics and Astronomy, Johns Hopkins University, 3400 N. Charles Street, Baltimore, MD 21218, USA
             }

   \date{Received Month XX, XXXX; accepted Month XX, XXXX}

 
  \abstract
   {
   In this work, we present the results from a study of X-ray normal galaxies (i.e. not harbouring active galactic nuclei; AGN) using data from the first complete all-sky scan of the \erosita{} X-ray survey (eRASS1) obtained with eROSITA on board the {\it Spectrum-Roentgen-Gamma} ({\it SRG}) observatory. eRASS1 provides the first unbiased X-ray census of local normal galaxies allowing us to study the X-ray emission (0.2-8.0 \energyunits{}) from X-ray binaries (XRBs) and the hot interstellar medium in the full range of stellar population parameters present in the local Universe.
   }
   {By combining the updated version of the Heraklion Extragalactic Catalogue (HECATE v2.0) value-added catalogue of nearby galaxies (\distance{}${\lesssim}200$Mpc) with the X-ray data obtained from the eRASS1, we study the integrated X-ray emission from normal galaxies as a function of their star-formation rate (\sfr{}), stellar mass (\stellarmass{}), Metallicity, and stellar population age.}
   {After applying stringent optical and mid-infrared activity classification criteria, we constructed a sample of 18790 bona-fide star-forming galaxies (HEC-eR1 galaxy sample) with measurements of their integrated X-ray luminosity (using each galaxy's D$_{25}$), over the full range of stellar population parameters present in the local Universe. 
   By stacking the X-ray data in \sfr{}-\stellarmass{}-\distance{} bins we study the correlation between the average X-ray luminosity and the average stellar population parameters. We also present updated \lx{}-\sfr{} and \lx{}/\sfr{}-Metallicity scaling relations based on a completely blind galaxy sample and accounting for the scatter dependence on the \sfr{}.}
   {The average X-ray spectrum of star-forming galaxies is well described by a power-law ($\Gamma = 1.75^{+0.12}_{-0.07}$) and a thermal plasma component ($kT = 0.70^{0.06}_{-0.07} \, \energyunits{}$).
   We find that the integrated X-ray luminosity of the individual HEC-eR1 star-forming galaxies is significantly elevated (reaching $10^{42}\rm{erg\,s^{-1}}$) with respect to that expected from the current standard scaling relations. The observed scatter is also significantly larger.
   This excess persists even when we measure the average luminosity of galaxies in \sfr{}--\stellarmass{}-\distance{} and metallicity bins and it is stronger (up to ${\sim}2$dex) towards lower \sfr{}s. 
   Our analysis shows that the excess is not the result of the contribution by hot gas, low-mass X-ray binaries, background AGN,  low-luminosity AGN (including tidal disruption events), or stochastic sampling of the X-ray binary  X-ray luminosity function.
   We find that while the excess is generally correlated with lower metallicity galaxies, its primary driver is the age of the stellar populations. 
   }
   {Our analysis reveals a sub-population of very X-ray luminous starburst galaxies with higher sSFRs, lower metallicities, and younger stellar populations. This population drives upwards the X-ray scaling relations for star-forming galaxies, and has important implications for understanding the population of X-ray binaries contributing in the most X-ray luminous galaxies in the local and high-redshift Universe.
   These results demonstrate the power of large blind surveys such eRASS1 which can provide a more complete picture of the X-ray emitting galaxy population and their diversity, revealing rare populations of objects and recovering unbiased underlying correlations.
   }

   \maketitle
%

\section{Introduction}


The X-ray emission from normal galaxies (i.e. not harbouring active galactic nuclei AGN) is a prime tool for studying the endpoints of stellar evolution and their effect on their galactic and intergalactic environment. 

Since this emission is mainly produced by X-ray binaries (XRBs), it is the most efficient way to probe the effectively invisible compact object populations and study their formation and evolutionary paths. As the potential predecessors of gravitational wave sources (GWs) and short $\gamma$-ray bursts (sGRBs) \citep[e.g.,][]{marchant16}, the study of XRB populations and their demographics can provide key insights into the physical processes that drive these phenomena. By constraining the evolutionary channels of XRBs and their formation rates in different galactic environments, we can predict the formation rates of GWs, and sGRBs over cosmic history, and model the high energy output of galaxies as a function of redshift. 

Another aspect of XRBs is their effect on their host galaxy and its surrounding intergalactic medium. Given the large mean free path of high-energy photons, they can affect large volumes around their source. This is of high importance, especially in the high-redshift Universe (z$\sim30-10$) where the first XRBs can heat the intergalactic medium affecting the formation of the first galaxies \citep[e.g.,][]{das17,abdurashidova23}. XRBs can also energise the interstellar medium of galaxies through photoionization and transfer of momentum via jets \citep[e.g.][]{fender05,schaerer19}. The formation of XRBs is inextricably connected with the characteristics of the stellar populations in their host galaxies. The study of the association between XRBs and stellar populations of different ages and metallicities is critical for understanding their formation and evolution over cosmic history \citep[e.g.,][]{zezas21,gilfanov22}. 

For these reasons, resolved XRB populations of \textit{local} galaxy samples have been extensively studied for the characterization of this correlation. Thanks to the availability of sensitive X-ray observations with the \textit{XMM-Newton} and \textit{Chandra} X-ray observatories, it has been shown that the number of XRBs and their total X-ray luminosity, can be parametrized in terms of their star-formation rate (SFR), stellar mass (M$_{\star}$), and a combination of both: the specific SFR (sSFR = SFR/M$_{\star}$), which is a proxy to the intensity of star-forming activity \citep[e.g.,][]{mineo14,lehmer19,riccio23}. Similar scaling relations also hold for the X-ray emitting hot gas in galaxies \citep[]{mineo12b,lehmer16} which is the result of supernovae (SNe) and massive stars. Recent studies also showed that metal-poor galaxies tend to host more luminous sources \citep[e.g.,][]{brorby16,fornasini20,kouroumpatzakis21}, which has important implications for the high-z universe \citep{madau17}. In addition, \citet[]{gilbertson22} showed that the X-ray output of normal galaxies (XRB populations and hot-gas emission) can vary as a function of the galaxy's stellar-population age.

A common characteristic of these scaling relations is the increased scatter in the dwarf star-forming galaxies regime \citep[e.g.,][]{gilfanov22,lehmer19}. This has been attributed to stochastic effects, variations in the stellar population ages and/or metallicity between galaxies, or even variability of individual XRBs \citep{kouroumpatzakis20}. Quantifying and addressing the origin of this scatter is very important, particularly for low SFR galaxies where a few XRBs are expected, both for predicting their X-ray luminosity and for inferring the scaling relations between the X-ray luminosity and the stellar populations of galaxies. Knowledge of these scaling relations provides the formation rate of XRBs \citep{antoniou19}, addresses their cosmological evolution (via comparisons with the \lx{}/SFR measured in high-z galaxies \citep[e.g.,][]{lehmer16,aird17}), and tests their evolution channels via comparisons with XRB population synthesis models \citep[e.g.,][]{fragos13}.

However, all these studies so far are based on a small sample (a few hundred galaxies as most) covering only a small fraction of the conditions (SFR, M$_{\star}$, metallicity) in the local Universe. This provides only a partial view of the X-ray scaling relations and their scatter (e.g. lacking rare and luminous sources that an all-sky survey can locate).

\erosita{} \citep{predehl}, the first 0.2--8 keV all-sky X-ray survey, provides an unbiased census of X-ray galaxies in the local universe \citep{basuzych20}. Its results enable the study of these scaling relations, by using for the first time a robust and unbiased statistical sample of star-forming galaxies. A first glimpse of its potential is given by the pilot eFEDS field (140 $\rm{deg^{2}}$) which detected 37 spectroscopically confirmed normal galaxies translating to over 10000 galaxies in the full eRASS survey \citep{vulic22}. These results show that there are galaxies with significant excess with respect to the \lx{}/\sfr{}--sSFR scaling relation for local galaxies \citep{lehmer16}. This excess is stronger in lower \sfr{} and lower metallicity galaxies. This indicates that dwarf galaxies with high sSFR and lower metallicity are more X-ray luminous. 


In this work, we use the data from the eRASS1 survey which allows us to study the \lx{}--\sfr{}--\stellarmass{}--Metallicity scaling relation for star-forming galaxies using an unprecedented sample of available galaxies (a few thousand). This survey in combination with the Heraklion Extragalactic CATalogue \citep{hecate} allows us to define an X-ray unbiased sample by including in our analysis not only X-ray detections but also constrains on the X-ray emission for all galaxies in our parent sample.  

This paper is organized as follows. In Sect.~\ref{sec:sample_construction} we present the X-ray and the galaxy sample used in this work. In Sect.~\ref{Sec:X-ray analysis} we describe the analysis of the X-ray data and in Sect.~\ref{sec:final sample} we present the final sample of star-forming galaxies. Section~\ref{sec:x-ray stacking analysis} describes the stacking analysis of the X-ray data and the average X-ray template spectrum of the eRASS1 galaxies. In Sect.~\ref{sec:Results} we present the results of our analysis as well as the results of the fit of the scaling relations. Finally in Sect.~\ref{sec:Discussion} we discuss our results and their implications and in Sect.~\ref{sec:conclusion} we present our conclusions.

\section{Sample construction}\label{sec:sample_construction}

\subsection{The \erosita{} X-ray survey}
 \erosita{} is an 0.2--8 keV X-ray instrument onboard the \textit{Spectrum Roentgen Gamma Observatory} \citep[\textit{SRG};][]{sunyav21}, which was launched on July 13$^{th}$, 2019. It consists of seven identical X-ray telescopes each one with a circular field of view (FoV) of ${\sim}1^{\circ}$. The on-axis point spread function (PSF) at 1.5 \energyunits{} is $15''$ (half-energy width, HEW) and it increases with off-axis angle resulting in an average angular resolution of ${\sim}30''$ over the FoV \citep{merloni24}. The \erosita{} All Sky Survey (eRASS) will scan the entire sky 8 times in 6-month intervals providing the deepest view of the X-ray sky to date by the completion of its final scan (eRASS1-8). The eRASS survey provides an unprecedented view of the X-ray sky, free of sample selection biases, which allows studies of source populations and the discovery of rare types of sources.  
 
The X-ray data used in this work are obtained from the first \erosita{} all-sky scan (eRASS1) which has been completed in June 2020 reaching a point source average sensitivity of ${\sim}10^{-14}$ \fluxunits{} (at the ecliptic equator) in the 0.2--2.3 \energyunits{} soft X-ray band and ${\sim}2.5\times10^{-13}$ \fluxunits{} (at the equator) in the 2.3--8 \energyunits{} hard X-ray band, for a median time exposure (Merloni et al. 2023, in press). 
In particular, our team has analyzed the eRASS1 raw data which belongs to the west hemisphere of the sky covering a region of Galactic longitudes between $180^{\circ}$ and $360^{\circ}$.

\subsection{Galaxy sample}\label{sec:galaxy_sample}
We use the Heraklion Extragalactic CATalogue, a reference catalogue that contains all known nearby galaxies ($204,733$ objects) within a distance of $D{\lesssim}200\,\mathrm{Mpc}$ ($z{\lesssim}0.048$) compiled by \citet{hecate}. In terms of the $B$-band luminosity (L$_{B}$) density, HECATE is almost complete (${>}75 \%$) up to distances of 100 Mpc including galaxies with L$_{B}$ down to L$_{B}{\sim}10^{10}L_{B,\odot}$, while even farther it reaches a completeness of ${>}50 \%$ up to a distance of 170 Mpc. By incorporating and homogenizing data from the HyperLEDA \citep{hyperleda}, NED \citep{ned} databases, and infrared \citep[\textit{WISE}; \textit{2MASS}; \textit{IRAS};][]{wise,2mass,iras} and optical \citep[\textit{SDSS};][]{sdss1,sdss2,sdss3} surveys, this value-added catalogue provides a broad array of information such as positions, sizes, distances, and stellar population parameters (\sfr{}, \stellarmass{}, gas-phase metallicity), and activity classifications. For a detailed description of the catalogue compilation see \citet{hecate}. Despite the wealth of information included in the first publicly available version of HECATE, the availability of \sfr{}, \stellarmass{}, gas-phase metallicity, and activity classification is limited to $46{\%}$, $65{\%}$, $31{\%}$, and $32{\%}$, of the full sample respectively. Obtaining a more complete coverage for these parameters is critical for the characterization of the galaxy sample observed by eRASS1. 

For that reason in this work, we use an updated version of the forthcoming catalogue HECATE v2.0. This new version includes the addition of multi-aperture photometric information from mid-infrared (\textit{WISE}) and optical (\textit{SDSS}, \textit{Pan-STARSS}; \citealp[]{panstarrs}) surveys, as well as spectroscopic data from the SDSS-DR17 spectroscopic database. Furthermore, spectroscopic information from our custom observational campaigns with the Skinakas and TNG telescope is included. In more detail, the HECATE v2.0 Catalogue used in this work provides:
\newline
\textit{Screening of galaxies with incorrect distances}: In the first release of the HECATE catalogue there are a few galaxies (${<}0.5\%$) with wrongly assigned velocities due to misassociatons with foreground stars. In HECATE v2.0 all these cases are flagged allowing the user to exclude these objects from further analysis.
\newline
\textit{Updated estimations of the \sfr{} and \stellarmass{}}: The \sfr{} and \stellarmass{} provided by the publicly available version of HECATE are based on the homogenization of data from different infrared/optical surveys and calibrations. Although this kind of analysis is very useful for large statistical samples, the combination of different calibrations can result to overestimation or underestimation of the \sfr{} and \stellarmass{} of individual galaxies depending on the dust content and the stellar population age of each galaxy. By using the new calibrations of \citet{sfr_stellar_mass_kouroumpatzakis} HECATE v2.0 provides \sfr{} and \stellarmass{} estimates using a combination of the \textit{WISE} bands 1 and 3 (W1,W3 ; $3.4\,\mathrm{\mu m}$,$12\,\mathrm{\mu m}$) for the \sfr{}, and \textit{WISE} band 1, with the optical \textit{u-r} or \textit{g-r} \textit{SDSS} colours for the \stellarmass{}, respectively. These new calibrations account for the contribution of old stellar populations in the dust emission for galaxies with low \sfr{}, and for the effect of the extinction providing more reliable estimations for objects over a broad range of star-forming activity. 
These new calibrations are applied to all HECATE v2.0 galaxies with available and good-quality mid-infrared (\textit{WISE}) and optical (\textit{SDSS}) data, while for the rest we used the standard mid-infrared calibration relations from \citet{cluver17} and \citet{salim14} for the \sfr{} and from \citet{wen13} for the \stellarmass{}. Special care is taken for the most nearby HECATE galaxies ($D{\lesssim}50\,\mathrm{Mpc}$), where the use of the mid-infrared calibration is not possible because of the low reliability of the catalogue data due to aperture effects. For all these galaxies, the \sfr{}, and \stellarmass{} values from \citet{leroy19} catalogue are used. This catalogue provides stellar population parameter estimation for all the nearby galaxies following a more accurate customized analysis for the aperture photometry. Based on this updated analysis in the HECATE v2.0 the availability of \sfr{}, and \stellarmass{} is increased from $46\%$ and $65\%$ to $76\%$ and $90\%$ of the total sample, respectively.
\newline
\textit{Updated activity classifications}: The activity classification of galaxies in the first release of the HECATE is based on spectroscopic information from  
the \textit{MPA–JHU DR8} catalogue \citep[\textit{SDSS};][]{sdss1,sdss2,sdss3}. In particular, using simultaneously four emission-line ratios and an advanced data-driven version of the traditional BPT diagrams, which utilizes a soft clustering scheme, we have available classifications for a large fraction of emission-line galaxies in HECATE \citet{stampoulis19}. Although this method is very accurate and robust, since it is based on spectroscopic information, it is limited to the \textit{SDSS} footprint. To obtain activity classification for a larger fraction of HECATE galaxies, HECATE v2.0 provides updated activity classifications based on the mid-infrared/optical photometric classifier of \citet{daoutis23}. 
It is based on the Random Forest (RF) \citep[]{louppe04} machine learning method, employing the \textit{WISE} \textit{W1-W2} and \textit{W2-W3}, and \textit{SDSS} \textit{g-r} colors. The method is trained on a set of galaxies with high-quality spectra from the \textit{MPA–JHU DR8} catalogue characterized using the \citet{stampoulis19} diagnostics for the emission line objects and photometric data for the passive galaxies. 
This new classifier is able to discriminate galaxies in five activity classes: star-forming, AGN, LINER, Composite, and Passive with very high accuracy especially for the star-forming galaxies (${>}80\%$). At the same time, since it is based on photometric data it is applicable to a much larger volume of galaxies for which there are \textit{WISE} and \textit{SDSS} (or \textit{Pan-STARSS}) photometric data.
The application of this new method on the HECATE v2.0 galaxies using the additional multi-aperture information from the \textit{WISE} and \textit{Pan-STARSS} surveys increased the availability of the activity classification from $30\%$ to $60\%$ of the total sample.
\newline
\textit{Updated gas-phase metallicities}: Gas-phase metallicities in the HECATE v2.0 are based on two methods. For all the galaxies with available emission-line fluxes provided by the \textit{MPA–JHU DR8} catalogue, the metallicity is adopted from the available HECATE version \citep{hecate}. In particular, following the O3N2 method of Pettini \& Pagel (2004, equation 3) (henceforth, PP04 O3N2), \citet{hecate} calculated the 12 + log(O/H) using the [O$\mathrm{III}$], [N$\mathrm{II}$], H$\alpha$, and H$\beta$ emission line fluxes with $S/N > 3$ for $62\,778$ HECATE galaxies within the \textit{SDSS} footprint. \citet{kewley_elison08} showed that the PP04 O3N2 method is robust enough being able to trace a wide range of metallicities, with lower scatter, and it is less sensitive to extinction effects. For the galaxies which do not have available spectroscopic metallicities the 12 + log(O/H) is calculated based on the mass-metallicity relation using the best-fit results from \citet{kewley_elison08} for the same spectroscopic calibrator (Table 2). A comparison between the spectroscopically derived metallicity and the one from the mass-metallicity relation shows that within the errors the two methods produce consistent results (Kyritsis et al. 2024, in prep.). As a result, given the availability of the updated \stellarmass{} measurements, HECATE v2.0 provides metallicities for a much larger fraction of its galaxies increasing the available values from $30\%$ to $90\%$ of the total sample. 

\subsection{HECATE-eRASS1 galaxy sample}\label{sec:HEC-eR1 sample}
By combining the HECATE v2.0 value-added information with the eRASS1 data we can study the relation between the X-ray emission of normal galaxies with their stellar population parameters (i.e. \sfr{}, \stellarmass{}, metallicity) for the largest galaxy sample so far. To that end, we measured the integrated X-ray flux (see Sect.~\ref{sec:x-ray photometry}) for all the HECATE galaxies within the eRASS1 footprint using the coordinates and the size of each galaxy. The provided galaxy coordinates are very accurate with astrometric precision ${<}1''$ for ${\sim}92\%$ of the galaxies and ${<}10''$ for the remaining objects. The completeness of HECATE in terms of galaxy angular size is $97.6\%$.
For ${\sim}80\%$ of the galaxies, the size information is based on the $D_{25}$ isophote in the $B$ band, while for the rest this information is obtained mainly from the \textit{2MASS} and \textit{SDSS}, as well as, other catalogues (e.g. \textit{2dFG},\textit{WINGS} etc.). All the supplementary sizes have been rescaled based on the reference semi-major axis provided by HyperLEDA. For a detailed description of the galaxy position and size information see \citet{hecate}. To avoid galaxies with problematic distances, due to misassociations with foreground stars, we removed 956 cases flagged as 'Star\_contamination = Y' (see Sect.~\ref{sec:galaxy_sample}). This results in an initial sample of 93806 HECATE galaxies which have been observed by eRASS1 (hereafter HEC-eR1) in the west hemisphere of the sky.

\subsection{Selection of bonafide star-forming galaxies}\label{sec:selection of bonafide sfg}
To select all the normal galaxies within the HEC-eR1 sample we utilized
the activity classifications provided by HECATE v2.0 (see Sect. \ref{sec:galaxy_sample}). 

First, we selected all galaxies within the HEC-eR1 sample classified as star-forming based on both the spectroscopic (SP) and the photometric (RF) classification ($\mathrm{SP}=0\,\&\,\mathrm{RF}=0$). Afterwards, we included in our sample also star-forming galaxies with available spectroscopic classification, but without available photometric classification ($\mathrm{SP}=0\,\&\,\mathrm{RF}\,\text{not available}$). Since we consider the spectroscopic classification as the ground truth, we also included star-forming galaxies with available spectroscopic classification and a contradicting photometric classification ($\mathrm{SP}=0\,\&\,\mathrm{RF}\neq0$). Finally in order to increase the completeness of our final sample we considered also galaxies for which there is no available spectroscopic classification but the photometric activity diagnostic tool classifies them as star-forming ($\mathrm{SP}\,\text{not available}\,\&\,\mathrm{RF}=0$). Given that the two methods predict the probability of a galaxy to belong in each activity class (\citealp[see;][]{stampoulis19,daoutis23}), we included in our final sample only the star-forming galaxies with a probability $P_{SFG}{>75\%}$. This probability threshold is well-calibrated for both methods ensuring the star-forming nature of these galaxies. This resulted in a final sample of 20392 star-forming galaxies within HEC-eR1 sample. 
Table \ref{tab:sample selection criteria} summarizes the construction of the star-forming galaxies HEC-eR1 sample based on the different criteria described above. The majority of the star-forming galaxies in our sample (14739/20392) have spectroscopic confirmation ($P_{SFG}{>75\%}$) while for the remaining 5653 outside the \textit{SDSS} footprint, the star-forming classification is robust given the very high performance of the photometric classifier (RF) on predicting star-forming galaxies based on their mid-IR and optical colors \citep{daoutis23}.  

To estimate the contamination in the HEC-eR1 galaxy sample, for which we do not have any spectroscopic information, by other galaxy types we performed the following analysis. First, we defined an initial sample for which we have photometric (RF) activity classifications and for comparison we also have spectroscopic (SP) activity classifications, without setting a probability threshold. This results in 16564 galaxies of all activity types. Afterward, by considering the spectroscopic classification as the ground truth, we calculated how many spectroscopically confirmed non-star-forming galaxies were classified as star-forming with $P_{SFG}{>75\%}$ by the photometric diagnostic. We found that only 212/16564 galaxies have been misclassified as star-forming by the photometric diagnostic. Based on this we consider that the photometric only classification of the star-forming galaxies in our final HEC-eR1 sample is robust with a false positive rate of ${\sim}1\%$.

\begin{table}[h]
\centering
\caption{Sample selection criteria for the star-forming galaxies within HEC-eR1 sample. We consider only star-forming galaxies with probability $P_{SFG}{>75\%}$.}\label{tab:sample selection criteria}
\begin{tabular}{cc}
\hline
  Criteria & Number of galaxies \\
\hline
SP = 0 \& RF = 0  & 7228 \\
SP = 0 \& RF not available  & 5441 \\
SP = 0 \& RF $\neq$ 0  & 2070 \\
SP not available \& RF = 0 & 5653\\
\hline
Total star-forming within HEC-eR1 & 20392\\
\hline
\end{tabular}
\tablefoot{ SP and RF refer to the spectroscopic classification and the photometric classification, respectively. 0 indicates all the galaxies classified as star-forming.}
\end{table}

\section{X-ray data analysis}\label{Sec:X-ray analysis}

\subsection{eRASS1 data reduction }\label{sec:x-ray data reduction}
Since we are interested in the integrated X-ray emission of the galaxies, we extracted the eRASS1 spectra and auxiliary files using version 1.72 of the
\texttt{srctool} task contained in version eSASSuser211214 of the eROSITA
Science Analysis Software System \citep[\texttt{eSASS},][]{brunner22_eSSAS} from event data of processing
version 010. The source regions were defined using the $D_{25}$ B-band isophote size of the galaxies which is defined as an ellipse with the semi-major axis (R1), the semi-minor axis (R2), and the position angle (PA) provided by the HECATE catalogue. For less than $10\%$ of the galaxies the extraction radius was smaller than the FoV average HEW of the \erosita{} PSF. We did not attempt to perform extractions for a larger aperture in order to avoid excess contamination by background emsission that might be present in the larger aperture. In any case, the aperture corrections for this small extraction region are included when we calculate the source flux through the \texttt{ARF}. For the HEC-eR1 galaxies without size information (PGC2807061, PGC24634), we considered a circular aperture with a radius of $1'$. The first galaxy is a low-mass dwarf irregular galaxy that contains a modest \ion{H}{II} region \citep{silva05} at a distance of 3.36 Mpc. The second galaxy is a typical SB galaxy at a distance of 22.65 Mpc. Visual inspection of their optical DSS images guarantees that we do not omit flux from the galaxy's main body by selecting this circular aperture. Background X-ray spectra were extracted from a circular annulus around the central position of each galaxy. We set the inner radius of this annulus to $60''$ if the semi-major axis of the respective galaxy is smaller, and $10''$ larger than the semi-major axis otherwise. For the outer radius we adopted values $150''$ larger than the inner radius. All detections listed in the main source catalogue of eRASS1 were excluded from the background regions. In order to avoid the removal of any potential X-ray emitters that belong to the galaxy (i.e. Ultra Luminous X-ray sources; ULXs) we did not mask any detected source into the source aperture. However, in Sect.~\ref{sec:bkg-AGN-contamination} we estimate the contamination from background AGN in HEC-eR1 sample based on a statistical approach. 

\subsection{X-ray source photometry}\label{sec:x-ray photometry}
In order to measure the observed counts we summed up the measured counts in the source and the background spectrum of each galaxy in the soft (S) 0.6-2.3 \energyunits{} and hard (H) 2.3-5.0 \energyunits{} \erosita{} bands. We used the \texttt{Sherpa v.4.15.1} package \citep{sherpa1,sherpa2,sherpa3}. Because more than $95\%$ of our sources have $\leq5$ counts above the background the source net counts could not be simply calculated by subtracting the estimated background counts from the source counts. Instead, we calculated the posterior distribution of the source counts given the observed counts in the source and background apertures, and the Bayesian formalism of \citet{vandyk01} as implemented in the \texttt{BEHR}\footnote{\texttt{Bayesian Estimation of Hardness Ratios}; \url{http://hea-www.harvard.edu/astrostat/behr/}} code \citep{park06}. 
\texttt{BEHR} accounts also for differences between the source and background effective areas, and exposure times.
The selection of the prior distribution is important since it may drive the shape of the posterior distribution in this low-count regime. After a sensitivity analysis for different assumptions for source and background counts and priors, we found that a non-informative prior in the log scale has the least effect on the posterior distribution for sources with a few observed counts. The posterior distribution was calculated by using the Gibbs sampler drawing 20000 draws (burn-in: 4000). In the subsequent analysis we use the full set of draws from the posterior count distribution.  

\subsection{X-ray fluxes, luminosities, and X-ray color}\label{sec:x-ray flux luminosity}
To calculate the X-ray fluxes in the 0.6--2.3 \energyunits{} \erosita{}'s most sensitive \citep{predehl} band, the posterior count distribution (calculated as described in Sect.~\ref{sec:x-ray photometry}) was multiplied by the count-rate to flux conversion factor of each galaxy in our sample. Although this conversion depends on the spectral model which describes better the X-ray spectrum \citep[e.g.][]{zezas06} the low number of counts of our sources does not allow us to fit each of them separately. Since the Ancillary Response File (\texttt{ARF}) and the Redistribution Matrix Function (\texttt{RMF}) are different for each sky location we calculated the count-rate-to-flux conversion factor for each galaxy by convolving the average spectrum of the HEC-eR1 galaxies (see Sect.~\ref{sec:average_spectrum}) with the \texttt{ARF} and \texttt{RMF} file for each galaxy. The spectral model consists of a power-law component ($\Gamma = 1.75$), a thermal component (APEC; $\rm{kT}= 0.70$ keV, $\rm{Z}=\rm{Z_{\odot}}$) which contributes ${\sim}13\%$ of the total flux (see Sect.~\ref{sec:hot_gas_contribution}). Both are absorbed by cold gas modeled with the Tuebingen-Boulder (tbabs) ISM absorption model \citep{wilms00}. The neutral hydrogen column density, \coldens{} is calculated independently for each galaxy using the survey of \citet{dickey90} and the \texttt{colden} tool which is part of \texttt{proposaltools} provided by \texttt{CIAO} \citep{ciao06}. 

Throughout this paper, we use the soft 0.5--2.0 \energyunits{} X-ray bands for comparison with other works. The flux and the corresponding luminosity in this band were calculated by converting the measured flux in the 0.6--2.3 \energyunits{} \erosita{} band assuming the above-mentioned absorbed power-law + APEC model. For simplicity, and to facilitate future comparisons with other works we adopted the same absorbing \coldens{} for all galaxies, which is the median $\coldens{}=3.06\times 10^{20} \, \mathrm{cm^{-2}}$ of the star-forming galaxies in the HER-eR1 sample, given that the dispersion of the conversion factors (using the individual \coldens{}) is less than 2$\%$. The conversion factor from our analysis band (0.6--2.3 \energyunits{}) to the adopted soft (0.5--2 \energyunits{}) is, $c_{1}=\frac{F_{0.5-2}}{F_{0.6-2.3}} = 0.97$. For comparison with other works the conversion factor from the 0.6--2.3 \energyunits{} to the full 0.5--8 \energyunits{} band, assuming the same model is $c_{2}=\frac{F_{0.5-8}}{F_{0.6-2.3}} = 2.30$, respectively. When needed, we convert luminosities from other works to our adopted band based on the spectral models used in each corresponding study. All the conversion factors used throughout this work are presented in Table \ref{tab:conversion factors}. 

\begin{table*}[]
    \centering
    \caption{The conversion factors used throughout this work. Each of them was calculated by using the formula $c_{i}=\frac{F_{\text{adopted band}}}{F_{\text{initial band}}}$ where the adopted bands are the ones used in this work 0.5-2 \energyunits{}, and 0.5-8 \energyunits{}. The initial bands are the bands used from other works. }
    \label{tab:conversion factors}
    \begin{tabular}{ccccccc} \hline\hline 
    Initial band& \multicolumn{2}{c}{Adopted band} & Spectral model& Values& Reference &\\ 
         & 0.5-2 &0.5-8 & & & & \\
         \hline
         0.6-2.3&c$_{1}$ = 0.97 & c$_{2}$ = 2.30 &abs$\times$(pow+apec)&N$_{H}=0.03$, $\Gamma=1.75$, $\rm{kT}=0.70\,\energyunits{}$& This work\\
         \hline
         0.5-8& c$_{3}$ = 0.45& - & abs$\times$(pow+apec) &$\coldens{}=0.20$, $\Gamma=1.7$, $\rm{kT}=0.67\,\energyunits{}$& \citealp{mineo14}\\
         0.5-8& c$_{4}$ = 0.29& - &abs$\times$(pow)  &$\coldens{}=0.19$, $\Gamma=1.7$  & \citealp{lehmer19}\\  
         0.5-8& c$_{5}$ = 0.47& - & abs$\times$(pow) &$\coldens{}=0.034$, $\Gamma=2$  & \citealp{brorby16}\\
         2-10& c$_{6}$ = 0.64& - & abs$\times$(pow) &$\coldens{}=0.10$, $\Gamma=2$  & \citealp{fornasini20}\\
         0.5-8& c$_{7}$ = 0.37& - & abs$\times$(pow) &$\coldens{}=0.037$, $\Gamma=1.7$ &\citealp{lehmer21} \\
         0.5-8& c$_{8}$ = 0.32& - & abs$\times$(pow) &$\coldens{}=0.3$, $\Gamma=2$ &\citealp{mineo12a}\\
         0.2-2.3&c$_{9}$ = 0.67& - & abs$\times$(pow) &$\coldens{}=0.01$, $\Gamma=1.7$ &\citealp{riccio23}\\ \hline
         
    \end{tabular}
\tablefoot{ The $\coldens{}$ values are in units of $10^{22}\,cm^{-2}$.}
    
\end{table*}

In addition, we also calculated the X-ray color of the galaxies in our sample which is defined as $C = log_{10}(\frac{S}{H})$, where S and H are the source counts in the two bands. Given that our galaxies have only a few counts we calculated the posterior distribution of the X-ray color using again the \texttt{BEHR} code \citep{park06}. In addition, during the calculation we took into account the variations in the ancillary response files of each galaxy due to their different position on the detector.   

\subsection{Reliability of the X-ray flux measurements}\label{sec:flux reliability}
\begin{figure}
        \includegraphics[width=\columnwidth]{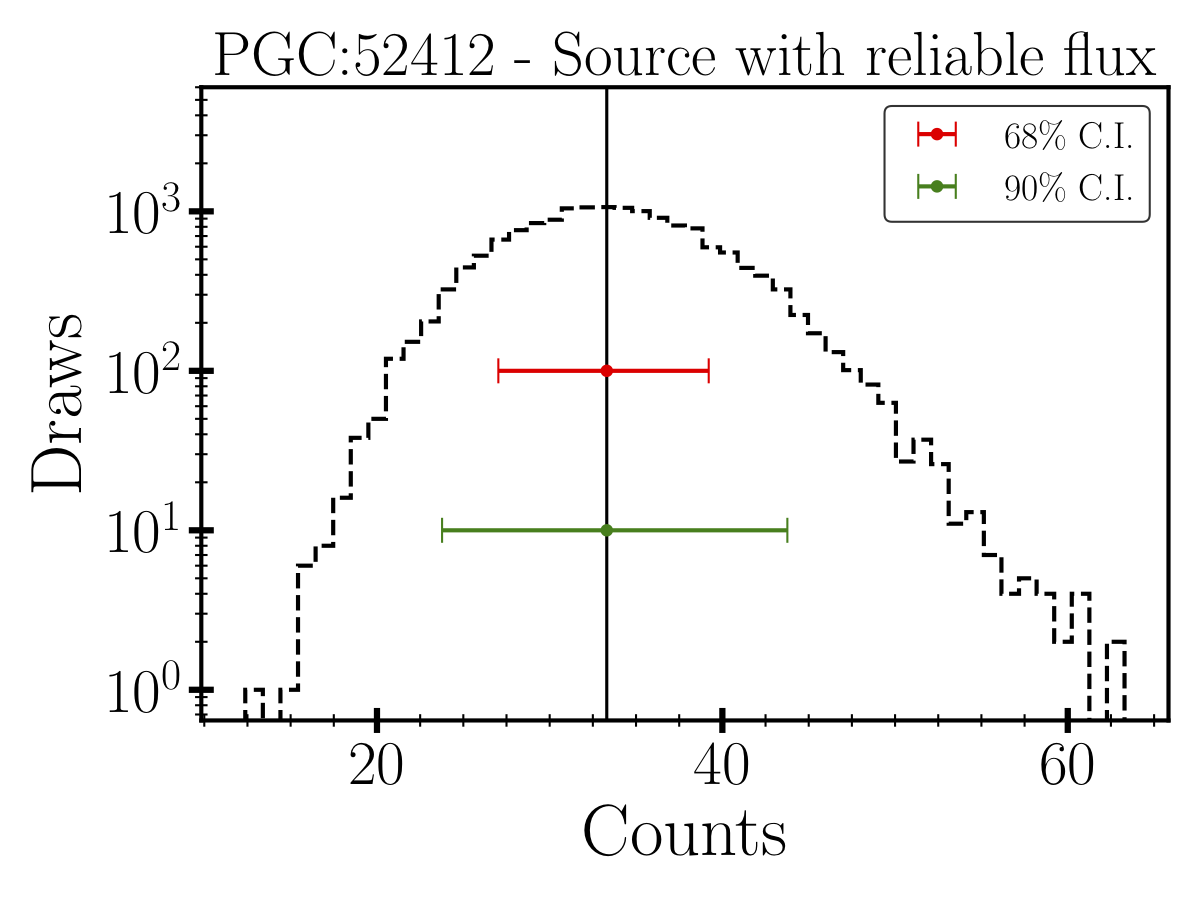}
        \includegraphics[width=\columnwidth]{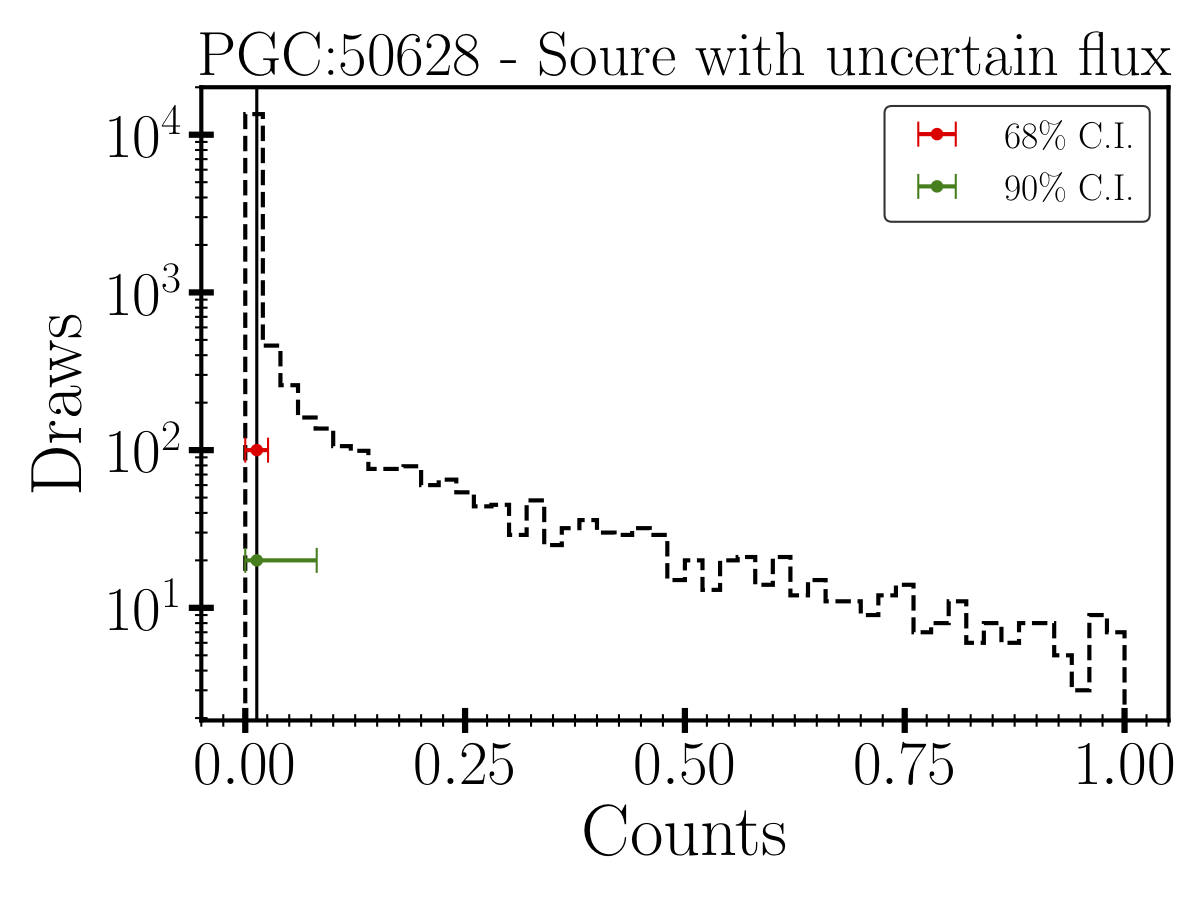}
    \caption{An example of the posterior count distribution of two sources in our sample for which we have reliable and unreliable flux measurement as defined based on the Eq. \ref{equat:reliable_formula}. The red and green errorbars indicate the upper and lower bounds of the distribution at the $68\%$ and the $90\%$ C.I., respectively. The vertical black solid line indicates the mode value.}
    \label{Fig:posterior_examples}
\end{figure}

\begin{figure}
        \includegraphics[width=\columnwidth]{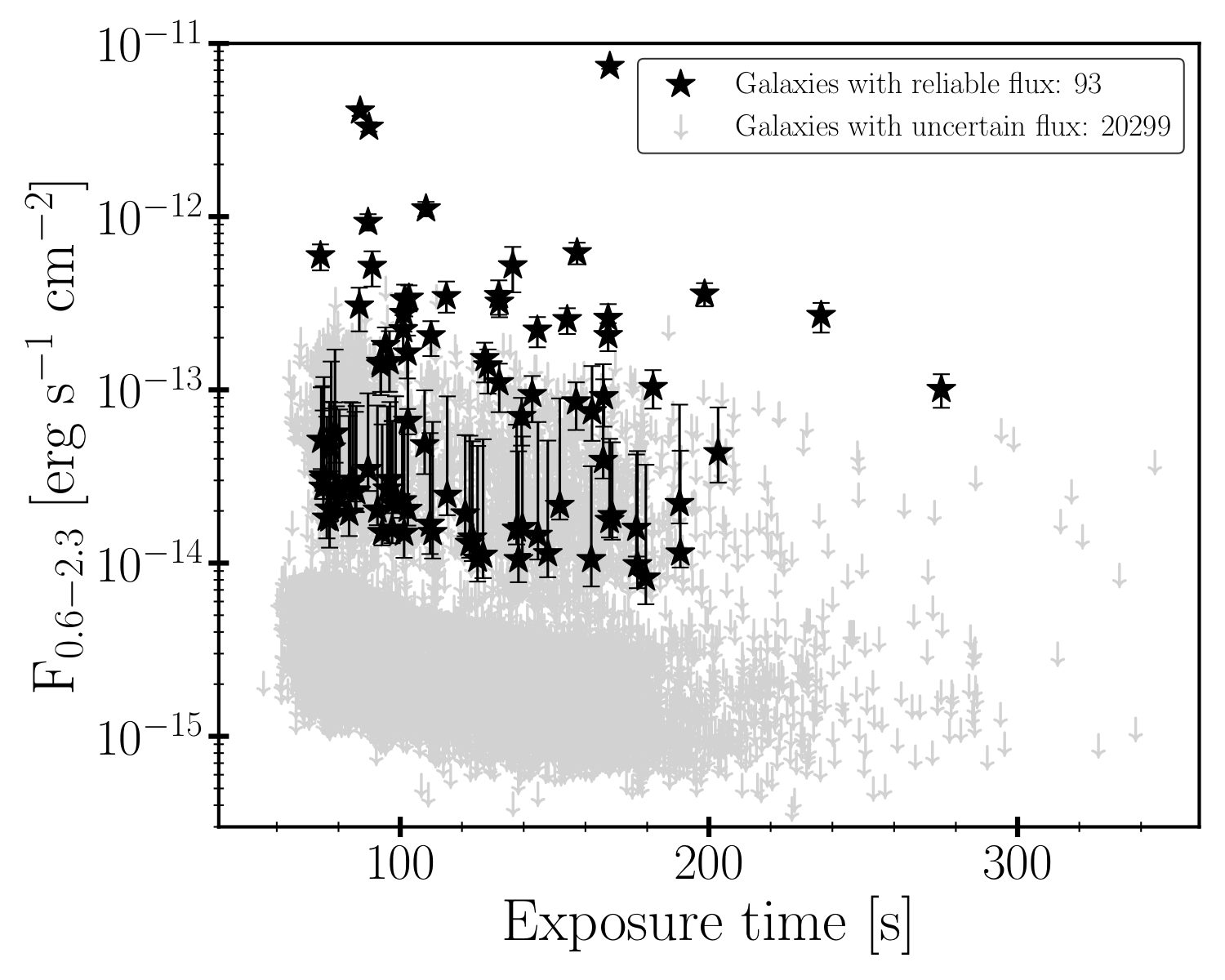}
    \caption{The flux distribution in the 0.6-2.3 \energyunits{} band of the star-forming galaxies in the HEC-eR1 sample as a function of the exposure time. Black stars show the mode value of the source's intensity for the galaxies with reliable fluxes and their 68$\%$ C.I. uncertainties. Gray down arrows show the upper 90$\%$ C.I. for the galaxies with uncertain flux measurements.}
    \label{Fig:Fx_exp_time}
\end{figure}

Although our approach of calculating the full posterior distribution of the source flux can deal with very faint sources, there are cases of extremely low-count sources where their posterior count distribution is very wide and positively skewed with a mode very close to zero. That means that the posterior intensity of such sources is almost zero not allowing us to handle them as point measurements given their very large uncertainties. In order to estimate which galaxies in our sample have reliable flux measurements we utilized the posterior distribution of the source counts in the 0.6--2.3 \energyunits{} \erosita{} band and we calculated its mode value $S_{mode}$ and its lower $S_{low,68\%}$, and upper bound $S_{up,68\%}$ at the 68$\%$ confidence interval (C.I.). Based on the shape of the posterior we considered as reliable measurements all the galaxies for which:
\begin{equation}\label{equat:reliable_formula}
\frac{S_{mode}}{S_{mode}-S_{low,68\%}} \geq 3\,
.\end{equation} 
Flux measurements for which this criterion is not satisfied are considered as uncertain. 
In Fig.~\ref{Fig:posterior_examples} we present two examples of sources in our sample with reliable flux measurement and with an uncertain flux measurement. The red and green errorbars indicate the upper and lower bounds of the distribution at the $68\%$ and the $90\%$ confidence interval (C.I.), respectively. The vertical black solid line indicates the mode value of the distribution. The posterior distribution of the reliable source (top panel) is more symmetric and its mode value is ${\gtrsim} 30$ counts. On the other hand, the posterior distribution of the source with uncertain flux measurement is strongly positively skewed with a mode of almost zero. This example demonstrates that our criterion of assessing the reliability of the flux measurement for a source (Eq.~\ref{equat:reliable_formula}) is robust and can characterize well the galaxies in our sample. 
In this way, we found 93 secure star-forming galaxies with reliable flux point estimated while the remaining 20299 have uncertain flux point estimates. For visualization purposes, throughout this paper, we use as a point estimate the mode of the flux distribution for the reliable galaxies and the error bars correspond to the 68$\%$ C.I. On the other hand, we adopt the upper 90$\%$ C.I. of the flux distribution for the galaxies with uncertain flux measurements. However, when performing the maximum-likelihood fit for the scaling relations (see Sect. \ref{sec:max_like_fit_scaling_relations}), we consider the complete posterior flux distribution for each galaxy (reliable and uncertain) rather than relying on a single value measurement.  
In Fig. \ref{Fig:Fx_exp_time} we show the distribution of our reliably measured fluxes (black stars) and the galaxies with uncertain fluxes (gray down arrows) in the 0.6--2.3 \energyunits{} band as a function of the exposure time. The reliable flux measurements are spanning a range from 10$^{-11}$ \fluxunits{} to $\sim 10^{-14}$ \fluxunits{}, while the exposure time for all the galaxies in our sample is in the order of a few hundred seconds. We do not see a trend for longer exposures to be associated with more reliable flux measurements. This is because the posterior distribution depends on the local background and the size of the extraction aperture. 

\begin{figure}
        \includegraphics[width=\columnwidth]{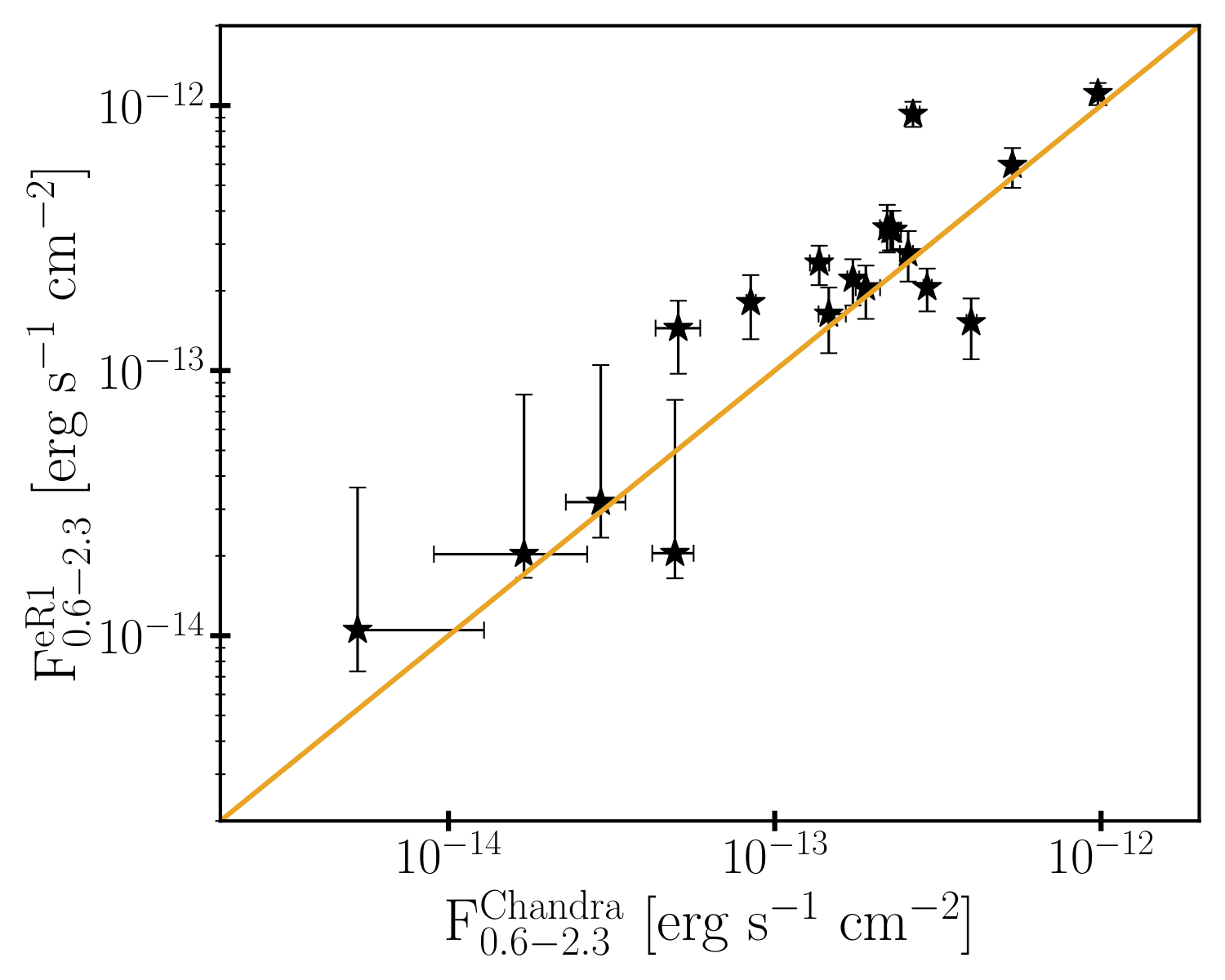}
    \caption{Comparison between the X-ray fluxes as they measured from the eRASS1 extractions (\S\ref{sec:x-ray flux luminosity}) and those derived from the existing \chandra\ observations. The orange line corresponds to the line of equality. We find excellent agreement supporting the robustness of the eRASS1 extractions.}
    \label{Fig:Fx_eR1_chandra_comp}
\end{figure}

In order to assess the robustness of these flux measurements, we cross matched the sample of HEC-eRASS1 galaxies with reliable X-ray fluxes with the \chandra\ and \xmm\ observation catalogues. We found 16 galaxies with available observations that could be used to measure the integrated X-ray emission of the galaxies. In all cases we used the \chandra\ data because of the lower background, resulting in higher SNR measurements for the entire galaxy. The analysis was performed with the Ciao v4.15 software package. After the application of standard screening criteria, we measured the count rate in the 0.6--2.3\,keV band within the galaxy aperture using the same apertures as in the eRASS1 analysis. The background regions were adjusted to fit within the limited field of the observations. The net source counts were calculated using the \texttt{BEHR} code with the same parameters as for the eRASS analysis (Sect.~\ref{sec:x-ray photometry}). For each observation we also calculated the count-rate to flux conversion factors based on \texttt{ARFs} derived for each observation individually, and the average spectrum described in Sect.~\ref{sec:average_spectrum}. In Fig. \ref{Fig:Fx_eR1_chandra_comp} we show a comparison between the eRASS1 fluxes (Sect.~\ref{sec:x-ray flux luminosity} and those derived from the existing \chandra\ observations. We find excellent agreement supporting the robustness of the eRASS1 extractions.

\section{Final sample of star-forming galaxies}\label{sec:final sample}
\subsection{Stellar population properties for the HEC-eR1 secure star-forming galaxies}\label{sec:sfr mstar cuts}
In order to study the correlation between the integrated X-ray emission and the stellar population parameters of the host galaxy, we need accurate and realistic estimations of the \sfr{}, \stellarmass{}, and metallicity of the galaxies in our sample. For that reason, we utilized the values provided by the HECATE v2.0 catalogue (see Sect.~\ref{sec:galaxy_sample}). In particular, we selected only the star-forming galaxies within the HEC-eR1 sample with $\sfr{} > 10^{-3} \, \rm{M_{\odot}}\,\rm{yr^{-1}}$ because for lower values the incomplete sampling of the stellar initial mass function (IMF) will lead to large fluctuations \citep[e.g.][]{kennicutt12}. We also opted to use only galaxies with $M_{\star}>10^{7} \,\rm{M_{\odot}}$ because lower estimations can be unrealistic indicating that they most likely were derived from unreliable photometry, and suffer from aperture effects. These cuts result in 18894 secure star-forming galaxies (91 with reliable X-ray flux measurements, and 18803 with uncertain X-ray flux measurements). 

For the majority of the star-forming HEC-eR1 galaxies the \sfr{}s and \stellarmass{}s were based on the calibrations of \citet{sfr_stellar_mass_kouroumpatzakis} ($58\%$ or 10960/18994, and $62.8\%$ or 11859/18894, respectively). For $27\%$ of the galaxies, the \sfr{}s are based on the calibrations of \citet{cluver17} and for the remaining $15\%$ the \sfr{} values from \citet{salim14} and \citet{leroy19} were adopted ($14.8 \%$ and $0.2 \%$, respectively). On the other hand, the \stellarmass{}s for the rest $37.1\%$ (7019/18894) are based on the calibrations of \citet{wen13} and only for a $0.1\%$ (16/18894) the \stellarmass{}s proposed by \citet{leroy19} were adopted. The use of these different calibrators is dictated by the available photometric data for each galaxy. 
As discussed in \citet{sfr_stellar_mass_kouroumpatzakis} and Kyritsis et al. 2024 there is a systematic offset of about 0.2\,dex between the two main calibrations we use for the \sfr{} and \stellarmass{} calculations, which is consistent with the scatter of these calibrators.
Finally, the vast majority of the gas-phase metallicities in our sample ($70\%$ or 13206/18894 ) were derived using the PPO3N4 spectroscopic method \citep{kewley_elison08} while for the rest of the galaxies ($30\%$ or 5688/18894) are based on the mass-metallicity relation for the same metallicity calibration (see Sect.~\ref{sec:galaxy_sample}). The systematic offset between the two metallicity calibrations for the spectroscopically confirmed star-forming galaxies in our sample is less than ${\sim} 0.1 \, \rm{dex}$ which does not introduce additional scatter on the metallicity estimations due to different calibrations.

\subsection{AGN screening}\label{sec:agn screening}
In Sect. \ref{sec:selection of bonafide sfg} we selected all the star-forming galaxies in the HEC-eR1 sample based on stringent activity classification criteria. Nevertheless, the measured X-ray flux can be still contaminated by photons originating from a background AGN that is in projection with the galaxy. To exclude such cases from our final sample of secure star-forming galaxies, we cross-matched them with the all-sky AGN catalogue from \citet{zaw19}. This catalogue includes spectra from the \textit{SDSS} and \textit{6dF} Galaxy surveys \citep[\textit{6dF};][]{jones04,jones09}, supplemented by private spectra obtained with the long-slit FAst Spectrograph for the Tillinghast telescope \citep[][FAST]{fabricant98} and the R-C grating spectrograph on the V.M. Blanco 4.0 m telescope at the Cerro Tololo Inter-American Observatory (CTIO). This makes it one of the largest, all-sky spectroscopic AGN catalogue in the nearby universe (z${<}$ 0.09). The sample consists of 1929 broad-line AGNs and 6562 narrow-line AGNs. 
However, we notice here that this catalogue is based on lower resolution spectra than the \textit{SDSS} spectra used in the HECATE catalogue. For that reason, we did not crossmatch the entire HECATE with this AGN catalogue. Instead, we handled it carefully after a detailed inspection of its provided classifications. The classification of the AGNs is based on emission line ratio measurements using the standard two-dimensional diagnostics from \citet{kewley01} and \citet{kauffmann03}. The crossmatch returned 248 matches. For consistency with the spectroscopic classifications in HECATE, and given that the 4-D diagnostics of \citet{stampoulis19} are more robust than the 2-D projections used in \citet{zaw19}, we removed only galaxies for which the two activity classifications agreed. For the classification, we used the same emission line quality cuts as the ones applied in the HECATE catalogue. We found that 41 out of 248 galaxies were also classified as AGNs from our diagnostic and we removed them from our final sample. More than $90\%$ of the remaining galaxies were classified either as star-forming or composites which are closer to the star-forming locus of the emission-line diagnostic diagram. 

To further ensure the star-forming nature of our galaxies, we visually inspected the X-ray/optical images and optical spectra (where available) of all the 91 star-forming HEC-eR1 galaxies with reliable flux measurements. We removed the galaxies PGC4125797 and PGC78895, 
since their X-ray and optical images showed that their X-ray emission is coincident with a spectroscopically confirmed background AGN.
In addition, we excluded two galaxies (PGC28415, PGC36648) for which the inspection of their optical spectra indicates a clearly broad-line AGN. We also cross-matched all the 91 star-forming HEC-eR1 galaxies with reliable flux measurements with the NED database to investigate for additional evidence for AGN activity. We removed two galaxies (PGC36126, PGC1809432) which belong to the work of \citet{liu19} who presented a comprehensive and uniform sample of broad-line AGN using spectra from the \textit{SDSS DR7}. Furthermore, we discarded the galaxy PGC38964 which is characterized as a Compton-thick AGN in the work of \citet{tanimoto22} who performed a systematic broadband X-ray spectral analysis of 52 Compton-thick AGN (CTAGN) candidates selected by the Swift/Burst Alert Telescope all-sky hard X-ray survey \citep{ricci15}. 

Given the very large number of star-forming galaxies with uncertain flux measurements (20299) it was not possible to inspect the images/spectra and bibliography for all of them. Instead, we randomly selected a subsample of 50 galaxies and we repeated the visual inspection and the literature search only for them. We removed the galaxy PGC2082767, since it is classified as a Seyfert galaxy based on its emission line measurements from \citet{hopp00}. In addition, although its optical spectrum indicates a star-forming galaxy, we did not include PGC1139676, a spectroscopically confirmed background AGN, which could potentially contaminate the measured X-ray emission projected on the body of the galaxy. This analysis suggests that the AGN contamination of the overall population of the star-forming galaxies with uncertain flux measurements (20299) is $\sim 4\%$. This is of the same order as the false positive rate estimated in the Sect.~\ref{sec:selection of bonafide sfg}. 

In addition, another source of potential contamination of the measured X-ray flux could be a time-variable (or transient) nuclear accretion episode not accounted for in the spectroscopic or photometric classification. This possibility is discussed in Sect.~\ref{Sec:low-AGN or TDE}.

So far, we screened our sample for all the galaxies associated with an AGN based on the visual inspection of their X-ray/optical images, and an extensive literature search. However, heavily obscured AGNs or unobscured low luminosity AGN \citep{merloni14} can be still contaminating the measured X-ray emission. To find such cases, we compared the calculated X-ray color $C$ (see Sect.~\ref{sec:x-ray flux luminosity}) of our secure star-forming galaxy sample with the expected X-ray color from a heavily obscured AGN, assuming an absorbed power-law model with photon index values $\Gamma$ = 0, or 1 and a typical $\coldens{}= 10^{20} \, \mathrm{cm^{-2}}$. This analysis resulted in the removal of 2 galaxies (PGC55410, PGC52042) with colors consistent with a very hard spectrum similar to those of an obscured AGN. In the end, based on our screening process we rejected 52 potentially AGN-contaminated galaxies.

X-ray variability is a tell-tale signature of AGN activity. The recent study of \citet{arcodia23} identified a sample of AGN based on X-ray variability detected in the eRASS data. A cross-correlation of our sample of star-forming galaxies with the variable sources in this analysis did not yield any matches, indicating that there are no strongly variable known AGN within our sample. 

\subsection{Screening for star-forming galaxies with galaxy cluster associations}\label{sec:clusters assoc}
Another known contaminant that could affect the measured X-ray fluxes of normal galaxies is the association of our star-forming galaxies with foreground or background galaxy clusters. For that reason, we cross-matched the HEC-eR1 sample with the eRASS1 galaxy cluster catalogue (Bulbul et al. 2023, submitted). For the crossmatch we used the \texttt{Sky with Errors} matching algorithm provided by \texttt{TOPCAT} \citep{taylor05}. Given the positions of the HEC-eR1 galaxies and the positions of the detected X-ray clusters we searched for matches using as positional errors the galaxy's semi-major axis (R1) and two times the radius of the detected cluster, respectively. In this way, we ensured that each galaxy is within the extent of the cluster. This resulted in 415 matches 12 of which are star-forming HEC-eR1 galaxies, and for that reason we removed them from our final sample. Given that the eRASS1 is a flux-limited survey, it is possible to not detect the soft X-ray emission of extended nearby clusters (e.g. Virgo, Fornax, Coma etc.). Since this may contaminate the signal in the stacked data (Sect.~\ref{sec:SFR-M-D stacks}) and the average X-ray spectrum of the star-forming galaxies (Sect.~\ref{sec:average_spectrum}) we also cross-matched the HEC-eR1 galaxy sample with the Abell and Zwicky Clusters of Galaxies catalogue \citep{abell74}. By using again the same matching algorithm (\texttt{Sky with Errors}) and as positional errors the R1 and the radius of the cluster, we removed 30 secure star-forming galaxies positionally coincident with clusters. X-ray emission from compact galaxy groups may also contaminate the X-ray emission we measured if the galaxy coincides with the extended X-ray emission from a galaxy group. For this reason we cross-matched the HEC-eR1 galaxies with the compact group catalogue of \citet{hickson82}. Using \texttt{Sky with Errors} and as positional errors the galaxy's R1 and the radius of the group, we removed 10 secure star-forming galaxies associated with compact groups of galaxies. Following the screening process described above, we removed 52 star-forming galaxies from our sample which were associated either with galaxy clusters or with compact groups of galaxies. In Table~\ref{tab:screening cuts} we summarize all the steps we followed for the construction of the final HEC-eR1 sample of star-forming galaxies.

\begin{table*}[]
\centering
\caption{Sample size in each step of the screening process for the construction of the final HEC-eR1 sample of star-forming galaxies.}
\label{tab:screening cuts}
\begin{tabular}{cc}
\hline\hline
Steps & Number of galaxies (removed objects)\\\hline
Initial HEC-eR1 galaxy sample (Sect.~\ref{sec:HEC-eR1 sample}) & 93806 \\
Star-forming based on spectroscopic and photometric activity diagnostics (Sect.~\ref{sec:selection of bonafide sfg})  & 20392\\ 
\sfr{} and \stellarmass{} cuts (Sect.~\ref{sec:sfr mstar cuts}) & 18894 (1498) \\
Additional AGN screening (Sect.~\ref{sec:agn screening}) & 18853 (41)\\
Visual inspection of optical/X-ray images and literature search (Sect.~\ref{sec:agn screening})& 18844 (9)\\
Hard X-ray colour cuts (Sect.~\ref{sec:agn screening}) & 18842 (2)\\
Galaxy clusters and compact group associations (Sect.~\ref{sec:clusters assoc}) & 18790 (52)\\
\hline
Final HEC-eR1 sample of star-forming galaxies (Sect.~\ref{sec:final sample of sfg}) & 18790\\
\hline
\end{tabular}
\tablefoot{ Numbers in parenthesis indicate the number of galaxies removed in each step of the screening process.}
\end{table*}

\subsection{HECATE - eRASS1 final sample of star-forming galaxies}\label{sec:final sample of sfg}
\begin{figure}
        \includegraphics[width=\columnwidth]{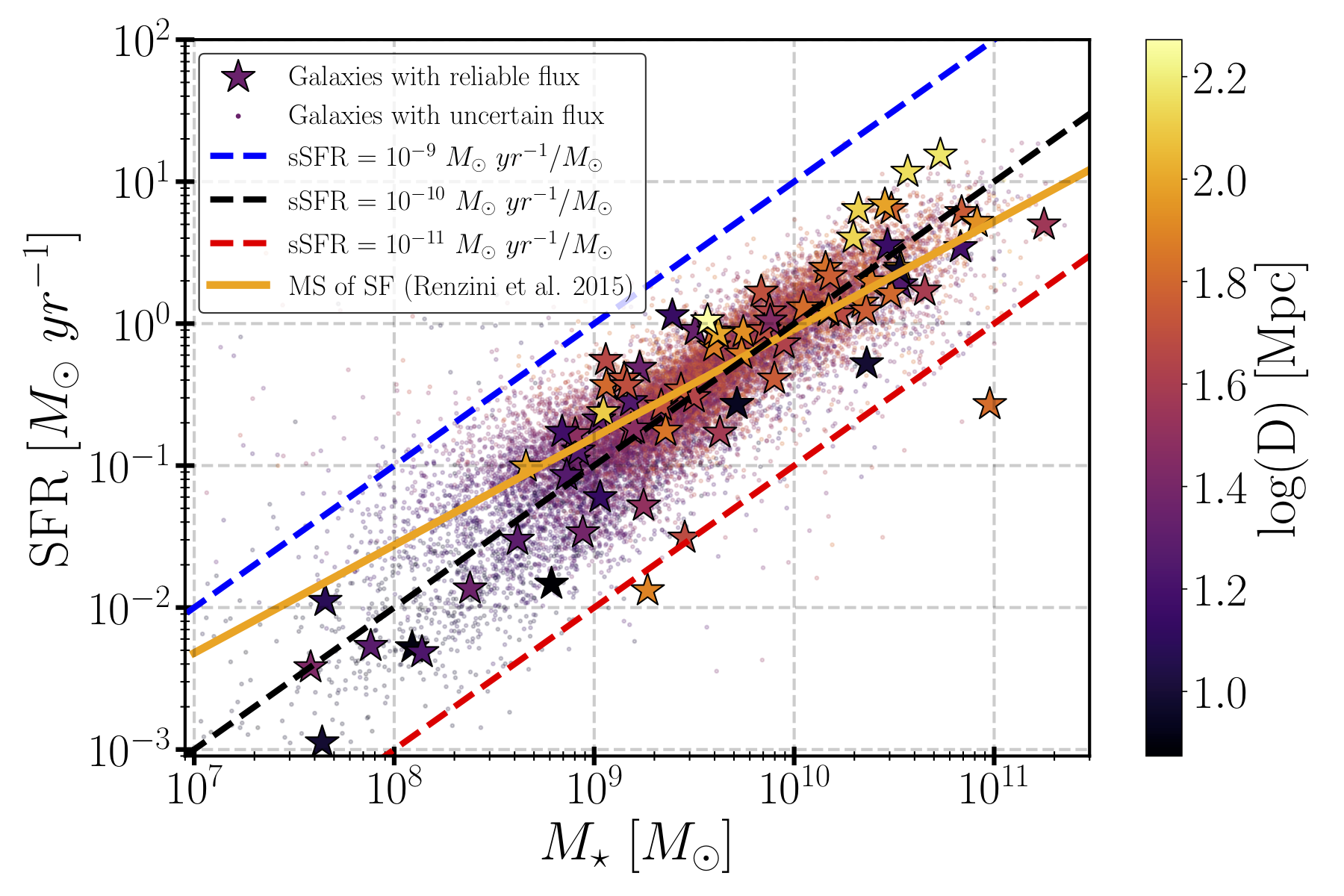}
         \includegraphics[width=\columnwidth]{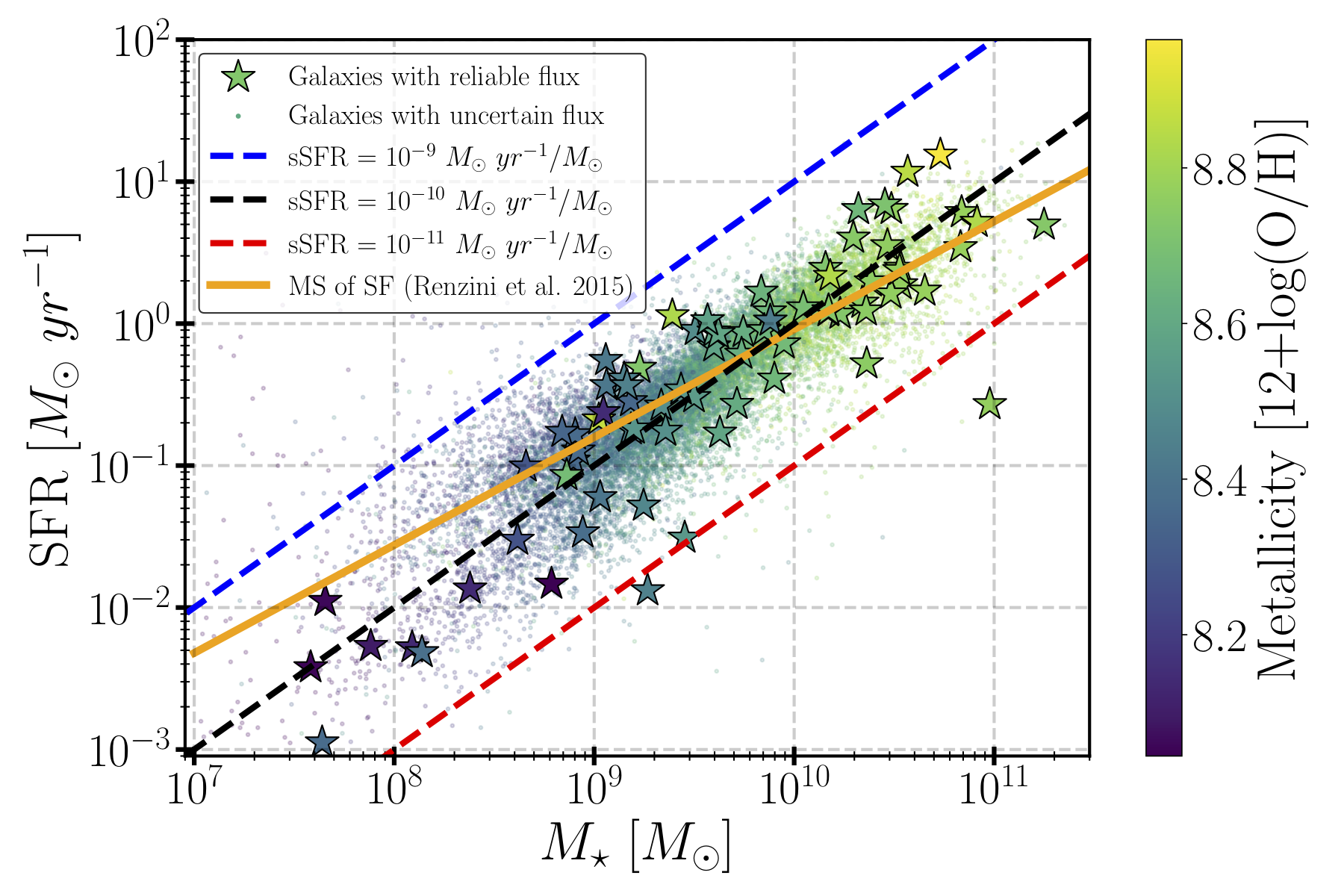}
    \caption{The distribution of the final clean sample of the secure star-forming HEC-eR1 galaxies in the main sequence plane. In both panels, stars indicate the galaxies with reliable X-ray flux measurements, and the dots the galaxies with uncertain fluxes. The color code in the top panel corresponds to the log(\distance{}), while in the bottom panel corresponds to the gas-phase metallicity, 12+log(O/H). The diagonal dashed lines indicate three different sSFR values (i.e. sSFR: $10^{-9}, 10^{-10},$ and $10^{-11}$ $\rm{M_{\odot}}\,\rm{yr^{-1}}/\rm{M_{\odot}}$).  The orange solid line shows the main sequence of the star-forming galaxies from \citet{renzini15}. Our final sample is well distributed along the \sfr{}-\stellarmass{} plane covering a wide range of distances and gas-phase metallicities.}
    \label{Fig:Parameter_space}
\end{figure}

Our final clean sample of secure star-forming HEC-eR1 galaxies is comprised of 18790 galaxies out of which 77 have reliable X-ray flux measurements and the remaining 18713 have uncertain fluxes. Despite that the majority of the galaxies that we removed during the AGN and galaxy cluster screening were characterized as uncertain X-ray flux measurements, we followed this approach in order to avoid any contamination during the stacking of the X-ray spectra (see Sect.~\ref{sec:SFR-M-D stacks}). In addition, although the number of galaxies with uncertain X-ray fluxes dominates our final sample, we include them in our analysis in order to avoid biasing our results. In fact, our methodology for the fitting of the scaling relations utilizes the posterior distribution of the X-ray luminosity of each galaxy (as it was derived from the flux posterior distribution and the corresponding distance; see Sect.~\ref{sec:x-ray flux luminosity}) instead of the point estimation of the luminosity (see Sect.~\ref{sec:max_like_fit_scaling_relations}). 
In Fig. \ref{Fig:Parameter_space} we present the distribution of the final HEC-eR1 sample of secure star-forming galaxies in the \sfr{}-\stellarmass{}-\distance{}-metallicity parameter space. In both panels, stars indicate the galaxies with reliable X-ray flux measurements, and the dots the galaxies with uncertain fluxes. The color code in the top panel depicts the logarithm of the distance of each galaxy while in the bottom panel the gas-phase metallicity, 12+log(O/H). The diagonal dashed lines indicate three different sSFR values (i.e. sSFR: $10^{-9}, 10^{-10},$ and $10^{-11}$ $\rm{M_{\odot}}\,\rm{yr^{-1}}/\rm{M_{\odot}}$). The orange solid line shows the main sequence of the star-forming galaxies from \citealp{renzini15}. Our sample is well distributed in the \sfr{}-\stellarmass{} plane (i.e. main sequence plane) spanning a range of \sfr{} $= 10^{-3}-25 \, \rm{M_{\odot}}\,\rm{yr^{-1}}$ and a range of \stellarmass{} $= 10^{7}- 5\times10^{11} \, \rm{M_{\odot}}$ from dwarf star-forming galaxies to large starburst galaxies. All the galaxies are symmetrically distributed around a sSFR of  $10^{-10} \rm{M_{\odot}}\,\rm{yr^{-1}}/\rm{M_{\odot}}$ while the reason that we find slightly more nearby galaxies in lower sSFRs is because the quality of our photometric data do not allow us to accurately calculate the \sfr{} and the \stellarmass{} of very distant galaxies. In addition, our sample covers a relatively wide range of metallicities from 12+log(O/H)$=$8.0 sub-solar to 12+log(O/H)$=$9.0 super-solar.
The distribution of our final sample in the \sfr{}-\stellarmass{}-\distance{}-Metallicity parameter space allows us to study the correlation between the widest range of host galaxy properties and the integrated \lx{} for the first unbiased X-ray sample of star-forming galaxies in the local Universe. In Table~\ref{Tab:Fluxes} we present the properties of the HEC-eR1 sample of star-forming galaxies with reliable flux measurements. 

\section{X-ray stacking analysis}\label{sec:x-ray stacking analysis}
\subsection{Average X-ray spectrum of the galaxy sample}\label{sec:average_spectrum}
\begin{figure}
        \includegraphics[width=\columnwidth]{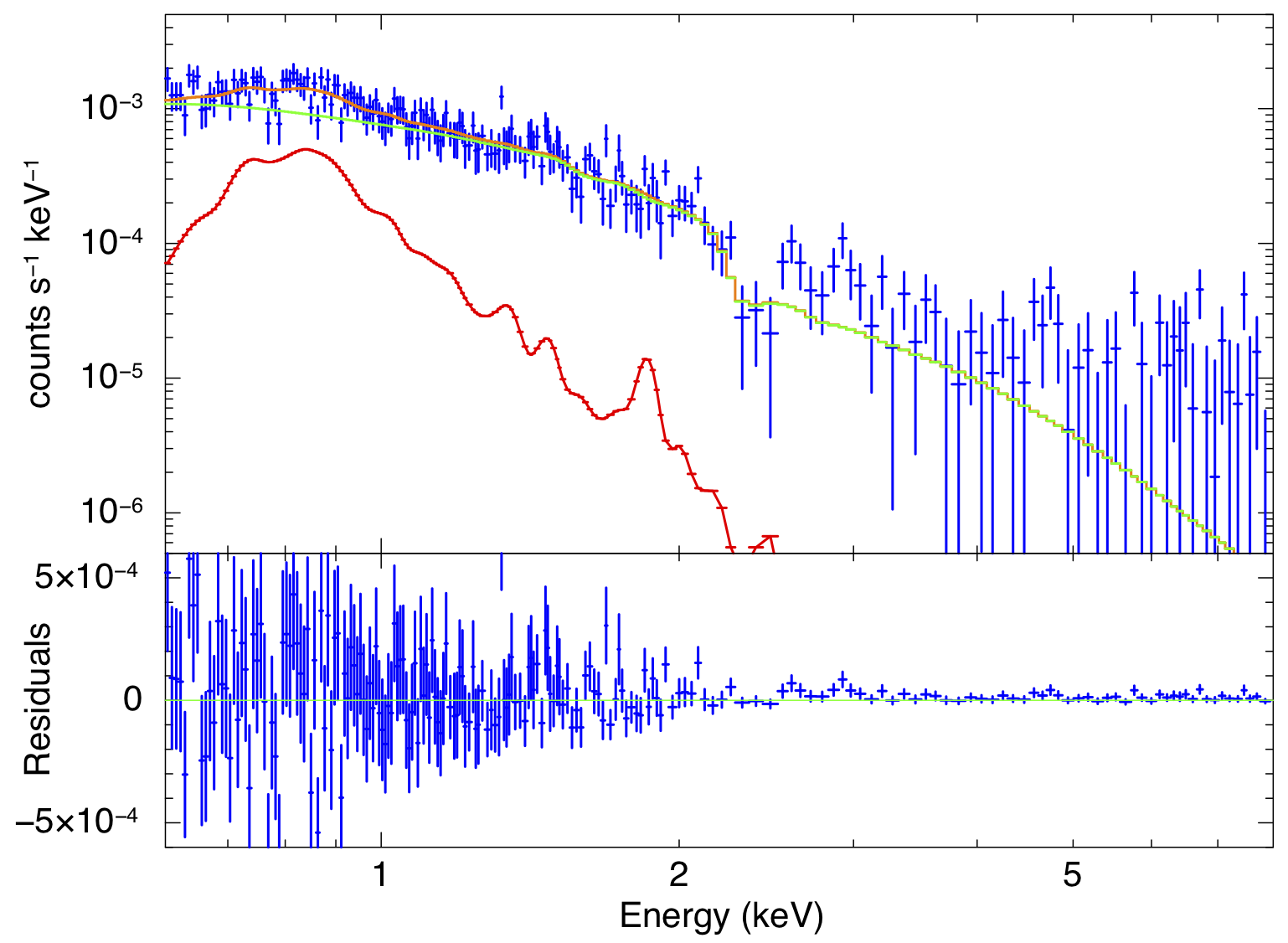}
    \caption{The stacked X-ray spectrum of all the secure star-forming galaxies within the HEC-eR1 sample. The green and red lines show the power-law and hot plasma model components, respectively. The orange line corresponds to the best-fit model.}
    \label{Fig:master_stacked_spectrum_model_2}
\end{figure}

As it is seen in Fig. \ref{Fig:Fx_exp_time}, eRASS1 is a relatively shallow survey with an average exposure time of the order of a few hundred seconds. This means that for the vast majority of the galaxies in our sample, the low number of observed counts does not allow us to calculate a count-rate to flux conversion factor based on the X-ray spectrum of each galaxy.
Instead, we stacked the X-ray spectra of all the 18790 star-forming galaxies within our final HEC-eR1 sample. In this way, we can study the average X-ray properties of the galaxy population by using their combined X-ray spectrum (which is representative of all the galaxies in our sample and it has a much higher S/N). Using the \texttt{combine\_spectra} command provided by the \texttt{Sherpa v.4.15.1} package and by setting the parameter \texttt{method='sum'} we summed all the source spectra, the associated response files (\texttt{ARF} and \texttt{RMF}), and the background spectra. The output combined spectrum contains the sum of all source counts. To combine the background counts and to compute the background to source area scaling (backscal) values, we set the parameter \texttt{bscale\_method='time'}. This method computes the total unweighted counts for the source and background spectra and provides the exposure time-weighted background scaling value which corresponds to the fraction of the background counts included in the source spectrum. In addition, the combined response files and the combined background are also provided. In Fig. \ref{Fig:master_stacked_spectrum_model_2} we present the final average X-ray spectrum of the secure star-forming galaxies in our sample. 

By using again the \texttt{Sherpa v.4.15.1} package we fitted the X-ray spectrum with a spectral model which includes an absorbed power-law and a thermal plasma \citep[APEC;][]{smith01} component. The former describes the X-ray emission produced by typical XRBs populations, and the latter accounts for the diffuse emission due to hot gas in the galaxy. Foreground absorption was modeled through the tbabs ISM absorption model \citep{wilms00}. The average spectral model was tbabs$\times$(power-law + APEC). The fit was performed within the \erosita{}'s most sensitive band (i.e. 0.6--2.3 \energyunits{}) since above $\sim$ 2.5 \energyunits{} the spectrum is background dominated. However, for plotting purposes we also show the extrapolation of the best-fit model up to 8.0 \energyunits{}. In order to subtract the background and to use the $\chi^{2}$ statistics the spectrum was binned to have at least 10 counts per bin. The best-fit model parameters for the integrated average spectrum are presented in Table \ref{tab:X-ray best-fit spectral parameters}. In order to estimate the uncertainties in the spectral parameters, we used the \texttt{confidence} Sherpa task. This task computes confidence interval bounds by varying a given parameter's value over a grid of values while all the other thawed parameters are allowed to float to new best-fit values. We adopted as an error in the best-fit parameters the 1$\sigma$ (68$\%$) confidence interval. The best-fit model is overplotted with an orange line, while the red and green lines indicate the absorbed power-law and the thermal plasma components, respectively. 

We see that the average X-ray spectrum of all the galaxies in our sample is dominated by a power-law component with a photon index $\Gamma = 1.75^{+0.12}_{-0.07}$, and a hot plasma temperature, $kT = 0.70^{0.06}_{-0.07} \, \energyunits{}$ which is typical for the sub-\energyunits{} temperatures of hot-gas in the star-forming galaxies \citep{owen09}. In addition, the fact that we did not detect a flat, hard X-ray tail in the spectrum or the presence of the Fe-K${\alpha}$ 6.4 \energyunits{} line strongly indicates that our sample is fairly clean of obscured AGNs \citep[c.f.][]{ballo04}. Furthermore, by examining the average spectrum of different individual \sfr{}-\stellarmass{}-\distance{} bins we found that they show a similar behaviour although with lower quality spectra. 

\begin{table}
\centering
\caption{Best-fitting model spectral parameters, for the average X-ray spectrum of all the HEC-eR1 star-forming galaxies.}
\label{tab:X-ray best-fit spectral parameters}
\begin{tabular}{ccc} 
\multicolumn{3}{c}{Spectral model: Absorption $\times$ (power law + APEC)} \\
\hline\hline
Spectral parameter & Value & Units\\
\hline
$\Gamma$ & $1.75^{+0.12}_{-0.07}$ & \\
$kT$ & $0.70^{+0.06}_{-0.07}$ & \energyunits{}\\
$N_{H} $ & $0.001^{+0.016}_{-}$ &10$^{22}$ $\rm{cm}^{-2}$\\
power-law norm & $1.91^{+0.12}_{-0.07}$ &10$^{-6}$ phs/s/\energyunits{}/cm$^{2}$\\
APEC norm$^\ast$ & $2.09^{+0.36}_{-0.36}$ &10$^{-7}$ \\
\hline 
Reduced - $\chi^{2}$ (0.6-2.3 \energyunits{})& 0.653 &\\
Total exposure time & 3686 & ks\\
\hline
\end{tabular}
\begin{tablenotes}
 \footnotesize 
 \item $^\ast$The APEC normalization is in units of $\frac{10^{-14}}{4\pi[D_{A}(1+z)]^{2}}\int n_{e}n_{H}dV$, where $D_A$ is the angular diameter distance to the source (cm), $dV$ is the volume element (cm$^3$), and $n_e$ and $n_H$ are the electron and H densities (cm$^{-3}$), respectively.
\end{tablenotes}
\end{table}

\subsection{X-ray stacking analysis per \sfr{} - \stellarmass{} - \distance{} bins}\label{sec:SFR-M-D stacks}
\begin{figure}[h]
        \includegraphics[width=\columnwidth]{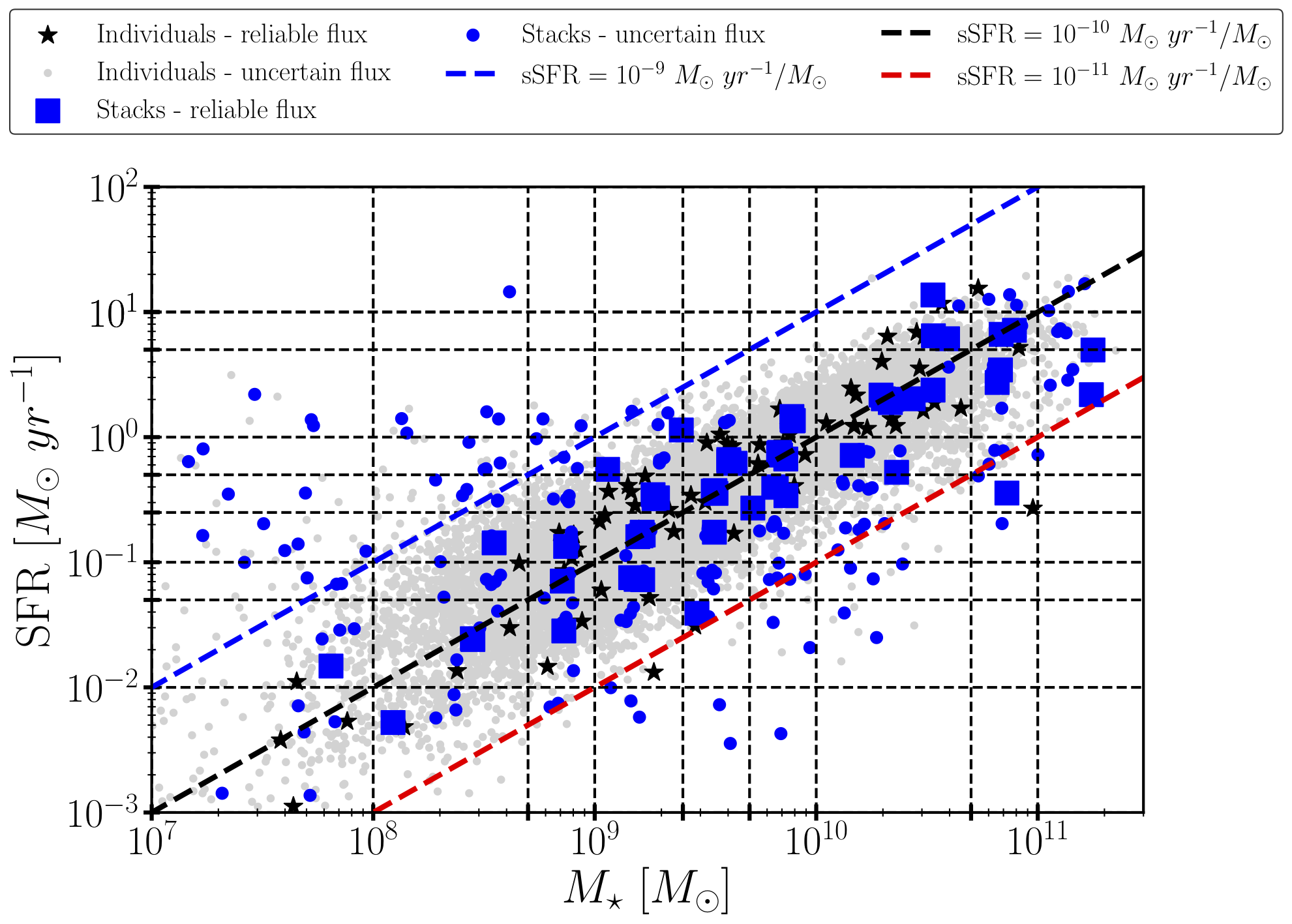}
    \caption{The distribution of \sfr{}-\stellarmass{}-\distance{} bins on the main sequence plane. Each bin is characterized by the mean \sfr{} and \stellarmass{} of the galaxies contained in each bin. Blue squares correspond to the stacks with reliable X-ray flux measurements based on the X-ray stacking analysis and the blue circles to those with uncertain flux measurements. Black stars show the individual galaxies with reliable X-ray flux measurements and the gray circles the galaxies with uncertain fluxes. The black dashed lines indicate the \sfr{} and \stellarmass{} range of the bins. The diagonal dashed lines indicate three different sSFR values (i.e. sSFR: $10^{-9}, 10^{-10},$ and $10^{-11}$ $\rm{M_{\odot}}\,\rm{yr^{-1}}/\rm{M_{\odot}}$). }
    \label{Fig:SFR_M_parameter_space_SFR_M_D_stacks}
\end{figure}

In order to compute the mean X-ray luminosity of our population of star-forming galaxies we stacked our sample in \sfr{}-\stellarmass{}-\distance{} bins.
Given that the population of the 18790 individual galaxies is not uniformly distributed along the main sequence plane (Fig. \ref{Fig:Parameter_space}), we followed an adaptive binning approach in order to ensure an adequate number of galaxies in each bin. The binning scheme is presented in Table \ref{tab:SFR-M-D bins} and it is shown on the \sfr{}-\stellarmass{} plane in Fig. \ref{Fig:SFR_M_parameter_space_SFR_M_D_stacks}. This resulted in 239 occupied \sfr{}-\stellarmass{}-\distance{} bins and 107 of them include more than 10 galaxies. By following the same stacking procedure as in Sect.~\ref{sec:average_spectrum}, we stacked the source spectra of all the individual galaxies in each one of these 239 \sfr{} - \stellarmass{} - \distance{} bins, and we obtained one combined spectrum per bin. In this way, we increased the X-ray signal even for the bins which were dominated by galaxies with uncertain flux measurements. Afterward, by performing the same analysis as described in Sect.~\ref{sec:x-ray photometry} we derived the posterior X-ray count distribution of each stacked spectrum and based on this we calculated the X-ray fluxes and luminosities as in Sect.~\ref{sec:x-ray flux luminosity}. These are the average X-ray flux and X-ray luminosity of the galaxies in each \sfr{}-\stellarmass{}-\distance{} in the 0.6--2.3 \energyunits{} energy band. For the calculation of the luminosities in the 0.5--2 \energyunits{} and the 0.5--8 \energyunits{} energy bands, we used the median distance of the galaxies in each \sfr{} - \stellarmass{} - \distance{} bin and the conversion factors c$_{1}$ and c$_{2}$ (see Table \ref{tab:conversion factors}). In addition, we used the count-rate to flux conversion factors based on the average galaxy spectrum (Table \ref{tab:X-ray best-fit spectral parameters}) and the response files calculated from the stacking analysis for each bin (see Sect.~\ref{sec:x-ray flux luminosity}). Although the stacking of a large number of individual galaxies per bin with uncertain flux measurements will lead to a combined posterior distribution with a larger number of counts, for a number of bins this does not necessarily lead to a reliable flux measurement. This is the case in the low or the high end of the \sfr{} and \stellarmass{} parameter range where the number of individual galaxies in each bin is very small resulting in a small increase of the total number of counts in comparison to the individual galaxies. To characterize the flux measurements of the stacked data we used the shape of the final posterior count distribution as it was described in Sec. \ref{sec:flux reliability}. We found 58 stacks with reliable flux measurements and 181 with uncertain flux measurements. 
In Fig. \ref{Fig:SFR_M_parameter_space_SFR_M_D_stacks} we present the final distribution of the stacks per \sfr{}-\stellarmass{}-\distance{} bin. Blue squares correspond to the stacks with reliable flux measurements and the circles of the same color to those with uncertain flux measurements. We note here that for visualization purposes we show the bin scheme only in the \sfr{}-\stellarmass{} plane. As a result, each region defined by the dashed lines is not a single bin but it further is splitted into 9 \distance{} bins. Therefore within each \sfr{}-\stellarmass{} bin (regions defined by the dashed black lines) there are more than one stacks.  For comparison, we overplot the individual galaxies with reliable flux measurements (black stars) and the galaxies with uncertain fluxes (gray circles). The black dashed lines indicate the \sfr{}-\stellarmass{} bins. To study the correlation of the mean X-ray luminosity with the average stellar population parameters of the galaxies, we calculated the mean \sfr{} and the mean \stellarmass{}, as well as the median metallicity of the galaxies in each \sfr{}-\stellarmass{}-\distance{} bin.
\begin{table}
\centering
\caption{Adopted bins in the \sfr{}, \stellarmass{}, \distance{} dimensions.}
\begin{tabular}{ccc}
\hline\hline
\sfr{} & \stellarmass{} & \distance{}\\

M$_{\odot}\,yr^{-1}$ &M$_{\odot}$ & Mpc \\
\hline
$0.001 - 0.01$ & $10^{7} - 10^{8}$ & 1-2.5\\
$0.01 - 0.05$ &$10^{8} - 5\times10^{8}$&2.5-5 \\
$0.05 - 0.1$  &$5\times10^{8} - 10^{9}$&5-10 \\
$0.1 - 0.25$ &$10^{9} - 2.5\times10^{9}$&10-25 \\
$0.25 - 0.5$&$2.5\times10^{9} - 5\times10^{9}$&25-50\\
$0.5 - 1$&$5\times10^{9}-10^{10}$&50-100\\
$1 - 5$&$10^{10}-5\times10^{10}$&100-150\\
$5 - 10$&$5\times10^{10}-10^{11}$ &150-200\\
$10 - 30$&$10^{11}-10^{12}$&200-300 \\
\hline
\end{tabular}
\label{tab:SFR-M-D bins}
\end{table}

\subsection{X-ray stacking analysis with bootstrap sampling}\label{sec:SFR-M-D stacks bootstrap}
In Sect.~\ref{sec:SFR-M-D stacks} we discussed how we calculated the average X-ray luminosity of the galaxies in our sample as a function of their \sfr{} and \stellarmass{} properties by stacking the X-ray spectra of the individual galaxies in \sfr{}-\stellarmass{}-\distance{} bins. However, in order to assess the variance of the calculated average X-ray luminosity we utilized the bootstrap sampling technique which is widely used in similar works \citep[e.g.][]{lehmer16,fornasini19}. Bootstrap sampling allows us to quantify the uncertainty of the mean X-ray luminosity, and assess how the individual galaxies affect the average stacked signal in each \sfr{}-\stellarmass{}-\distance{} bin. To do that, we started from the initial list of galaxies in each bin, and we created 100 "similar" galaxy lists by resampling the initial list. More specifically, each of the new lists was constructed by randomly replacing it with another one drawn from the initial list. In this way, all the resampled lists contain the same number of galaxies but they are not identical to the parent list since each of them contains a different combination of the original list of individual galaxies. Afterward, we stacked again the X-ray spectra from all the bootstrapped galaxy lists per \sfr{}-\stellarmass{}-\distance{} bin, following exactly the same methodology as in Sect.~\ref{sec:SFR-M-D stacks}. This resulted in 100 posterior X-ray count distributions for each bin. Given that it is very computationally expensive to calculate the response files of each bootstrapped sample, we followed a slightly different procedure to calculate the count-rate to flux conversion factor and the corresponding flux. We produced the response files for a smaller sample of bootstraps (i.e. 10) for each \sfr{}-\stellarmass{}-\distance{} bin and we found that the count-rate to flux conversion factor does not change significantly (typical error $\sim 3\%$). The larger change is observed in the \sfr{}-\stellarmass{}-\distance{} bins which contain only a few galaxies and as a result the statistical fluctuations of the count-rate to flux are larger. However, this happens in only 7 out of 239 bins. As a result, we can use the average count-rate to flux value (derived based on the 10 bootstraps) instead of the individual count-rate to flux conversion factors per bootstrap. This allows us to decrease the computational time of bootstrap stacking. By using the average count-rate to flux conversion factor for each \sfr{}-\stellarmass{}-\distance{} bin we produced the corresponding 100 flux posterior distributions per bin. Afterwards, we used the median distance of the galaxies contained in each of the bootstrapped lists and we calculated the corresponding 100 X-ray luminosity posterior distributions for each bin. Finally, we merged them into a master X-ray luminosity posterior distribution per \sfr{}-\stellarmass{}-\distance{} bin and we calculated the corresponding lower and upper 68\%, and 90\% C.I.  In this way, we measured the statistical standard error on the mean X-ray luminosity per \sfr{}-\stellarmass{}-\distance{} bin.

\subsection{Background AGN contamination}\label{sec:bkg-AGN-contamination}
One of the most common sources of contamination in surveys of extended objects, are background AGNs which fall within the area of the extraction aperture of the sample galaxies and increase the number of the observed counts. 

In our analysis, we measure the integrated flux of a galaxy sample, so we wish to account for the contribution of background AGN in the measured flux. We calculated the total background AGN contribution by integrating their number count distribution (log$N$-log$S$) down to the flux limit of the eRASS1 survey for the detection of a point source at the location of each galaxy:
\begin{equation}\label{equat:agn_contam_eq_a}
\ S_{bkg\,AGN}^{0.3-2.5} = \int_{S_{sens}}^{\infty} \frac{dN}{dS} \times S dS \times A \,\,\,\,\, [erg\, s^{-1} \, cm^{-2}],
\end{equation}
where the $\frac{dN}{dS}$ is the differential number counts in units of  $\rm{erg\,s^{-1}\,cm^{-2}\,deg^{-2}}$, and $S$ is the flux in units of $\rm{erg\,s^{-1}\,cm^{-2}}$. $S_{sens}$ is the limiting flux at the location of each galaxy 
given its local background and exposure time provided by the \texttt{eSASS} sensitivity maps.
The surface of the galaxy is given by $A=\pi\cdot R_{1}\cdot R_{2}$, where $R_{1}$ and $R_{2}$ are the semi-major and semi-minor axis in units of $\rm{deg}$. We considered the \textit{Chandra Multiwavelength Project} \citep[\textit{ChaMP;}][]{kim07} which covers a wide area and an energy band ('S':0.3--2.5 \energyunits{}) from their Table 3) very similar to that of the \erosita{} band (0.6--2.3 \energyunits{}) considered. 

\begin{figure}[h]
        \includegraphics[width=\columnwidth]{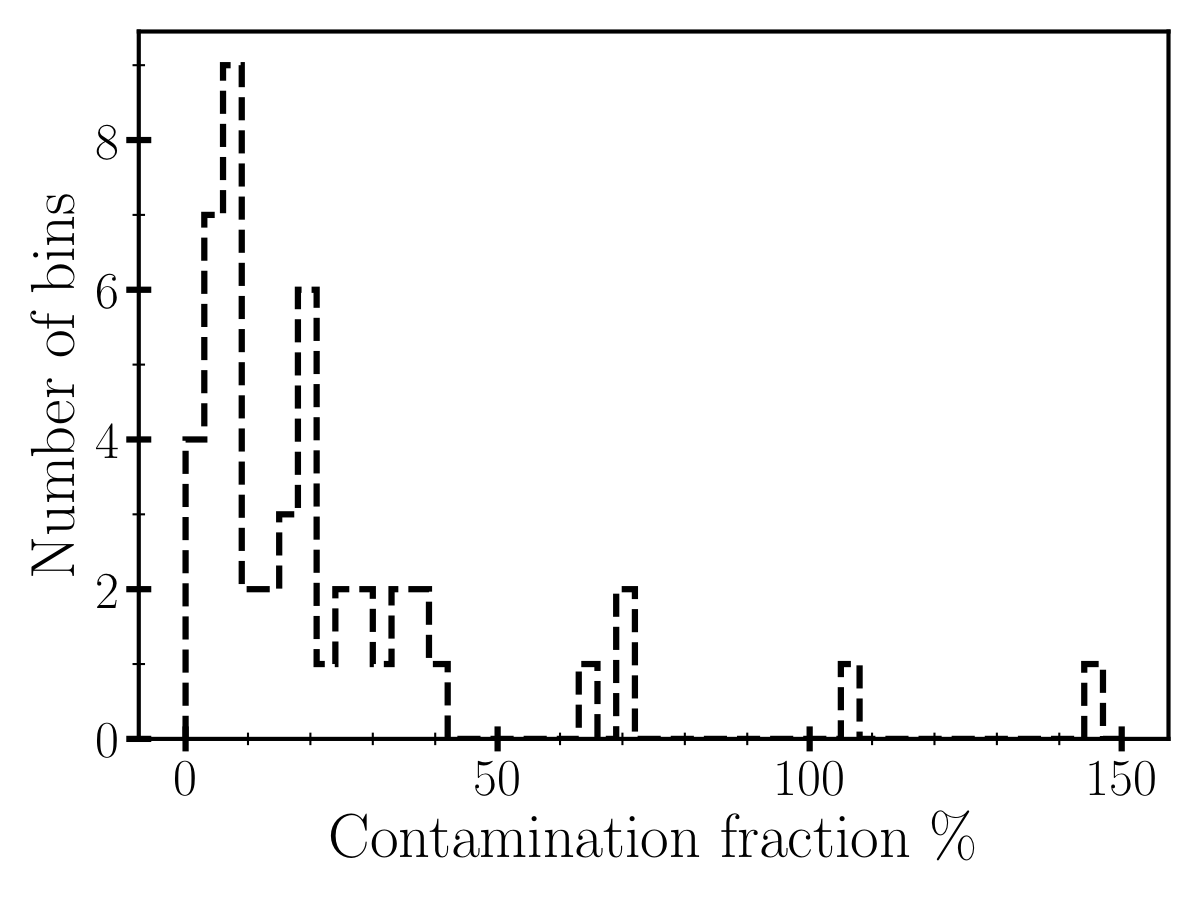}
    \caption{The distribution of the estimated background AGN contamination fraction in the \sfr{}-\stellarmass{}-\distance{} stack bin in our sample. The contamination fraction assuming the eRASS1 flux sensitivity limit of each individual galaxy. The median value of the distribution is $17\%$.
    The total surface of the 5 stacks with contamination fraction higher than $50\%$ is significantly higher than all the rest used in our analysis resulting in an overestimation of the background AGN contamination.}
    \label{Fig:Bkg AGN contamination hist}
\end{figure}

This calculation though cannot be performed on a galaxy-by-galaxy basis since an individual galaxy is affected by i) Poisson sampling of the background sources and ii) stochastic sampling of the fluxes drawn from the log$N$-log$S$. On the other hand, considering a population of galaxies with similar characteristics reduces the stochastic effects because of a better sampling of the background AGN distribution. For that reason, we perform this calculation for the galaxies in the \sfr{}-\stellarmass{}-\distance{} bins defined in Sect.~\ref{sec:SFR-M-D stacks}.
We took into account only the individual galaxies with at least 1 source net count in each bin because only for these cases the calculation of the background contamination is meaningful. This resulted in 146 \sfr{}-\stellarmass{}-\distance{} bins out of which 50 had reliable flux measurements using the criterion of Eq. \ref{equat:reliable_formula}. To calculate the expected average flux due to background AGNs in each of these 50 \sfr{}-\stellarmass{}-\distance{} bins we used the formula

\begin{equation}\label{equat:agn_contam_eq_b}
S^{\sfr{}-\stellarmass{}-\distance{}}_{bkg\,AGN} = \sum_{i=1}^{N}S_{bkg\,AGN}^{0.3-2.5}\cdot A_{i} \,\,\,\,\, [\rm{erg\, s^{-1} \, cm^{-2}}],
\end{equation}
where $N$ corresponds to the total number of individual galaxies included in each bin, and A$_{i}$ is the area (in $deg^{-2}$) of each galaxy contributed in the bin. 
We calculated the contamination fraction as 
\begin{equation}\label{equat:agn_contam_eq_c}
\textrm{Contamination fraction} \, \% = \frac{S^{\sfr{}-\stellarmass{}-\distance{}}_{bkg\,AGN}}{S^{\sfr{}-\stellarmass{}-\distance{}}_{0.6-2.3}} \times 100 \, ,
\end{equation}
where $S^{\sfr{}-\stellarmass{}-\distance{}}_{bkg\,AGN}$ is the expected flux due to background AGNs and $S^{\sfr{}-\stellarmass{}-\distance{}}_{0.6-2.3}$ is the total measured flux in each bin (considering only the galaxies with more than 1 count). In Fig. \ref{Fig:Bkg AGN contamination hist} we present the distribution of the estimated contamination fraction per \sfr{}-\stellarmass{}-\distance{} bin of our sample. For the vast majority of our bins, the background AGN contamination fraction is lower than $20\%$, while only few of them have contamination higher than $50\%$. Based on this distribution we considered that the median background AGN contamination in our sample is $17\%$.
It should be noted that this estimation is an upper limit on the background AGN contamination since they are observed through the body of each galaxy and their X-ray emission is attenuated by the ISM of each of our sources. 

\begin{figure}[h]
        \includegraphics[width=\columnwidth]{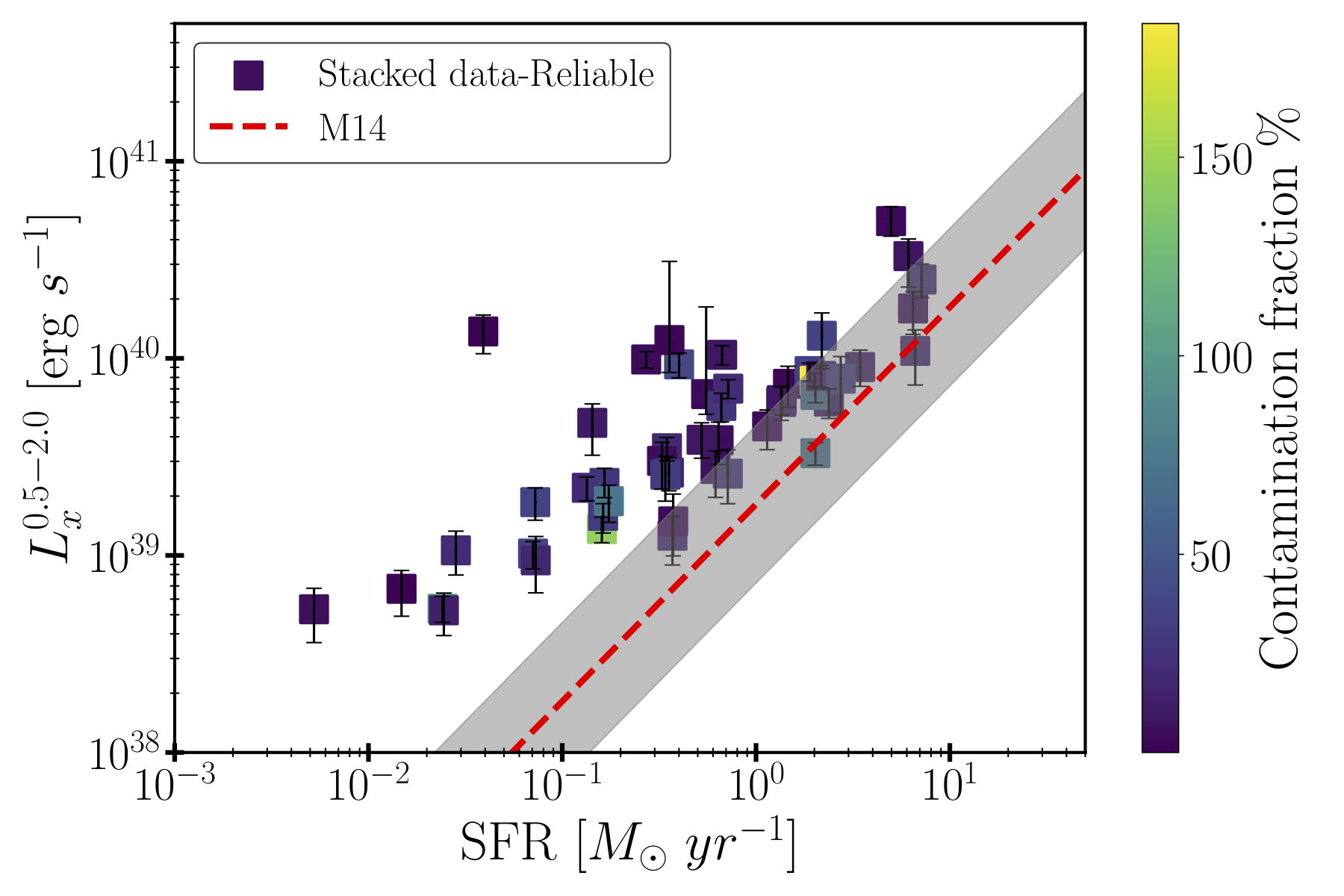}
  \caption{X-ray luminosity (0.5--2 \energyunits{}) as a function of the SFR for the stacked data with reliable flux measurements color-coded with the estimated contamination by the background AGN. Only the stacks with available estimation of the background contamination are shown. For comparison, we overplot the scaling relation from \citet{mineo14} (M14). We see that the contamination is independent of the \sfr{}-\stellarmass{}-\distance{} bins and there is no systematic trend with \sfr{} or \lx{}. The few stacks with very high contamination fraction (${>}50$\%) are due to their total surface which is systematically higher than all the rest used in our analysis resulting in an overestimation of the background AGN contamination.}
    \label{Fig:Bkg AGN contamination}
\end{figure}

In addition, we calculated the mean \lx{} of the stacks in the 0.5--2 \energyunits{} band, assuming the median distance of the galaxies in each \sfr{}-\stellarmass{}-\distance{} bin. In Fig. \ref{Fig:Bkg AGN contamination} we plot the \lx{} as a function of the mean \sfr{} color-coded with the contamination fraction. For comparison, we overplot the scaling relation from \citet{mineo14} (M14) after converting it to the adopted 0.5-2 \energyunits{} band by using the conversion factor c$_{3}$ (see Table \ref{tab:conversion factors}). As it is shown the contamination is independent of the \sfr{}-\stellarmass{}-\distance{} bins and there is no systematic trend with \sfr{} or \lx{}. 

For very few bins (i.e. 5) the contamination fraction is higher than $50\%$. This could be due to two reasons. The first is because the galaxies that contribute to these stacks have large angular sizes increasing the total surface and consequently the calculated background AGN contamination is overestimated. Indeed the total surface of these 5 stacks is significantly higher than all the rest that are used in our analysis. The second is the lack of strong XRBs populations in the galaxies that are included in these bins (their vast majority have no reliable flux measurements) resulting in a measured flux that is almost equal to or lower than the expected flux from background interlopers. This also explains the stacks that have ${>}100$\% AGN contamination. However, these statistical fluctuations for a small sub-population of galaxies do not change the main conclusion that the median AGN contamination of our final sample is negligible.

\section{Results}\label{sec:Results}
\subsection{eRASS1 sensitivity}\label{sec:eR1_sens}
\begin{figure}
        \includegraphics[width=\columnwidth]{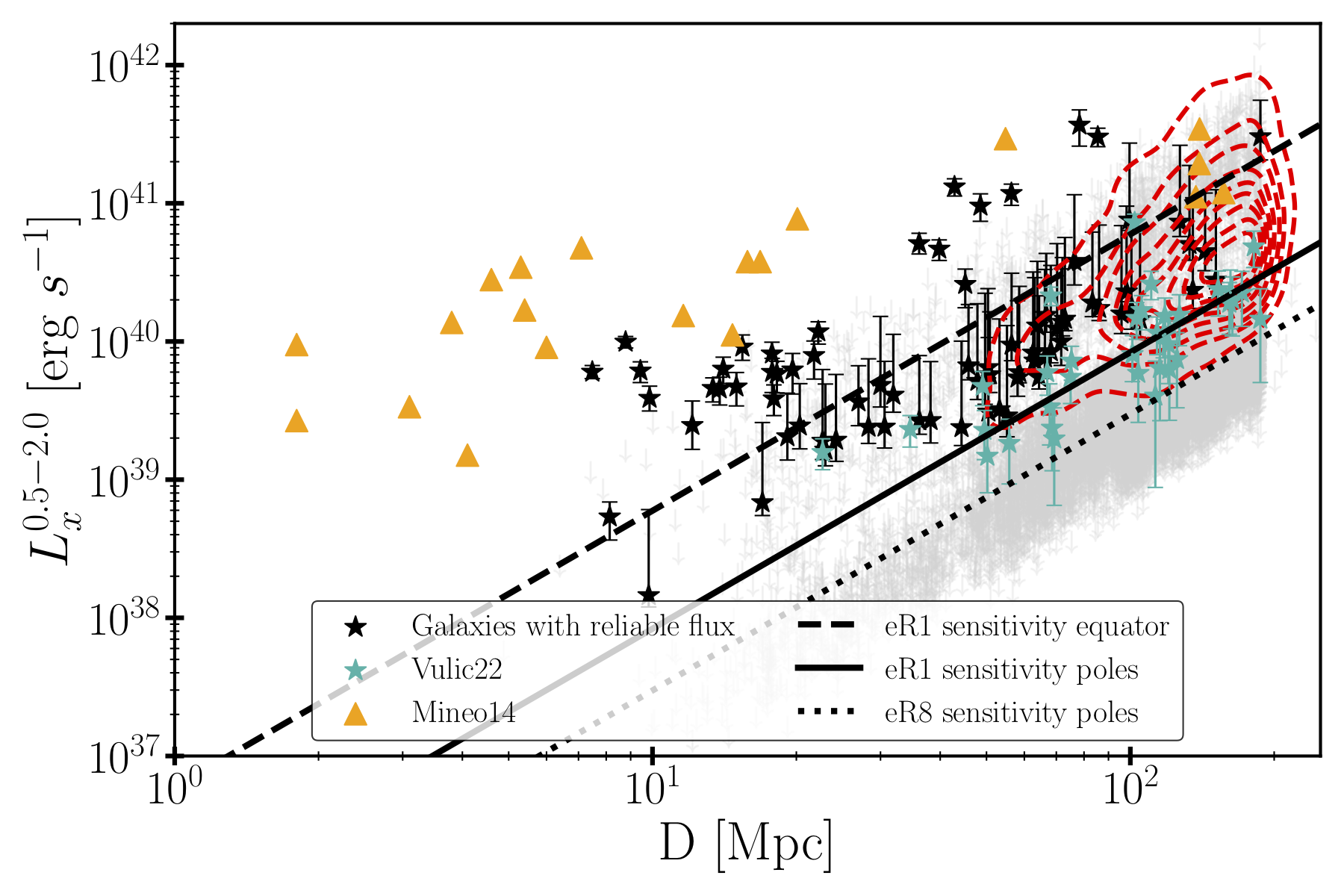}
    \caption{The \lx{} vs \distance{} for the HEC-eR1 sample of star-forming galaxies. Black stars show the galaxies with reliable flux measurements and lightgray down-arrows, galaxies with uncertain flux measurements. For comparison, we also overplot the normal galaxies from \citet{mineo14} and \citet{vulic22} as orange triangles and turquoise stars. We also overplot with red contours the galaxy population expected to be observed in the eRASS:8 survey based on the simulation study of \citet{basuzych20}. The dashed and solid lines mark the 
eRASS1 sensitivity limits at the ecliptic equator and poles respectively. The dotted line indicate the deepest sensitivity of the eROSITA survey which is expected to be reached at the poles of the eRASS:8. These lines are based on the  sensitivity values reported in \citet{predehl}}
    \label{Fig:Lx_05_2_D_sensitivity}
\end{figure}

The all-sky nature of the eRASS1 survey, as opposed to the other local galaxy surveys, allows us to study for the first time a completely blind and statistically significant sample of normal galaxies as a function of their stellar population parameters. 
In Fig. \ref{Fig:Lx_05_2_D_sensitivity} we present the measured X-ray luminosity of the HEC-eR1 star-forming galaxies as a function of their distance in comparison with previous normal galaxy surveys (\citet{mineo14} orange triangles; and \citet{vulic22}; lightseagreen stars).

As expected, the HEC-eR1 star-forming galaxies with reliable fluxes are consistent with the eRASS1 sensitivity limits at the ecliptic equator and poles \citep[$F_{sens}^{eq}=5{\times}10^{-14}\,\rm{erg\,s^{-1}\,cm^{-2}}$ and $F_{sens}^{poles}=7{\times}10^{-15} \, \rm{erg\,s^{-1}\,cm^{-2}}$ respectively;][]{predehl}. When we compare our sample with the previous study of nearby galaxies from \citet{mineo14}, we find that the HEC-eR1 galaxy sample probes an unexplored region of the \lx{}-\distance{} parameter space, that has not been studied before due to the limitations of the current and past observatories. 
The eRASS1 sample allows us to probe star-forming galaxies at luminosities as low as $10^{39}\,\rm{erg\,s^{-1}}$  out to distances of ${\sim}70$\,Mpc, and luminosities of as $10^{40}\,\rm{erg\,s^{-1}}$  out to distances of ${\sim}200$\,Mpc.  
Given the sensitivity of the eRASS1 survey we would expect to detect all the galaxies from \citet{mineo14} sample. However, because a large fraction of the latter galaxies, especially at very small distances, fall outside the west hemisphere of the sky they are not covered by eRASS1. 
The comparison with the results from the eFEDS survey \citep{vulic22} shows that with the full depth eRASS:8 survey we will be able to reach X-ray luminosities of typical dwarf galaxies such as those observed in our very local universe,  out to $\sim200$\,Mpc. The results from the eRASS1 survey are generally consistent with the simulation study of \citet{basuzych20}, who predicted the populations expected in the complete eRASS:8 survey (red contours), especially considering that this simulation is based on pre-launch sensitivities and detector background.

Since we are interested in the scaling relation between the X-ray luminosity and the \sfr{}, it is informative to explore the eRASS1 sensitivity on the \lx{}-\sfr{} plane for the HEC-eR1 galaxy sample. 
To that end, we utilize the \sfr{}-\stellarmass{}-\distance{} binning scheme presented at Sect.~\ref{sec:SFR-M-D stacks}, along with the \texttt{eSASS} sensitivity maps which provide the limiting flux given the local background and exposure time at the location of each galaxy. 
For each bin, we calculated the limiting luminosity (in the 0.5--2 \energyunits{} energy band), based on the average limiting flux and median distance of the galaxies. This gives us an indicative minimum luminosity above which we can detect point sources in each \sfr{}-\stellarmass{}-\distance{} bin. The sensitivity for the sample galaxies depends on their angular size but nonetheless, these lines give us an indication of the eRASS1 survey. However, the sensitivity calculated for the individual objects is expected to tend towards the indicative value calculated with the above method for the more distant galaxies. 

\begin{figure}
        \includegraphics[width=\columnwidth]{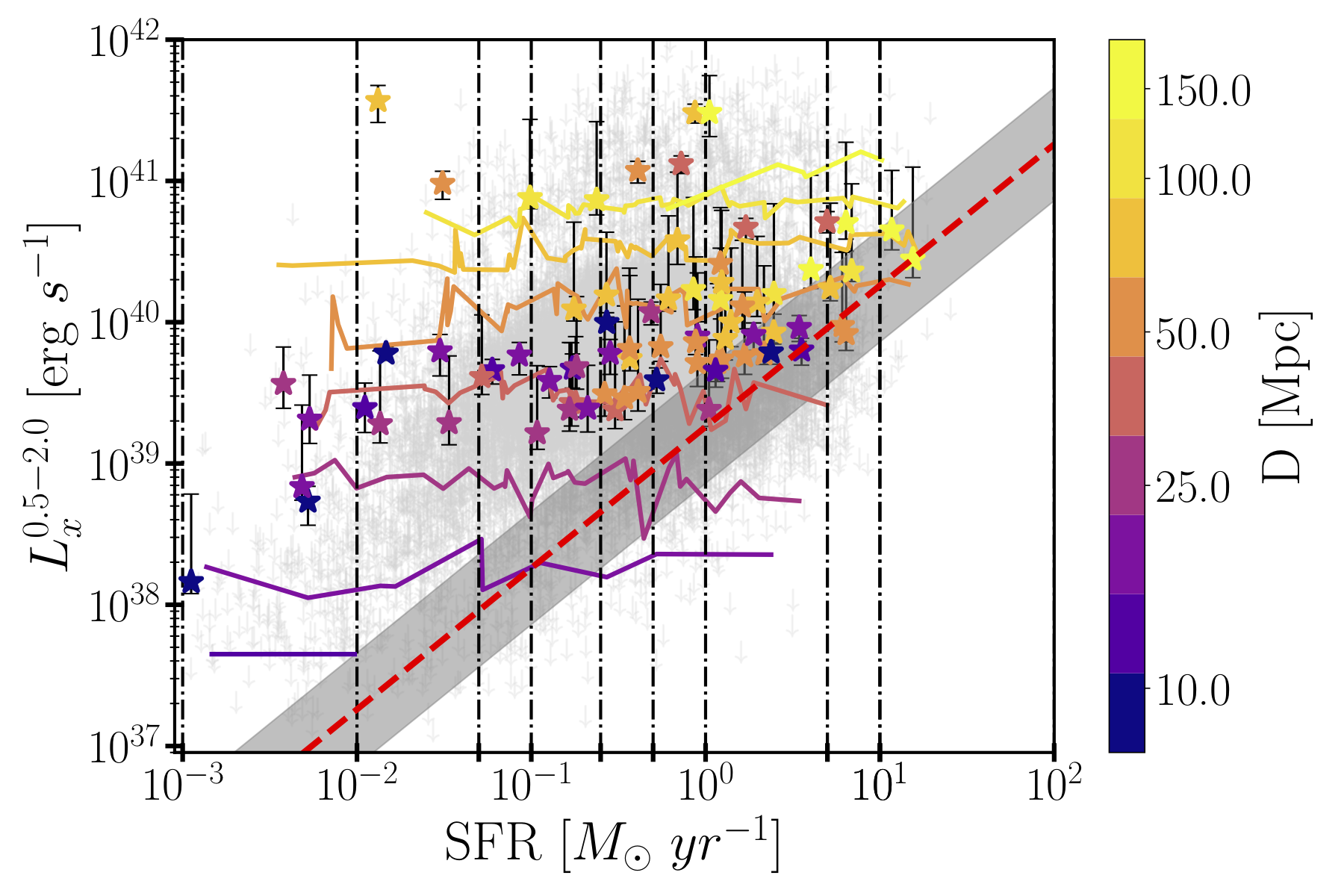}
    \caption{\lx{} vs \sfr{} for the HEC-eR1 sample of star-forming galaxies. Black stars show the galaxies with reliable flux measurements and lightgray down-arrows galaxies with uncertain flux measurements. The lines correspond to the limiting \lx{} in each \sfr{}-\stellarmass{}-\distance{} bin, color-coded by their median distance. Vertical black dashed lines indicate the \sfr{} bins presented in Table \ref{tab:SFR-M-D bins}. The red dashed line shows the standard \lx{}-\sfr{} scaling relation from \citet{mineo14} along with the $1\,\sigma$ scatter.}
    \label{Fig:M14_final_sample_sensitivity_mpe}
\end{figure}

In Fig.~\ref{Fig:M14_final_sample_sensitivity_mpe}, we present the result of this analysis, in the \lx{}-\sfr{} plane. The colored lines correspond to the limiting \lx{} in each \sfr{}-\stellarmass{}-\distance{} bin, color-coded by the median distance. Stars and gray down-arrows show the HEC-eR1 star-forming galaxies with reliable and uncertain flux measurements, respectively. The red dashed line corresponds to the standard \lx{}-\sfr{} scaling relation from \citet{mineo14} along with the $1\sigma$ scatter.

As it is shown, all the HEC-eR1 star-forming galaxies with reliable fluxes are above the sensitivity limit of the eRASS1 survey.  
This plot also indicates that in the very local universe ($\lesssim10$\,Mpc) the galaxies with reliable flux measurements are at least $\sim1$\,dex above the sensitivity limit of ${\sim}10^{38}$\,$\rm{erg\,s^{-1}}$,
while at larger distances we are sensitive to systematically high-luminosity galaxies. However, by including the constraints on the luminosity for the unreliable measurements (gray points) we remedy this bias. 

\subsection{Hot gas contribution}\label{sec:hot_gas_contribution}

It is well known that the integrated X-ray emission of star-forming galaxies is partly due to the diffuse X-ray emission of hot-ionized gas (with temperature $\sim$0.5--1.5 \energyunits{}) \citep{mineo12a,lehmer22}. This hot gas is the result of the cumulative effect of strong winds from massive stars and supernovae (SNe) and it is correlated with regions of recent star formation \citep[see][for a review]{fabbiano19}.

Our best-fit results obtained from the spectral analysis in Sect.~\ref{sec:average_spectrum} suggest that the temperature of the hot gas is $kT=0.70 \, \energyunits{}$ which is in excellent agreement (in the band that we measured the spectrum; 0.6--2.3 \energyunits{}) with the stronger hot gas components of \citet{mineo12b} and \citet{lehmer22}, who calculated $kT=0.67 \, \energyunits{}$, and $kT=0.70 \, \energyunits{}$, respectively.
By using the best-fit model, we calculated the flux of the total (XRB+hot gas) component ($F^{total}_{0.5-2} = 4.89\times10^{-15} \,\rm{erg\,s^{-1}\, cm^{-2}}$) and the flux of the thermal component ($F^{APEC}_{0.5-2} = 6.29\times10^{-16} \,\rm{erg\,s^{-1}\, cm^{-2}}$). The average hot-gas contribution is
\begin{equation}\label{equat:hot gas contribution}
\textrm{Hot-gas contribution} = \frac{F^{APEC}_{0.5-2}}{F^{total}_{0.5-2}}\times 100 \,\% =12.86 .
\end{equation}
As a result, the average diffuse X-ray emission of the HEC-eR1 star-forming galaxies, is only a small fraction of the total average integrated flux, indicating that the measured X-ray emission is dominated by a power-law component associated with XRBs.

\subsection{Scaling relations derived from the eRASS1 galaxy sample}\label{sec:scaling_relations_individ}

\begin{figure*}
        \includegraphics[width=\columnwidth]{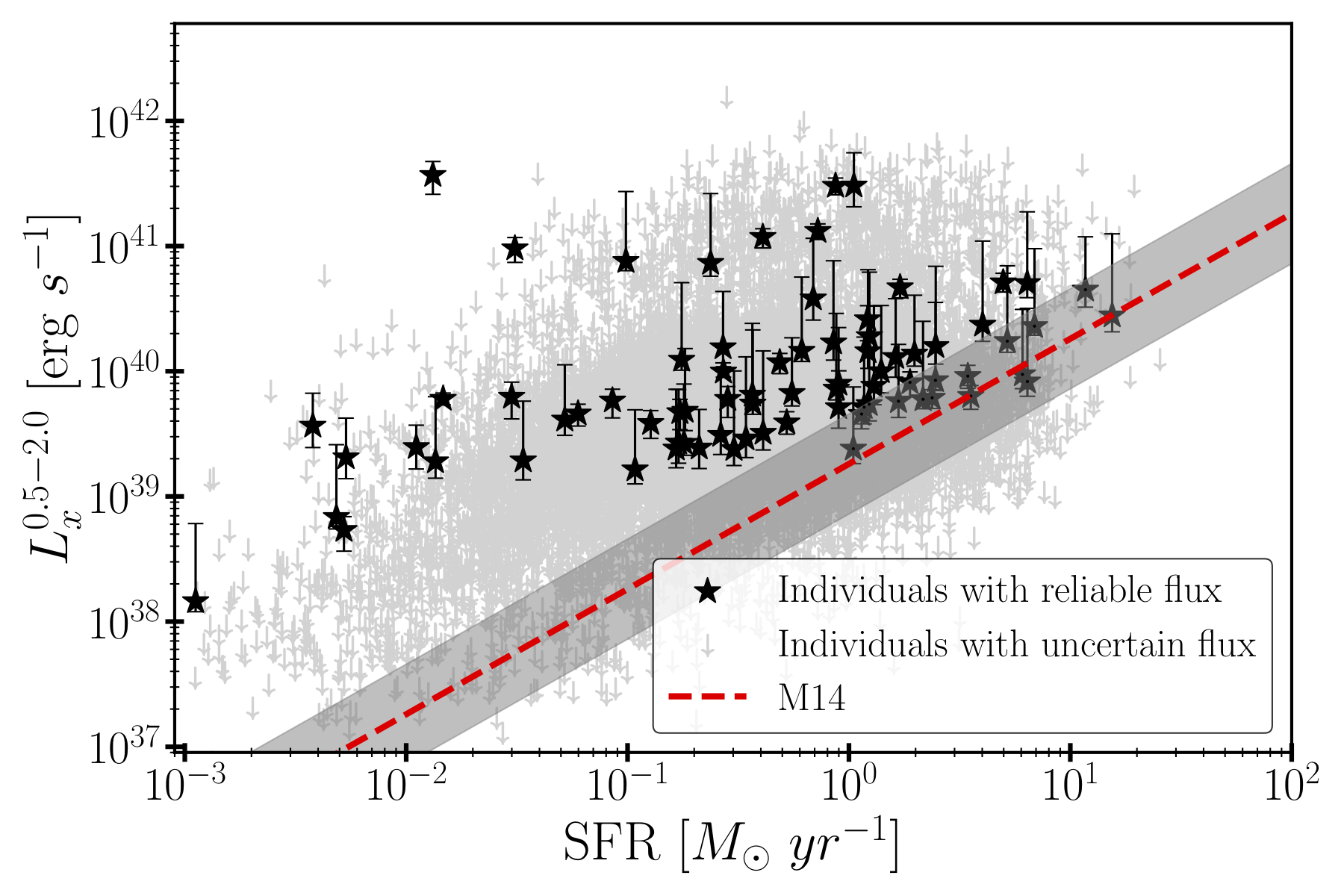}
        \includegraphics[width=\columnwidth]{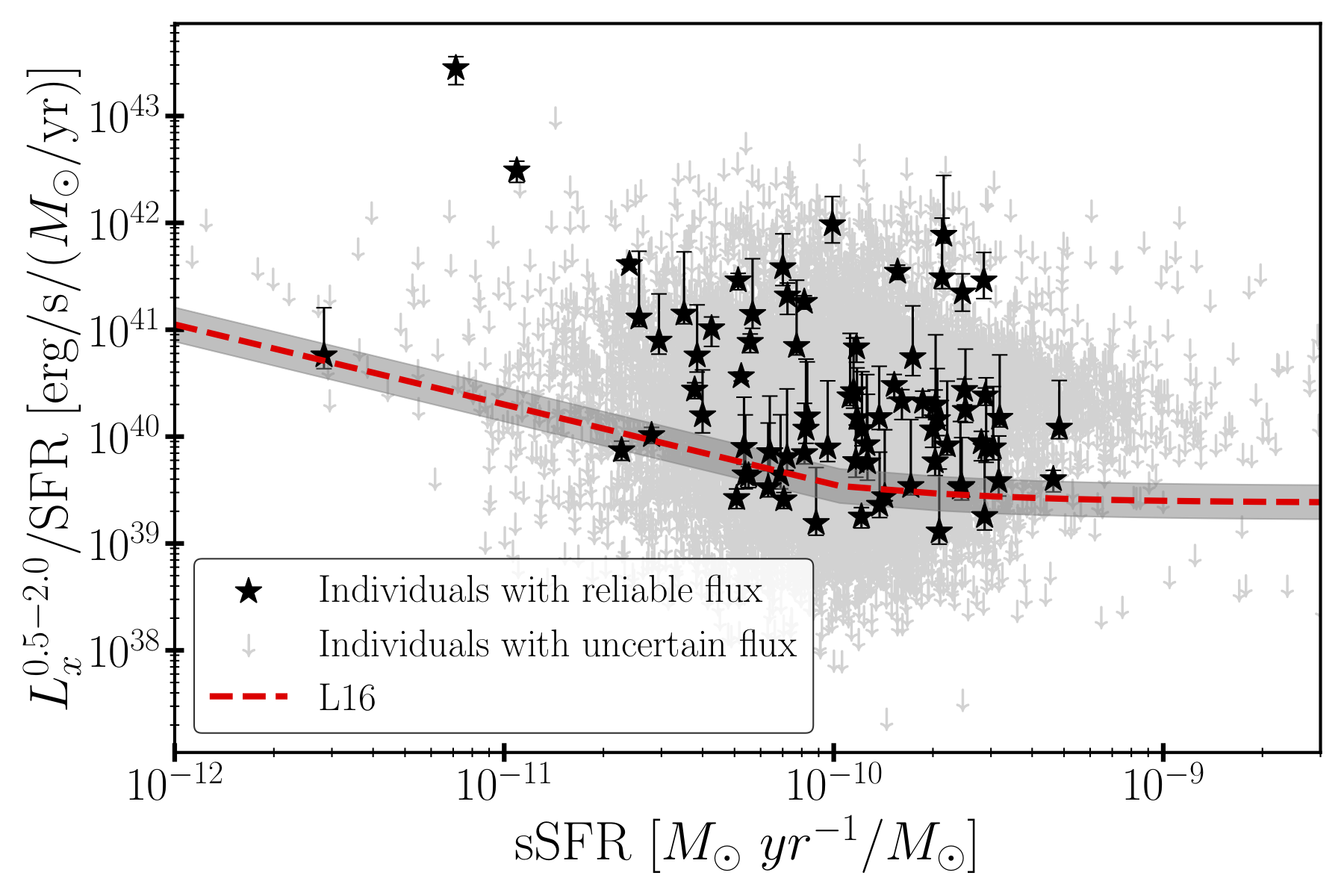}
    \caption{Left panel shows the distribution of all the HEC-eR1 secure star-forming galaxies in the \lx{}$^{0.5-2}$-\sfr{} plane. Red line depicts the standard scaling relation from \citealp[]{mineo14} (M14). Right panel shows the same distribution in the \lx{}$^{0.5-2}$/\sfr{} - sSFR plane and the red line the corresponding scaling relation from \citealp[]{lehmer16} (L16). Black stars show the secure star-forming galaxies with reliable flux measurements with their corresponding uncertainties, and the gray down-arrows the galaxies with uncertain flux measurements.}
    \label{Fig:M14_L16_final_sample_individuals}
\end{figure*}

In Fig. \ref{Fig:M14_L16_final_sample_individuals} we present the correlation between the integrated X-ray luminosity and the stellar population parameters of the star-forming galaxies for the HEC-eR1 sample. In the left panel, we plot the integrated \lx{} in the 0.5--2 \energyunits{} energy band as a function of the \sfr{}. The right panel shows the \lx{} per \sfr{} in the 0.5-2 \energyunits{} energy band as a function of the sSFR. The black stars indicate the individual galaxies with reliable flux measurement and their corresponding uncertainties at the $68\%$ C.I. of the flux distribution. The gray down arrows show the galaxies with uncertain flux measurements at the upper $90\%$ C.I. For comparison, we overplot the scaling relations from \citet[]{mineo14} (M14) and \citet{lehmer16} (L16), respectively. The M14 scaling relation was converted from the the reference 0.5--8 \energyunits{} band to the adopetd 0.5--2 \energyunits{} band, using the conversion factor c$_{3}$ (see Table \ref{tab:conversion factors}).

We interestingly find that our population of star-forming galaxies shows elevated X-ray luminosity in comparison to the standard scaling relations of M14 and L16. The majority of galaxies with reliable flux measurements show systematically higher luminosities which can reach $10^{42}\,\rm{erg\,s^{-1}}$. In particular, galaxies with \sfr{}${<}5\times10^{-1}\,\rm{M_{\odot}yr^{-1}}$ exhibit significant excess from the standard scaling relations while this excess is decreasing as we go towards higher \sfr{} regimes. In addition, a sub-population of very actively star-forming galaxies with $\rm{sSFR} {>} 10^{-10}\,\rm{M_{\odot}}\,\rm{yr^{-1}}/\rm{M_{\odot}}$ reaches very high X-ray luminosities per \sfr{}, up to $10^{42}\,\rm{erg\,s^{-1}\, /M_{\odot}\,yr^{-1}}$. 

We also see that although all the reliable galaxies are above the scaling relation, a significant fraction of the galaxies with uncertain fluxes has 90$\%$ upper confidence intervals well below the M14 best-fit line. This can be explained by the fact that eRASS1 is a flux-limited survey resulting only in the detection of the most X-ray luminous star-forming galaxies. However, even with this relatively shallow survey, we can set useful limits on the X-ray emission of galaxies below the expected scaling relations. The same behavior is also observed in the \lx{}/\sfr{}-sSFR scaling relation. The vast majority of the star-forming galaxies with reliable measurements show a significant excess from the L16 relation while the galaxies with uncertain flux measurements are symmetrically distributed along the L16 best-fit line.

From the figure, we also see that the scatter of the X-ray emission in our blind galaxy sample (including reliable and uncertain flux measurements) is significantly larger than the 1$\sigma$ scatter from the best-fitted lines of M14 and L16. The high scatter is also confirmed by the two individual extremely luminous galaxies with $\lx{}/\sfr{}{>}10^{42}\,\rm{erg\,s^{-1}\,cm^{-2}\, /M_{\odot}\,yr^{-1}}$ (PGC24071, PGC1133474). A detailed inspection of their X-ray and optical images revealed that they are clean of X-ray contaminants (e.g. background AGN etc.) implying that all the X-ray emission is produced by XRB populations with a small contribution of hot gas. While they follow the linear form of the L16 in the low-sSFR, their dispersion is much higher than the expected value. 

\subsection{Average scaling relations of the eRASS1 galaxy sample}
\begin{figure*}
        \includegraphics[width=\columnwidth]{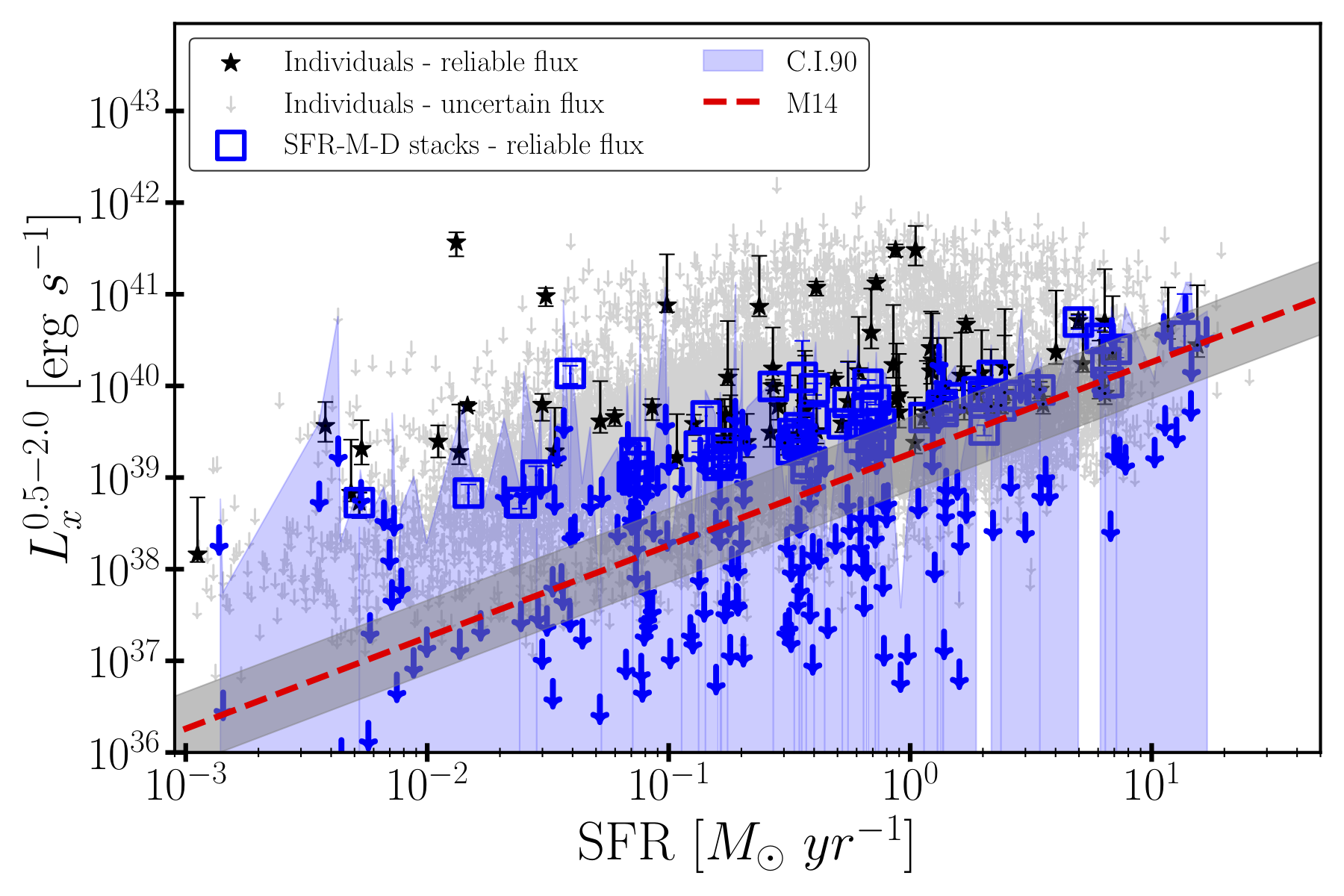}
        \includegraphics[width=\columnwidth]{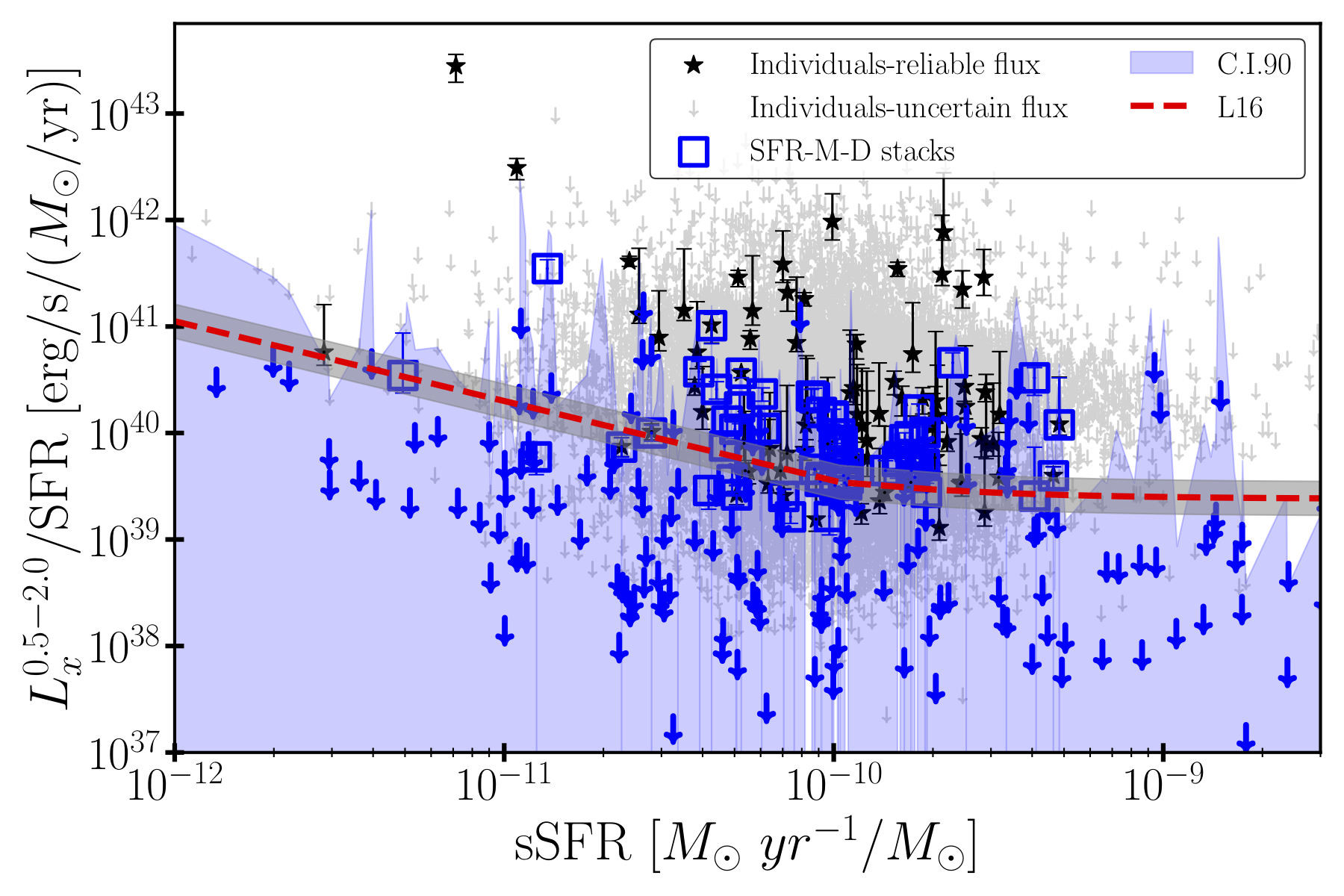}
        
    \caption{Left panel shows the distribution of the \sfr{}-\stellarmass{}-\distance{} stacks in the \lx{}$^{0.5-2}$-\sfr{} plane. Red line depicts the standard scaling relation from \citealp[]{mineo14} (M14). Right panel shows the same distribution in the \lx{}$^{0.5-2}$/\sfr{} - sSFR plane and the red line the scaling relation from \citealp[]{lehmer16} (L16). Cyan errorbars correspond to the 90$\%$ C.I. of the X-ray luminosity based on a bootstrap sampling. Black stars show the secure star-forming galaxies with reliable flux measurements with their corresponding uncertainties, and the gray down-arrows the galaxies with uncertain flux measurements.}
    \label{Fig:M14_L16_SFR_M_D_stacks_bootstraps}
\end{figure*}

In order to study the average X-ray luminosity of the HEC-eR1 star-forming galaxies and its connection with their stellar populations we used the results of the \sfr{}-\stellarmass{}-\distance{} stacks (see Sec. \ref{sec:SFR-M-D stacks}). In Fig. \ref{Fig:M14_L16_SFR_M_D_stacks_bootstraps} we present the distribution of the \sfr{}-\stellarmass{}-\distance{} stacks in the \lx{}-\sfr{} and the \lx{}/\sfr{}-sSFR parameter space, respectively. The open blue squares indicate the stacks with reliable flux measurements and the blue down-arrows the stacks with uncertain flux measurements. For comparison, we overplot the population of individual galaxies following the same notation as in Fig. \ref{Fig:M14_L16_final_sample_individuals}. The scaling relations from M14 and L16 are also plotted. The blue shaded region indicates the upper and lower 90\% C.I. of the master posterior distribution of the X-ray luminosity per stack based on our bootstrapping analysis (see Sect.\ref{sec:SFR-M-D stacks bootstrap}). 
As we see the stacked data are more symmetrically distributed and closer to the scaling relations than the individual galaxies. This is expected since the large number of individual galaxies with uncertain fluxes per bin tends to bring the stacks closer to the best-fit lines of M14 and L16, respectively. Based on the results of the bootstrap analysis, we see that the upper 68$\%$ C.I. of the master posterior X-ray luminosity is very close to the upper error of the actual measurement. This indicates that the high X-ray luminosities of the average population of galaxies are not driven by a few luminous individual sources. 

\begin{figure}
        \includegraphics[width=\columnwidth]{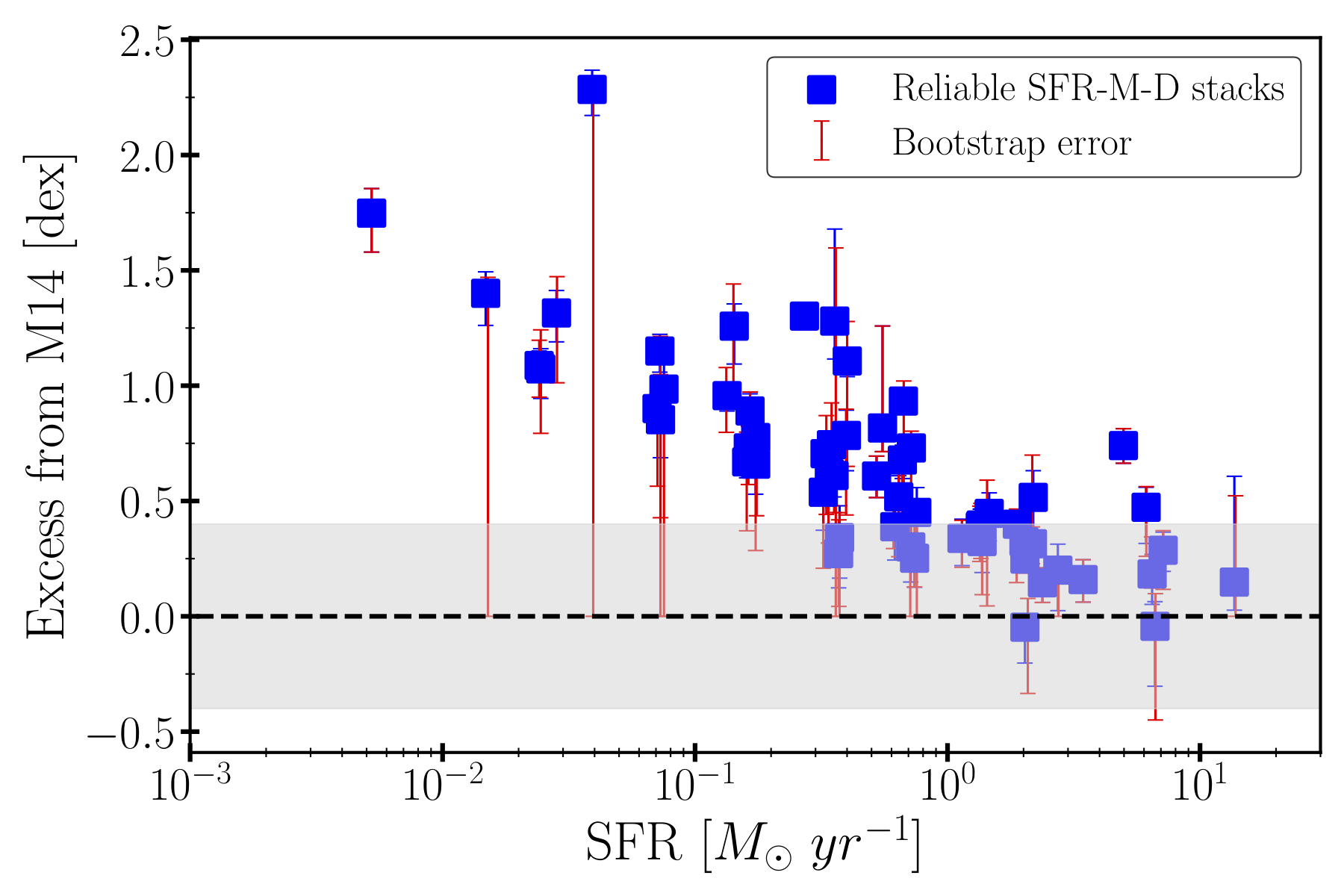}
    \caption{Excess of the \sfr{}-\stellarmass{}-\distance{} stacks with reliable flux measurements from the \lx{}$^{0.5-2}$-\sfr{} scaling relation of \citealp[]{mineo14} (M14). The blue error bars show the 68$\%$ C.I of the posterior X-ray luminosity distribution for each stack. The red error bars indicate the excess of the upper and lower 68$\%$ C.I. of the master X-ray posterior luminosity from the bootstrap analysis with respect to the M14 relation.}
    \label{Fig:M14_excess_SFR_M_D_stacks}
\end{figure}

However, there is still an excess of our stacked data with reliable flux measurements (i.e. 58 stacks) from the M14 standard scaling relation, especially for the lower \sfr{}s and the higher sSFRs. In order to quantify the average excess per \sfr{} bin, in Fig. \ref{Fig:M14_excess_SFR_M_D_stacks} we present the excess (in units of dex) from the M14 standard scaling relation (in the 0.5--2 \energyunits{} energy band) as a function of \sfr{}.
The error bars were calculated based on the excess of the upper and lower 68$\%$ of the X-ray luminosity distribution from the reference relation. The light gray shaded region is the 1$\sigma$ scatter around the M14 relation. As it is shown the average excess for a galaxy population within the \sfr{} bin $0.001-0.01\,\rm{M_{\odot}\,yr^{-1}}$ is up to ${\sim} 2 \, \rm{dex}$ while as the \sfr{} increases the excess decreases. For the \sfr{} bins $1-5\,\rm{M_{\odot}\,yr^{-1}}$, $5-10\,\rm{M_{\odot}\,yr^{-1}}$, and $10-30\,\rm{M_{\odot}\,yr^{-1}}$  the excess is consistent with the scaling relation. 

In addition, in Fig. \ref{Fig:M14_excess_SFR_M_D_stacks} we also show with the red errorbars the errors derived from the boostrap analysis. These are calculated based on the excess of the upper and lower 68$\%$ C.I. of the master X-ray posterior luminosity with respect to the M14 relation (see Sect.~\ref{sec:SFR-M-D stacks bootstrap}). This demonstrates that the observed excess is real and not due to strong statistical fluctuations of a sub-population of extreme X-ray luminous individual galaxies that dominates the stacked signal in each bin. Furthermore, the scatter of the stacked data seems to correlate with the \sfr{}. 
For \sfr{} $>3\times\,\rm{M_{\odot}\,yr^{-1}}$ the stacks are consistent with the 1$\sigma$ scatter of M14, while for lower \sfr{}s the dispersion around the relation is increased. In Sect.~\ref{sec:max_like_fit_scaling_relations} we quantify this SFR-dependent scatter by taking it into account during the fits of the scaling relations. 

We find that both the individual HEC-eR1 star-forming galaxies and the stacked data per \sfr{}-\stellarmass{}-\distance{} bins show significant excess from the standard scaling relations. This excess is very high and cannot be due to the background AGN contamination (${\sim}10\%$ of the total measured X-ray flux; see Sect.~\ref{sec:bkg-AGN-contamination}) or hot-gas contribution (${\sim}13\%$ to the total X-ray flux; see Sect.~\ref{sec:hot_gas_contribution}). As a consequence, this excess can be the result of the effect of i) the gas-phase metallicity of the stellar populations \citep{fornasini19}, ii) the age of the stellar populations \citep{gilbertson22}, or iii) the stochastic sampling of the X-ray luminosity function (XLF) of XRBs \citep{lehmer21}. In Sect.~\ref{sec:Discussion} we discuss all these possible explanations in detail.

\subsection{Fitting of the scaling relations}\label{sec:max_like_fit_scaling_relations}
In order to study the connection between the integrated X-ray luminosity of HEC-eR1 star-forming galaxies with the \sfr{}, \stellarmass{}, and metallicity, and to quantify the observed excess, we employed the maximum likelihood fitting methodology developed in \citet{kouroumpatzakis20}. In contrast to other standard fitting techniques, this method utilizes the full probability distribution for the dependent and indipendent variables data which can be different for each data point and of any form even non-parametric. This is particularly important given the wide range of X-ray intensities in our sample which can be described neither by a Gaussian nor by a Poisson distribution \citep[c.f.][]{vandyk01} given the very small number of counts and the non-negligible background. 
In addition, since the fit is performed on the posterior distribution, we can use the entire dataset including all the galaxies, even those with uncertain flux measurements, alleviating in this way the bias due to X-ray selection effects. 

In our analysis, we simultaneously fitted the X-ray probability distributions as they were calculated in Sect.~\ref{Sec:X-ray analysis} for the stacked data per \sfr{}-\stellarmass{}-\distance{} bin. This allows us to examine the scaling relation between X-ray luminosity and the parameters of stellar populations in terms of average statistical measures and with a higher signal-to-noise ratio than the individual galaxies. 

\subsubsection{\lx{}-\sfr{}-\stellarmass{} scaling relations}\label{subsec:lx-sfr-m scaling relation}
Following previous works suggesting non-linear scaling of the X-ray luminosity and \sfr{} \citep[e.g.,][]{lehmer16,lehmer19,kouroumpatzakis20,riccio23} for the fit of the \lx{}-\sfr{} scaling relation, we used the following model 
\begin{equation}\label{equat:model1}
logL_{X} = A\cdot log(SFR) + B + \epsilon(log(SFR)) \, ,
\end{equation}
where $A$ is the power-law slope and $10^{B}$ is the \lx{}/\sfr{} scaling factor, and $\epsilon(log(SFR))$ is an intrinsic scatter term in log\lx{} that can depend on the \sfr{}. Our results (see Fig. \ref{Fig:M14_L16_final_sample_individuals}, Fig. \ref{Fig:M14_L16_SFR_M_D_stacks_bootstraps}) as well as previous studies \citep{lehmer19,kouroumpatzakis20} indicate that the scatter in the \lx{}-\sfr{} scaling relation is anti-correlated with the \sfr{}, especially in the lower \sfr{} regimes. For this reason, the term $\epsilon(log(SFR))$ is modeled as a Gaussian random variable with mean $\mu=0$ (i.e. around the best-fit line) and standard deviation $\sigma = \sigma_{1}logSFR+\sigma_{2}$. In particular, the intrinsic scatter due to the stochasticity (parameterised by $\sigma_{1}$) was allowed to vary linearly with the log(\sfr{}). 
The fit was performed using the Markov Chain Monte Carlo (MCMC) technique with uniform priors for the model parameters: $A$$\in$$[0,2]$, $B$$\in$ $[36,42]$, $\sigma_{1}$ $\in$ $[-1,2]$, and  $\sigma_{2}$ $\in$ $[0,2]$. We opted to use 100 walkers and 10000 iterations in order to ensure the convergence of the fit. 

In Table \ref{tab:scaling relations best fit} we present the best-fit results on the stacked data per \sfr{}-\stellarmass{}-\distance{} bin. We find that the slope is significantly sub-linear ($A=0.85^{+0.03}_{-0.09}$) indicating that the correlation between the \lx{} and the \sfr{} is not just a simple one-to-one linear relation. In addition, the dependence of the intrinsic scatter on the \sfr{} (parametrized as $\sigma_{1}$) is significantly lower than zero ($\sigma_{1}=-0.43^{+0.06}_{-0.09}$). The best-fit value for the free parameter $\sigma_{2}$ is $\sigma_{2}=0.13^{+0.05}_{-0.02}$.

\begin{figure*}
        \includegraphics[width=\columnwidth]{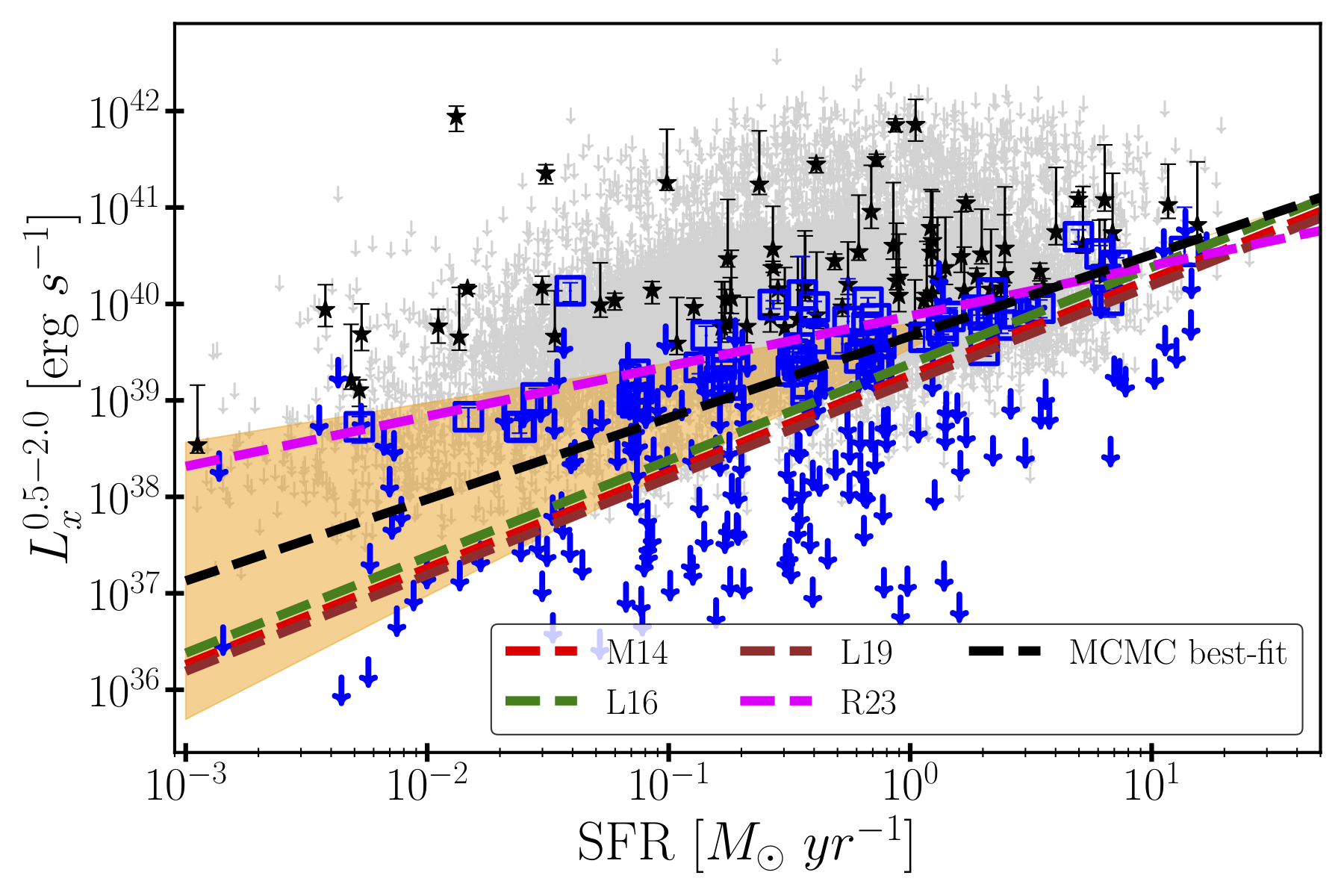}
        \includegraphics[width=\columnwidth]{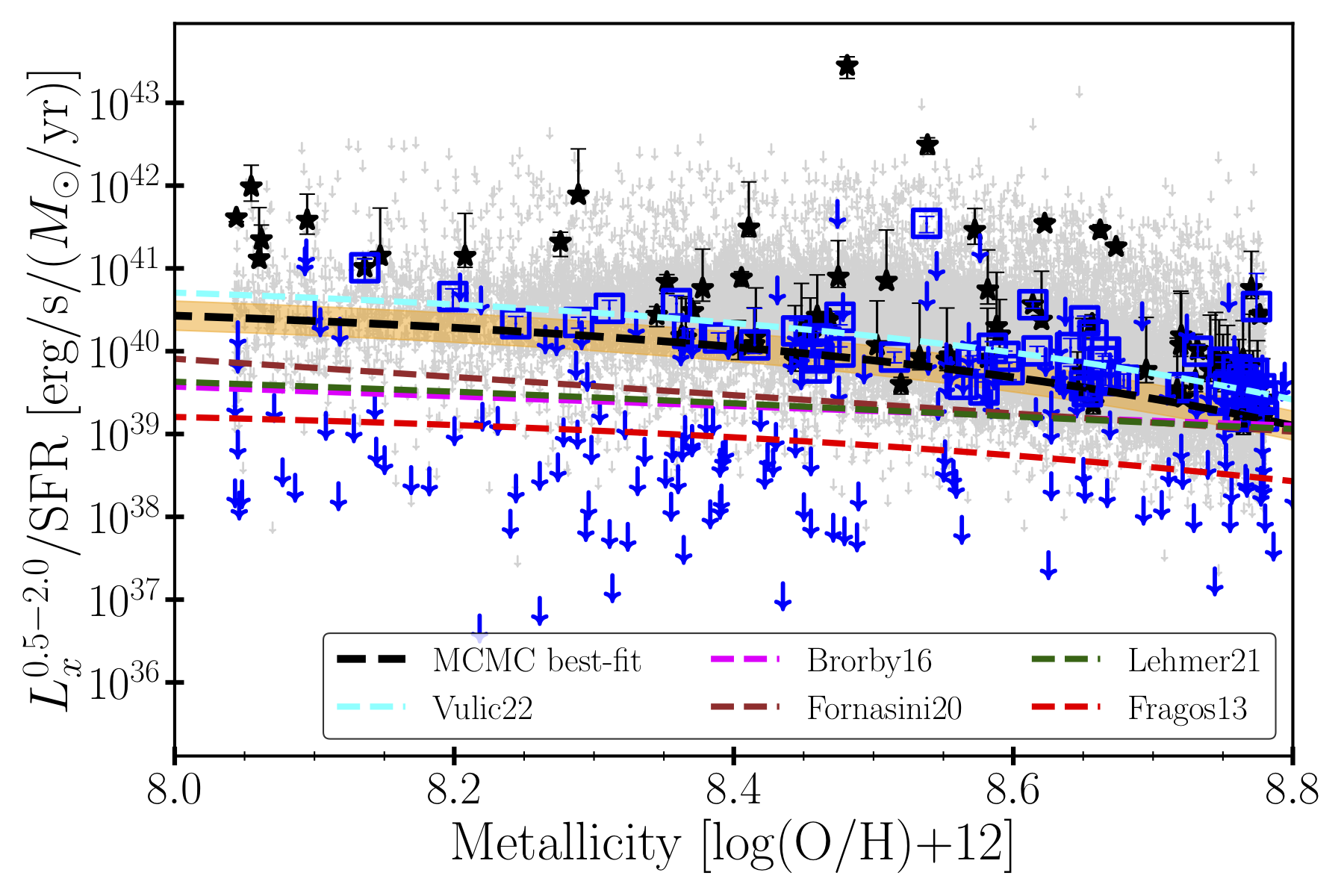}
        
    \caption{Left panel shows the distribution of the \sfr{}-\stellarmass{}-\distance{} stacks in the \lx{}$^{0.5-2}$-\sfr{} plane. Black stars show the secure star-forming galaxies with reliable flux measurements with their corresponding uncertainties, and the black down-arrows the galaxies with uncertain flux measurements. Our best-fit result is plotted with the black dashed line. The orange shade region indicates the $1\sigma$ \sfr{}-dependent scatter. Red dashed line depicts the standard scaling relation from \citet{mineo14} (M14). The green and brown dashed lines indicate the scaling relations from \citet{lehmer16} (L16) and \citet{lehmer19} (L19), respectively. Magenta dashed line shows the scaling relation from \citet{riccio23}.\newline
    Right panel shows the same distribution in the \lx{}$^{0.5-2}$/\sfr{} - Metallicity plane. Our best fit result is shown again with a dashed black line along with the $1\sigma$ scatter (orange shade region). The dashed magenta and brown line shows the best-fit line from \citet{brorby16} (B16) and \citet{fornasini20} (F20), respectively. In addition, the dashed darkgreen line indicates the best-fit line from \citet{lehmer21}. Finally, the cyan dashed line, shows the best-fit result from \citet{vulic22}. We also overlay the theoretical prediction from the XRB population synthesis best model from \citet{fragos13} (red dashed line)}
    \label{Fig:scaling relations}
\end{figure*}

In the left panel of Fig.\ref{Fig:scaling relations} we present the correlation between the X-ray luminosity and the \sfr{} following the same notation as in Fig. \ref{Fig:M14_L16_SFR_M_D_stacks_bootstraps} and our best-fit is shown with a black dashed line. The orange-shaded region indicates the $1\sigma$ \sfr{}-dependent intrinsic scatter. For comparison, we also plot the \lx{}-\sfr{} scaling relations which are widely used in the literature. The red dashed line shows the standard scaling relation M14. The green dashed line indicates the relation from L16 who fitted a combined sample of normal local galaxies and stacked sub-samples of the ${\sim}6$ Ms Chandra Deep Field-South (CDF-S). In particular, we used their best global fit (Table 3) which is redshift-dependent and accounts for the contribution of the LMXBs. However, we considered only the HMXBs component. The reason for that is that our sample is dominated by pure star-forming galaxies and the contribution of LMXB populations is negligible (see right panel of Fig. \ref{Fig:M14_L16_SFR_M_D_stacks_bootstraps}, and Sect.~\ref{sec:Excess from scaling relations}). In addition, we assumed $z=0$ given that our sample only includes galaxies within a distance of $200\, \rm{Mpc}$. The brown-dashed line shows the \lx{}-\sfr{} scaling relation from \citet{lehmer19} (L19) who presented updated constraints based on the X-ray luminosity functions of XRB populations based on a sample of 38 nearby galaxies. Finally, the magenta dashed line indicates the recent scaling relation from \citet{riccio23} who fitted the \lx{}-\sfr{} relation in a sample of normal galaxies detected in the eFEDS. Because each work performed in a different energy band we converted all of them in the adopted 0.5--2 \energyunits{} band using the conversion factors from the Table \ref{tab:conversion factors}. 

We find that our best-fit line confirms the correlation between the X-ray luminosity and the \sfr{} of the average population of star-forming galaxies in each \sfr{}-\stellarmass{}-\distance{} bin. This is in agreement with the results of the previous works. However, our relation is higher than almost all the other works, with a slightly shallower slope. This shallowness compared to the other scaling relations reflects the population of X-ray luminous galaxies, 
especially in the lower \sfr{} regimes. Only the R23 scaling relation is higher than our best fit. The reason for that is that \citet{riccio23} worked only with the eFEDs detections tracing only the positive fluctuation of the XRBs populations. On the other hand, in our analysis, we considered all galaxies including those with uncertain flux measurements. We also find that the M14, L16, and L19 relations are consistent with each other. 

We also find that the free parameter $\sigma_{1}$ which traces the intrinsic scatter is negative. This indicates for the first time that the \textit{intrinsic} scatter in the X-ray emission of the average population of star-forming galaxies anti-correlates with the \sfr{}s. The effect of the \sfr{}-dependent scatter is clearly shown for \sfr{}${<}2\,\rm{M_{\odot}}\,\rm{yr^{-1}}$ where the expected X-ray luminosity from star-forming galaxies is strongly affected by the stochastic sampling of the XRBs X-ray Luminosity Function (XLF) of the galaxies. In addition, we see that $\epsilon(log(SFR))=0$ for \sfr{}${>}2\,\rm{M_{\odot}}\,\rm{yr^{-1}}$. This behaviour has no physical origin, but instead it is the result of the linear parametrization of the $\epsilon(log(SFR))$.

As opposed to the previous studies of local galaxies, the eR1-HEC star-forming galaxy analysis is completely blind to selection biases. This is because we also consider objects that are not formally detected in the eRASS1 survey. As a result, it is more representative of the X-ray properties of normal galaxies in the local Universe, indicating that the correlation between the integrated X-ray luminosity and the \sfr{} is slightly different than the previous works which were limited to small galaxy samples.

\begin{table}
\centering
\caption{Scaling relations best-fit results.}
\label{tab:scaling relations best fit}
\begin{tabular}{cccc} 
\multicolumn{4}{c}{$log(L_{X}/\rm{erg}\,\rm{s}^{-1}) = A\cdot log(SFR/\rm{M_{\odot}}\,\rm{yr^{-1}}) + B + \epsilon(log(SFR))$} \\
\hline\hline
A & B & $\sigma_{1}$ & $\sigma_{2}$ \\
$0.85^{+0.03}_{-0.09}$ &$39.67^{+0.02}_{-0.04}$&$-0.43^{+0.06}_{-0.09}$ & $0.13^{+0.05}_{-0.02}$ \\
\hline \\
\multicolumn{4}{c}{$log(L_{X}/\rm{erg}\,\rm{s}^{-1}) = log(SFR/\rm{M_{\odot}}\,\rm{yr^{-1}}) + A\cdot (Z/Z_{\odot}) + B + \sigma$}\\
\hline\hline
A & B & \multicolumn{2}{c}{$\sigma$} \\
$-1.22^{+0.24}_{-0.10}$ &$40.68^{+0.41}_{-0.29}$&\multicolumn{2}{c}{$0.17^{+0.02}_{-0.16}$}\\
\hline 

\end{tabular}
\begin{tablenotes}
 \footnotesize 
 \item 1) $\epsilon(log(SFR))$ accounts for this intrinsic scatter and it is modeled as a Gaussian random variable with mean $\mu=0$ and standard deviation $\sigma = \sigma_{1}logSFR+\sigma_{2}$.
  \item 2) $\sigma$ accounts for the fixed scatter and it is modeled as a Gaussian random variable with mean $\mu=0$ and standard deviation $\sigma$.
\end{tablenotes}
\end{table}

Given that we are working with the integrated X-ray emission of each galaxy we have contributions from both HMXB and LMXB populations. In order to disentangle these two contributions we also performed a joint X-ray luminosity, \sfr{}, and \stellarmass{} maximum likelihood fit. Our model was of the form
\begin{equation}\label{equat:model2}
logL_{X} = log(10^{A + log(SFR)} + 10^{B+log(M_{\star})}) + \sigma \, ,
\end{equation} 
where the free parameters $A$, and $B$ trace the young stellar populations (associated with HMXBs) and the old stellar populations (associated with LMXBs), respectively. The term $\sigma$ was used in order to account for the intrinsic scatter in the data. 

However, the results of this fit were strongly unconstrained because as it is seen in the right panel of Fig. \ref{Fig:M14_L16_SFR_M_D_stacks_bootstraps}, our stacks are underpopulated in the low sSFR regime ($\rm{sSFR}{<}2\times10^{-10}\,\rm{M_{\odot}}\,\rm{yr^{-1}}/\rm{M_{\odot}}$) where the LMXB populations start to contribute significantly. Furthermore, the majority of these stacks did not yield reliable X-ray measurements. This indicates that our sample is dominated by pure star-forming galaxies. This is also confirmed by our analysis in Sect.~\ref{sec:Excess from scaling relations} which shows that the contribution of the LMXBs to the observed X-ray luminosity excess is negligible for the majority of the stacks.  As a consequence, the fitted X-ray luminosity scaling relation does not require any additional LMXB component at a statistically significant level. For that reason our sample does not allow us to make any conclusive statement about the correlation between the X-ray luminosity, \sfr{}, and the \stellarmass{}. 

In order to explore the effect of the \stellarmass{} in introducing scatter and driving the observed excess we repeated the stacking process in SFR-D bins (i.e. ignoring any dependence on the \stellarmass{}). This results in a tighter \lx{}-\sfr{} relation, as expected, but with still a flatter slope than the standard scaling relations. This indicates that stacking all available bins at low SFR does not bring the average X-ray luminosity to consistency with the standard relation, indicating that the excess is not an artifact of the binning scheme.

\subsubsection{\lx{}-\sfr{}- Metallicity scaling relation}\label{subsec:lx-sfr-metallicity scaling relation}

Several studies \citep[e.g.,][]{fornasini19,brorby16,vulic22} have shown that \lx{}/\sfr{} tends to be elevated in star-forming galaxies with lower-metallicity. Given that our sample spans about $1{dex}$ in metallicity from 8.0 subsolar to 9.0 super-solar, we can study the \lx{}-\sfr{}-Metallicity scaling relation for a wide range of the parameter space. In contrast to the observed log\lx{} scatter dependence on the \sfr{} (Sec.\ref{subsec:lx-sfr-m scaling relation}), in the right panel of Fig.\ref{Fig:scaling relations} we see that the log\lx{} scatter of our stacked data does not depend on the metallicity, since it remains constant across the entire metallicity range. Therefore, we consider a model with a fixed scatter that does not depend on metallicity:
\begin{equation}\label{equat:model2}
logL_{X} = log(SFR) + A\cdot Z + B + \sigma \, ,
\end{equation} 
where Z is the metallicity in solar units calculated from the log(O/H) metallicity: $Z=10^{(log(O/H)+12)-8.69}$ \citep[adopting $(log(O/H)+12)_{\odot}=8.69$;][]{asplund09}. The $A$ and $10^{B}$ are the power-law of the non-linear metallicity dependence and the \lx{}/\sfr{} scaling factor, respectively. The fixed scatter term $\sigma$ indicates a Gaussian random variable with $\mu = 0$ and standard deviation $\sigma$.
The fit was performed again using the MCMC technique assuming uniform priors for the model parameters $A$$\in$$[-3,1]$, $B$$\in$ $[39,44]$, and  $\sigma$ $\in$ $[0,5]$. The number of walkers and iterations was 100 and 10000, respectively. 

The best-fit results are presented in Table \ref{tab:scaling relations best fit}. We find that the index is negative and sub-linear ($A=-0.94^{+0.29}_{-0.12}$) which is consistent with the results from previous works \citep[e.g.,][]{fornasini19,brorby16,vulic22}. In addition, we find that the fixed scatter of the \lx{}/\sfr{} is $\sigma=0.17^{+0.02}_{-0.16}$ and is not correlated with the metallicity. 
In the right panel of Fig. \ref{Fig:scaling relations} we present the best-fit line for the \lx{}-\sfr{} - Metallicity relation (black dashed line) along with the metallicity-dependent scatter (orange shaded region). Following the same nomenclature as in Fig. \ref{Fig:M14_L16_SFR_M_D_stacks_bootstraps}, the blue squares and down arrows indicate the stacked data per \sfr{}-\stellarmass{}-\distance{} bin. The black stars and the lightgray down arrows represent the individual HEC-eR1 galaxies with reliable and uncertain fluxes, respectively. For comparison, we overplot the empirical relations from a number of works. The dashed magenta line shows the best-fit line from a sample of 10 Lyman break analogs (LBAs), as presented by \citet{brorby16}. The dashed brown line indicates the best-fit result from \citet{fornasini20} on stacked samples of galaxies spanning a redshift range ${z}\simeq 0.1-0.9$. In addition, the dashed darkgreen line indicates the best-fit line from \citet{lehmer21} who analyzed \chandra{} data from 55 actively star-forming nearby (${\distance{}}{\lesssim} 30\,\rm{Mpc}$) galaxies within a wide metallicity range 12+log(O/H)$\simeq 7-9.2$. For these works, we used the conversion factors c$_{5}$, c$_{6}$, and c$_{7}$ (see Table \ref{tab:conversion factors}), respectively, to convert the X-ray luminosity from their respective bands to the 0.5--2 \energyunits{} band. Finally, with the cyan dashed line, we present the best-fit result from \citet{vulic22} who studied the \lx{}/\sfr{} - Metallicity dependence for all the star-forming galaxies detected in the eFEDs field which reaches the full depth of \erosita{} survey (eROSITA:8). We also overlay the theoretical prediction from the XRB population synthesis best model from \citet{fragos13} (red dashed line). We find that our results are in general agreement with those works but they show higher X-ray luminosities. In Sect.~\ref{Sec:Metallicity dependence} we further discuss these results in more detail and compare them with the current \lx{}-\sfr{} - Metallicity scaling relations. 

\section{Discussion}\label{sec:Discussion}
\subsection{Excess from the standard scaling relations}\label{sec:Excess from scaling relations}
Our best-fit results from the \sfr{}-\stellarmass{}-\distance{} stacks, as well as the distribution of the individual galaxies, indicate that the local galaxy population shows a significant excess from the standard scaling relations of M14 and L16. In addition, this excess anti-correlates with the \sfr{} of the host galaxy confirming similar indications from previous studies which have been performed on much smaller galaxy samples \citep{lehmer16,kouroumpatzakis20}. 

\begin{figure*}
        \includegraphics[width=17cm]{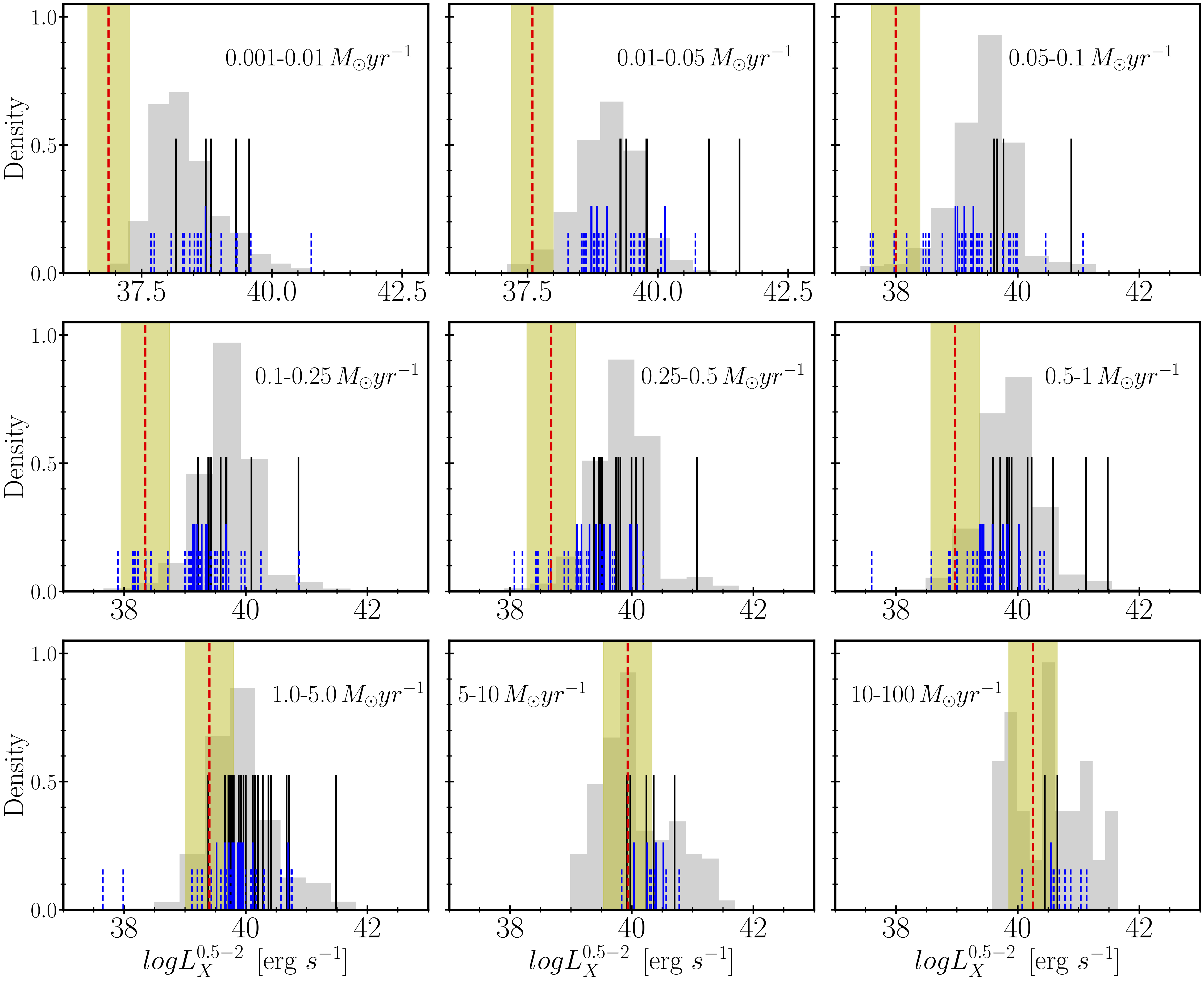}
    \caption{\lx{}$^{0.5-2}$ distribution of HEC-eR1 secure star forming galaxies in different \sfr{} bins. Black solid lines indicate the \lx{} from galaxies with reliable flux measurements, and light gray histogram the 90$\%$ C.I of the \lx{} from galaxies with uncertain flux measurements. Blue solid and dashed lines show the stacked \lx{} for the stacks with reliable and uncertain flux measurements, respectively. Red dashed line and yellow band, show the expected \lx{} and the $1 \sigma$ scatter from \citealp[]{mineo14} for the mean \sfr{} of the galaxies in a \sfr{}-\stellarmass{}-\distance{} bin.}
    \label{Fig:M14_individuals_SFR_M_D_stacks_sfr_bins_hist}
\end{figure*}

In order to assess the significance of this excess, in Fig. \ref{Fig:M14_individuals_SFR_M_D_stacks_sfr_bins_hist} we present the distribution of the 0.5--2 \energyunits{} X-ray luminosity of the galaxies in the different \sfr{} bins we considered. This plot can be seen as an alternative version of the \lx{}-\sfr{} relation (Fig. \ref{Fig:M14_L16_SFR_M_D_stacks_bootstraps}) where each subplot represents a vertical slice with respect to the \sfr{}. 
In each subplot, the black solid lines indicate the \lx{}$^{0.5-2}$ from individual galaxies with reliable flux measurements, and the light gray histogram the \lx{}$^{0.5-2}$ from individual galaxies with uncertain flux measurements. For the latter histograms, we used the upper $90\%$ C.I. of the X-ray luminosity, and as the result, the actual value of the \lx{} is definitely lower. The blue solid and dashed lines show the \lx{}$^{0.5-2}$ distribution for the stacks with reliable and uncertain flux measurements, respectively. The red dashed line, shows the expected \lx{} from the M14 for the mean \sfr{} of the galaxies in each bin, while the yellow shaded region indicates the 1$\rm{\sigma}$ scatter from the best-fit line of M14.

As we can see, the integrated X-ray luminosity of the individual HEC-eR1 star-forming galaxies is significantly elevated with respect to the expected X-ray luminosity for a given \sfr{}. This excess is very strong in the low \sfr{} regime (${\lesssim}\,1\,\rm{M_{\odot}\,yr^{-1}}$) and it is decreasing towards higher \sfr{}s. For the low \sfr{}s the individual galaxies with reliable flux measurements reach very high X-ray luminosities up to typically ${\gtrsim}10^{42} \, \rm{erg\,s^{-1}}$, well above the M14 scaling relation. On the other hand for high \sfr{}s the X-ray luminosity of the individual galaxies is more consistent with the M14 scaling relation. We should note here that although the M14 relation is measured for galaxies down to \sfr{}\,${\sim}0.1\,\rm{M_{\odot}\,yr^{-1}}$, \citet{kouroumpatzakis20} showed that it can be extended to lower \sfr{}s.

Although the eRASS1 sensitivity limit does not allow us to measure precisely the flux for the majority of the galaxies in our sample, the distributions of the average luminosities of the stacks indicate a noteworthy trend. While many stacks are consistent or slightly above the standard scaling relation, we see that the upper 90$\%$ C.I. of the X-ray luminosity distribution of galaxies with uncertain flux measurements falls consistently well below the M14 scaling relation, specifically for \sfr{} ${>}1\,\rm{M_{\odot}\,yr^{-1}}$. In particular, for the two highest \sfr{} bins the distributions are almost symmetric around the scaling relation. This is the result of two factors: (i) the integrated X-ray luminosity is high enough to be detectable given the higher photon statistics of the stacks and their sensitivity limit (given their total exposure time and background), (ii) adequate sampling of the XRBs luminosity function since for these very highly star-forming galaxies stochastic effects are minimized, resulting in a large number of XRB populations with X-ray luminosities close to that expected based on the average scaling relation for a given \sfr{}. 

These results alongside the behaviour of the individual galaxies imply that: i) by taking into account the stacks with uncertain fluxes, the current scaling relation from M14 can describe well the correlation between the average X-ray luminosity and the average \sfr{} of the HEC-eR1 star-forming galaxies for \sfr{} ${>}0.5\,\rm{M_{\odot}\,yr^{-1}}$, ii) for ${<<} 0.5\,\rm{M_{\odot}\,yr^{-1}}$ the measured average X-ray luminosity is much higher than expected from the M14 relation (Fig. \ref{Fig:M14_excess_SFR_M_D_stacks},\ref{Fig:scaling relations}), and iii) the scaling relation proves inadequate in accurately describing the individual galaxies since it does not account for the much larger intrinsic scatter (Fig. \ref{Fig:scaling relations}). 

Apart from the excess, another very interesting result in Fig. \ref{Fig:M14_individuals_SFR_M_D_stacks_sfr_bins_hist} is the increased scatter in comparison to the 1$\sigma$ scatter of the M14 relation. In particular, we see an anticorrelation of the scatter of the HEC-eR1 star-forming galaxies with the \sfr{}. This is clearly shown by the distribution of the X-ray luminosities of the individual galaxies around the scaling relation. As the \sfr{} increases, the scatter decreases and the \lx{} distribution of the individual galaxies with reliable or uncertain flux measurements becomes tighter and more consistent with the standard scaling relation. This is confirmed by the anti-correlation of the intrinsic scatter term with \sfr{} in the scaling relation fits (Sect. \ref{subsec:lx-sfr-m scaling relation}).

A possible explanation for the increased excess and scatter could be the contamination by background AGNs within the $D_{25}$ of the galaxies. However our results in Sect.~\ref{sec:bkg-AGN-contamination} suggest that the contamination in our sample is very low to significantly affect the measured fluxes and the corresponding luminosities, especially in the low \sfr{} regimes.

Another source that potentially contributes to the integrated X-ray luminosity (and as a result to the measured X-ray luminosities) is the diffuse emission from hot-ionized gas. \citet{lehmer22} calculated the scaling relation between the hot-gas luminosity and the \sfr{} by analyzing \chandra{} observations of a sample of 30 very actively star-forming low-metallicity galaxies. By taking into account the effect of metallicity in the XRB component they fitted a global spectral model considering an absorbed power-law component corresponding to the XRB component, and two absorbed thermal plasma components of different temperatures. By using the best-fit parameters from this work (Table 3), we calculated the relative contribution of the HMXB and the hot-gas components given the \sfr{} of the galaxies in each stack. We found that the hot-gas contribution is ${\sim}32\%$ in the 0.5--2 \energyunits{} energy band with the hot gas component being negligible above 2 \energyunits{}. In addition, \citet{mineo12b} measured the hot-gas emission from a sample of local galaxies which covered a broad range of \sfr{} and \stellarmass{} and they found that the X-ray luminosity of the hot-gas correlates with the \sfr{} as $L_{X}^{gas}/\sfr{}=5.2\times10^{38}\,\rm{erg\,s^{-1}}/\rm{M_{\odot}\,yr^{-1}}$ in the 0.5--2 \energyunits{} energy band. Furthermore, they found that the scaling relation for total X-ray luminosity (HMXBs + hot-gas) and the \sfr{} is $L_{X}^{tot}/\sfr{}=4.0\times10^{39}\,\rm{erg\,s^{-1}}/\rm{M_{\odot}\,yr^{-1}}$ in the 0.5--8 \energyunits{} energy band. By using the conversion factor c$_{3}$ from Table \ref{tab:conversion factors}
we converted the scaling relation from the 0.5--8 \energyunits{} band to 0.5--2 \energyunits{} and we calculated that the hot-gas contribution fraction is ${\sim}40\%$ of the total (HMXBs + hot gas) X-ray luminosity in the 0.5--2 \energyunits{} energy band.
In Sect.~\ref{sec:hot_gas_contribution} we calculated the hog-gas contribution by fitting the stacked X-ray spectrum of all the galaxies in our sample, and we found it ${\sim}13\%$ in the 0.5-2 \energyunits{} energy band. This result corresponds to an X-ray luminosity excess of $0.05\,\rm{dex}$ which is not enough to explain the observed excess and the scatter that we see in our HEC-eR1 sample. We see that our estimation of the hot-gas contribution is much lower than that inferred from the analysis of \citet{lehmer22} and \citet{mineo12b}. This difference could be due to two reasons: i) the galaxy samples of \citet{mineo12b} and \citet{lehmer22} are more focused on intensely star-forming galaxies resulting in higher hot gas contribution, ii) our average stacked X-ray spectrum, includes a large fraction of individual galaxies that have undetectable X-ray fluxes which do not allow us to account for the very weak hot gas component. On the other hand, since our galaxy sample is completely blind we estimate the hot-gas contribution from a more representative population of star-forming galaxies.

X-ray emission originating from LMXBs is another potential contributor to the overall integrated emission of galaxies. More specifically, the X-ray emission produced by LMXB populations usually dominates the X-ray output of galaxies in the very low sSFR regimes \citep[see][and references there in]{fabbiano19}. Given that our sample of galaxies reaches \sfr{} as low as ${\sim}10^{-3}\,\rm{M_{\odot}\,yr^{-1}}$ and sSFR values around sSFR${\sim}3\times10^{-12}\,\rm{M_{\odot}\,yr^{-1}\,M_{\odot}^{-1}}$ it is possible that the observed excess is due to emission from LMXB populations. 

In order to estimate the contribution of the LMXB populations we calculated the total luminosity by integrating their X-ray luminosity function (XLF):
\begin{equation}
L_{LMXB} = \int^{L_{c}}_{L_{min}}\frac{dN_{LMXB}}{dL}\cdot L \cdot dL .
\end{equation}
We adopted the XLF of \citet{lehmer19} (Eq.2) who parameterised it with a broken power-law and by using the best-fit parameters given in their Table 4 (column 6) we calculated the expected total 0.5--8 \energyunits{} X-ray luminosity from LMXBs for a given \stellarmass{}. 
As low limit in the integration, we assumed $L_{min}=10^{36}\rm{erg\,s^{-1}}$ and for $L_{c}$ we used the value $L_{c}=5\times10**{40}\,\rm{erg\,s^{-1}}$. Using the conversion factor c$_{4}$ (see Table \ref{tab:conversion factors}) we converted the X-ray luminosity from the 0.5--8 \energyunits{} energy band to the 0.5--2 \energyunits{} band adopted in this work. Then the LMXB contribution is
\begin{equation}\label{equat:lmxb_contam}
f_{\rm{LMXB}} = \frac{L^{0.5-2}_{LMXB}}{L^{0.5-2}}\times100\, \%, 
\end{equation}
where $L^{0.5-2}_{LMXB}$ is the expected luminosity based on the XLF of LMXBs and $L^{0.5-2}$ is the measured integrated X-ray luminosity. 

\begin{figure}
        \includegraphics[width=\columnwidth]{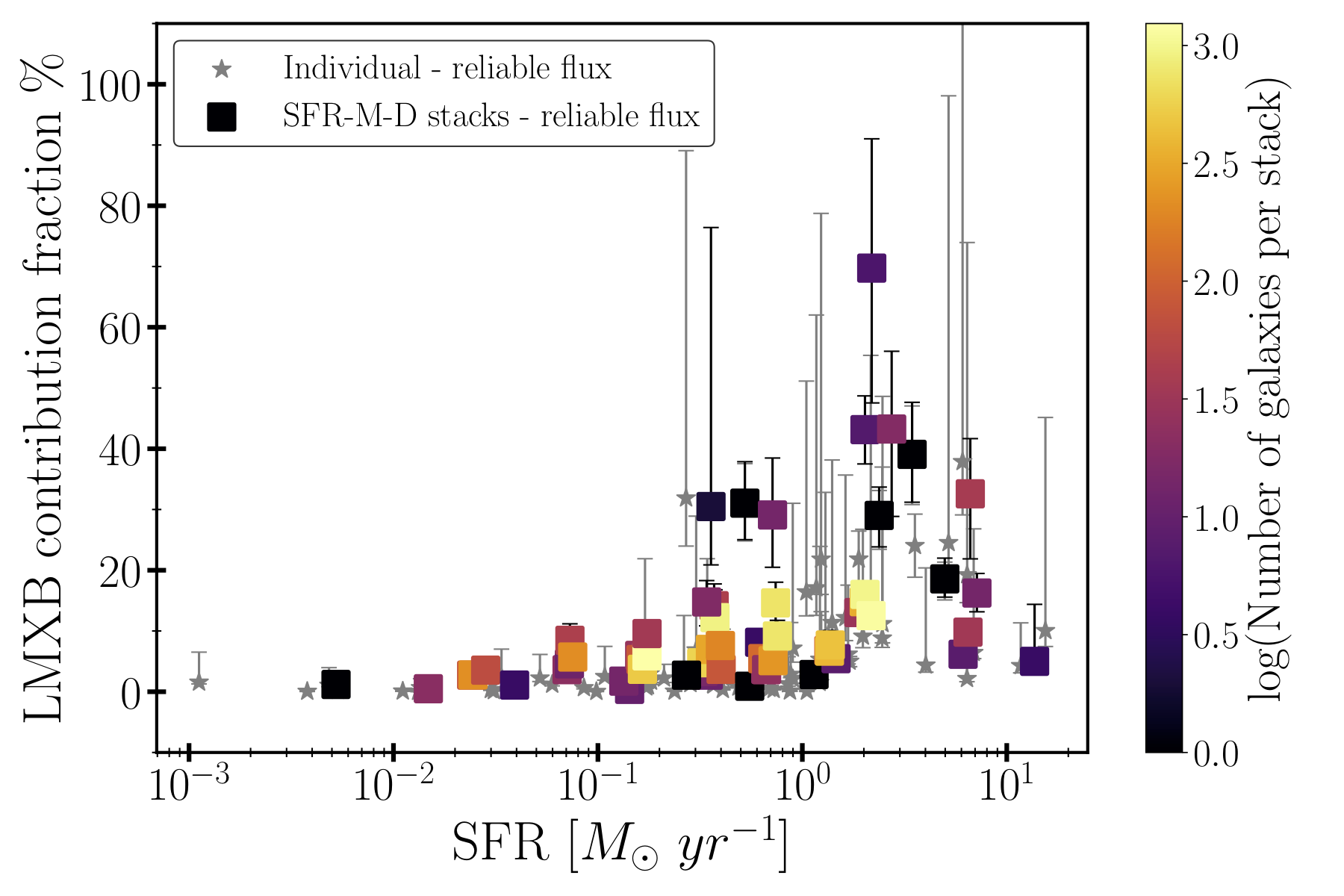}
        \includegraphics[width=\columnwidth]{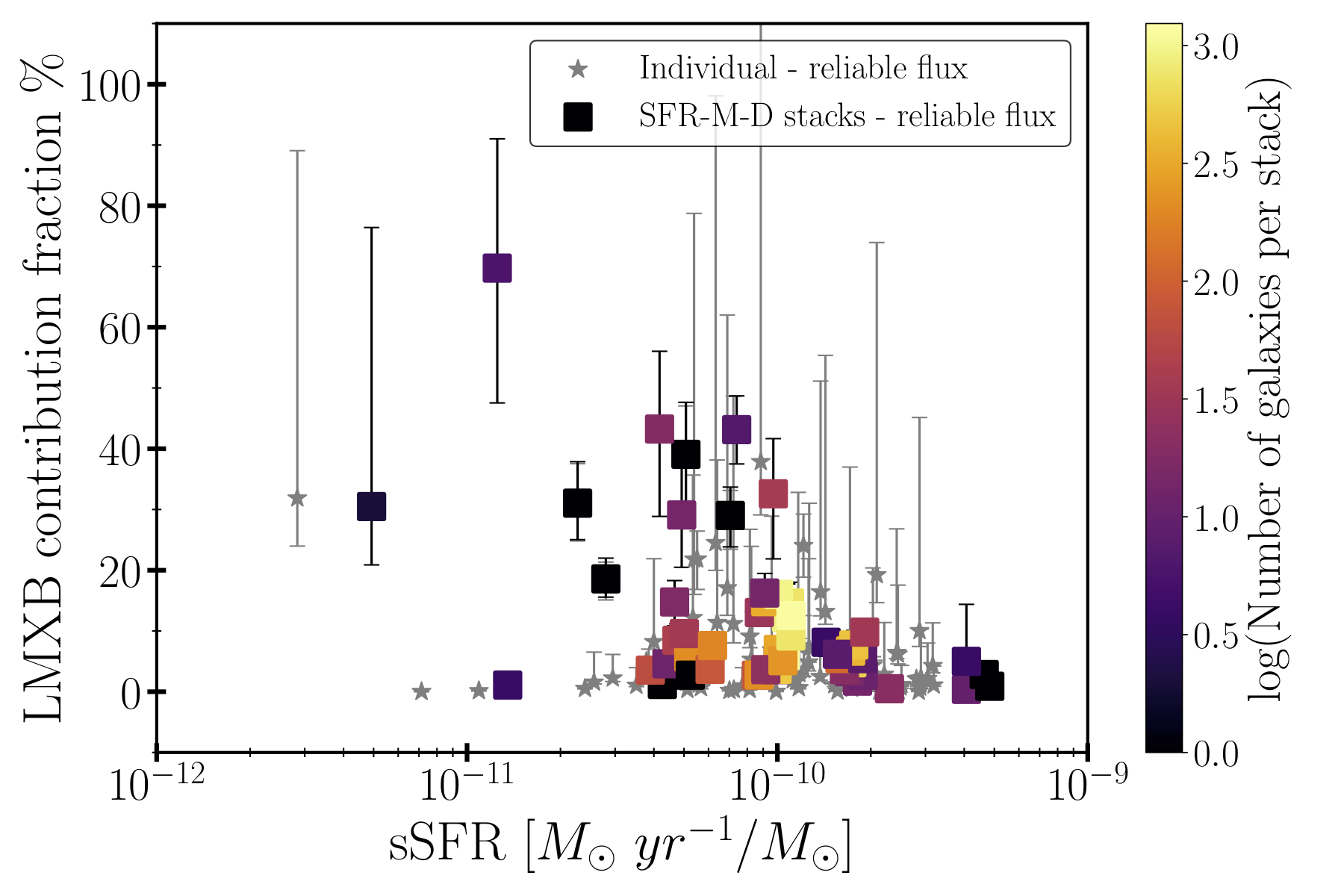}
    \caption{The percentage of LMXB contribution as a function of the \sfr{} (top panel) and the sSFR (bottom panel), respectively. The gray stars represent the individual galaxies with reliable flux measurements and the squares the reliable stacks. The color code indicates the logarithm of the number of galaxies that participate in each \sfr{}-\stellarmass{}-\distance{} stack. The errorbars were calculated assuming the lower and upper $68\%$ C.I. of the X-ray luminosity distribution.}
    \label{Fig:LMXBs_contribution}
\end{figure}

We performed the calculation for both, the individual galaxies and the \sfr{}-\stellarmass{}-\distance{} stacks with reliable flux measurements. In this way, we can estimate the impact of the LMXB populations on the observed excess on a galaxy by galaxy basis, as well as on the average galaxy population per \sfr{}-\stellarmass{}-\distance{} bin. In Fig. \ref{Fig:LMXBs_contribution} we plot the percentage of LMXB contribution as a function of the \sfr{} and the sSFR, respectively. The gray stars represent the individual galaxies with reliable flux measurements and the color-coded squares the reliable stacks. The color code indicates the logarithm of the number of galaxies that participate in each \sfr{}-\stellarmass{}-\distance{} bin. The errorbars were calculated assuming the lower and upper $68\%$ C.I. of the X-ray luminosity distribution for each measurement. 
We see that for both populations (individual galaxies, and stacks) the LMXB contribution fraction is negligible (${<}10\%$) for the low \sfr{}s and it is slightly increased (${\sim}$ a factor of 3) for higher \sfr{} values. This result implies that the excess of up to ${\sim}2\,\rm{dex}$ in galaxy populations with \sfr{}${<}0.1\,\rm{M_{\odot}\,yr^{-1}}$ (see Fig. \ref{Fig:M14_excess_SFR_M_D_stacks}) cannot be attributed to emission from LMXB populations. In addition, we find that the LMXB contribution anticorrelates with the sSFR as it is expected from the connection of the LMXBs with the old stellar populations of the host galaxy. For a few stacks, the estimated contribution is ${>}40\%$ and highly uncertain. This is attributed to two reasons: (i) as shown in Fig. \ref{Fig:SFR_M_parameter_space_SFR_M_D_stacks} the asymmetry in the \sfr{}-\stellarmass{} bins results in wider \stellarmass{} bins compared to \sfr{} bins. As a consequence, the computed mean \stellarmass{} is biased toward higher stellar masses resulting in lower mean sSFRs for these stacks and hence larger inferred LMXB contribution. This bias is also strengthened by the systematically smaller number of individual galaxies within these particular bins, which introduce larger statistical fluctuation in the calculation of the mean sSFRs; (ii) the small number of galaxies included in those bins coupled with the eRASS1 sensitivity limit, results in large uncertainties in the calculation of the integrated X-ray luminosity for the more distant bins.

It is clear from this analysis that the substantial excess and the large scatter, indicated by our findings, cannot be the result of contamination by background sources, contribution of hot gas emission, or the presence of LMXB populations. This suggests that the observed behavior is likely influenced by either the characteristics of the host galaxies' stellar populations, such as metallicity and stellar population age, or the inherent stochastic sampling of the HMXB XLF. In the next sections, we discuss in detail these factors and their role in the observed excess and dispersion in relation to the standard scaling relations.

\subsection{Metallicity dependence}\label{Sec:Metallicity dependence}
It is well known from previous studies \citep[e.g.,][]{basuzych13,fornasini19,lehmer21,vulic22} that the integrated X-ray emission produced by HMXB populations is strongly correlated with the gas-phase metallicity of star-forming galaxies. In particular, it has been found that the \lx{}/\sfr{} tends to be elevated in galaxies with lower metallicity when compared to those with nearly solar metallicity. Moreover, it has been established that lower metallicity plays a significant role in the observed scatter of the \lx{}-\sfr{} scaling relation in the lower \sfr{} regimes \citep{kouroumpatzakis20,kouroumpatzakis21,vulic22}. These results also extend to higher redshift galaxies indicating that metallicity is responsible for the redshift evolution of HMXBs \citep[e.g.,][]{basuzych13,fornasini19,fornasini20}. The observational constraints from the previous works are in general agreement with the theoretical predictions from binary population synthesis models which suggest that the observed \lx{} excess and scatter are driven by the metallicity \citep{fragos13,madau17}. This \lx{}-\sfr{}-Metallicity dependence arises from the fact that lower metallicity stellar populations are expected to have relatively weak stellar winds resulting in less angular momentum loss from the binary systems, and hence tighter orbits. Consequently, more Roche lobe overflow systems are formed resulting in higher accretion rates and thus in higher \lx{}/\sfr{} \citep[e.g.][]{linden10}. 
  
Given that our galaxy sample spans a metallicity range from sub-solar (down to [log(O/H)+12]${\sim}8$) to super-solar (up to [log(O/H)+12]${\sim}9$), in Sect. \ref{subsec:lx-sfr-metallicity scaling relation} we fitted the \lx{}-\sfr{}-Metallicity relation to the stacked HEC-eR1 galaxy data per \sfr{}-\stellarmass{}-\distance{}. In the right panel of Fig. \ref{Fig:scaling relations} we present our best-fit result (black dashed line) in the form of \lx{}/\sfr{} as a function of the gas-phase metallicity. The shaded orange region shows the $1\sigma$ intrinsic \lx{} metallicity-dependent scatter. 

We see that our best-fit line follows well the distribution of the stacked star-forming HEC-eR1 sample of galaxies confirming the anti-correlation between the gas-phase metallicity and the \lx{}/\sfr{}. As the metallicity of the host galaxy decreases, its total X-ray output increases, in agreement with the theoretical predictions and the previous observational results. Both our dataset of individual galaxies and the stacked data with reliable flux measurements (depicted as black stars and blue open squares, respectively) exhibit an excess of ${\sim}2.5\,\rm{dex}$ above the best-fit lines established by \citet{brorby16,fornasini19}, and \citet{lehmer21}. This is confirmed by our best-fit line which is higher than the current scalings relation by ${\sim}0.5$dex, indicating once again the effect of a previous overlooked population of X-ray luminous galaxies, particularly in lower metallicities. 
Interestingly, we find that the best-fit line for the individual galaxies detected in \citet{vulic22} is higher than our line (which represents the average population of galaxies, including stacks with reliable and uncertain flux). This is expected given the higher sensitivity limit of the eFEDs survey (eight times deeper than eRASS1), and the fact that \citet{vulic22} used only X-ray detections without taking into account upper limits. On the other hand, \citet{vulic22} lacks the exceedingly X-ray luminous galaxies detected in the eRASS1 survey because of the large difference in the area between the two surveys, which does not allow the detection of rare objects within the 140 $\deg^{2}$ region covered by the eFEDs survey. 

As discussed earlier, lower metallicity galaxies are more likely to produce X-ray luminous sources. In addition, lower-\sfr{} galaxies, which are more sensitive to stochastic effects, also tend to have lower metallicities. The combination of these two effects results in the observed increase of the excess towards to lower metallicities.
A quantitative analysis of the role of stochastic sampling as well as, the stellar population age, on the observed excess is presented in the following sections.

\subsection{Stellar age dependence}\label{Sec:Stellar age dependence}
Besides the effect of metallicity in the observed X-ray luminosity excess (see Sect.~\ref{Sec:Metallicity dependence}) another possible factor that contributes to the enhanced integrated X-ray luminosity of a star-forming galaxy, is the age of its stellar populations. 
Theoretical predictions of XRB population synthesis models \citep[e.g.,][]{fragos13} show that the X-ray emission of an XRB population is higher at very young ages (${<}20$\,Myrs), gradually declining up to ${\sim}100\,\rm{Myr}$.
This age dependence is also confirmed by a number of recent observational studies \citep{antoniou16,lehmer17,antoniou19,gilbertson22}.

In order to test the role of stellar population age in the enhanced X-ray emission of some of the \sfr{}-\stellarmass{}-\distance{} bins used in our stacking analysis we performed the following analysis. We obtained \ion{He}{I}\,5876 and \ion{H}{${\alpha}$}\,6563 flux measurements by cross-matching the spectroscopically confirmed star-forming galaxies of the eRASS1 sample with the \textit{MPE-JHU DR8} galaxy catalogue (based on the SPECOBJID) which provides stellar continuum subtracted emission-line measurements. This resulted in a sample of 13126 galaxies with available \ion{He}{I}\,5876 and \ion{H}{${\alpha}$} flux measurements. We then calculated the ratio $\frac{\ion{He}{I}}{\ion{H}{$\alpha$}}$ for the galaxies with \ion{H}{${\alpha}$} S/N ratio ${>5}$ , and \ion{He}{I}\,5876 S/N ratio ${>3}$ (8287 galaxies). The different S/N thresholds were chosen in order to maximize the sample size with available measurements given the weakness of the \ion{He}{I}\,5876 line. For the remaining 4837 galaxies with \ion{He}{I}\,5876 S/N ratio ${<3}$ we set the $\frac{\ion{He}{I}}{\ion{H}{$\alpha$}}=0$. In this way, we can quantify the relative contribution of the different age stellar populations in each galaxy. Higher flux ratios indicate stronger contribution of very young stellar populations (${\lesssim}\,8\,\rm{Myr}$) in the ionizing emission. On the other hand, when the ratio is close to zero, older (${\sim} 30 \,\rm{Myr}$) stellar populations dominate.

The exact age dependence depends strongly on the stellar evolution models assumed and especially the role of stripped stars \citep[e.g.,][]{gotberg20}. Nonetheless, the \ion{He}{I} lines trace younger stellar population than the Balmer lines. While XRB populations may also produce strong ionizing radiation, \citet{garofali23} showed that the contribution of XRBs in the ionization of the \ion{He}{I}\,5876 line is negligible indicating that this line is only due to the ionized gas around very young stellar populations and hence a good age tracer.

\begin{figure*}
        \includegraphics[width=\columnwidth]{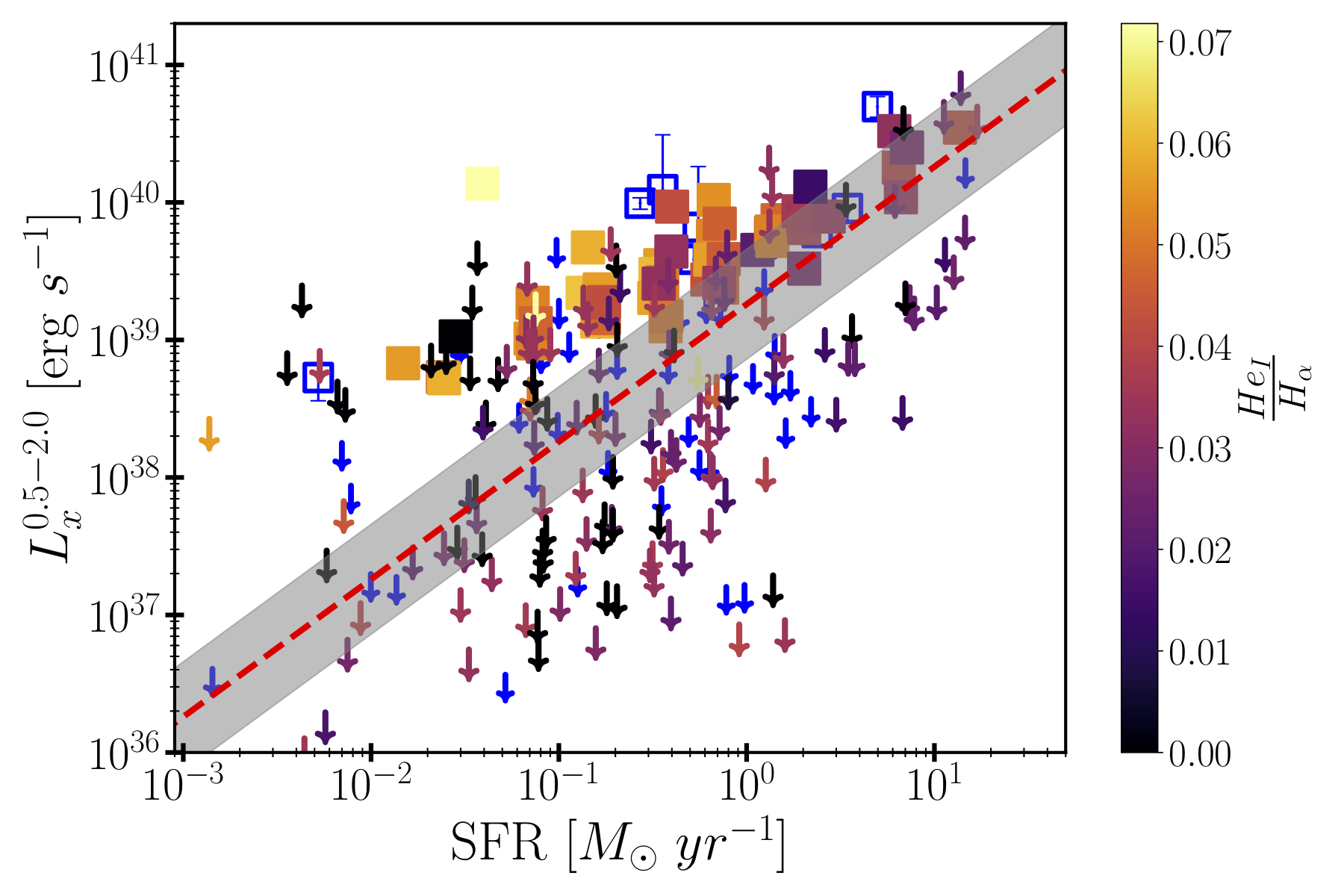}
        \includegraphics[width=\columnwidth]{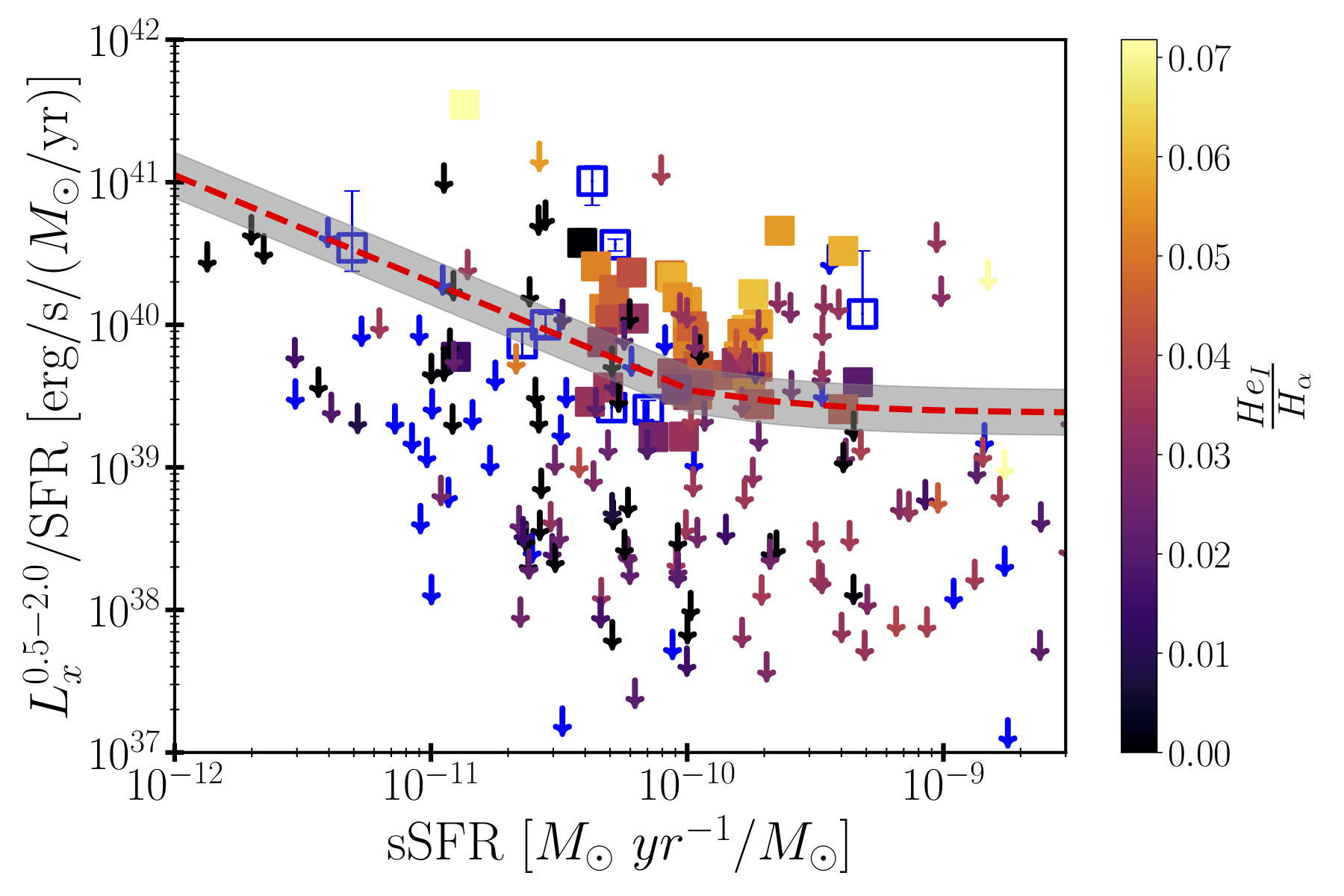}
    \caption{Left panel shows the distribution of the \sfr{}-\stellarmass{}-\distance{} stacks in the \lx{}$^{0.5-2}$-\sfr{} plane color coded with the \ion{He}{I}/\ion{H}{$\alpha$} ratio. Red line depicts the standard scaling relation from \citealp[]{mineo14} (M14). Right panel shows the same distribution in the \lx{}$^{0.5-2}$/\sfr{} - sSFR plane and the red line the scaling relation from \citealp[]{lehmer16} (L16).}
\label{Fig:M14_L16_SFR_M_D_stacks_color_code_with_HeIHa}
\end{figure*}

In order to connect this age information with the average X-ray luminosity of the galaxy population in our sample, we calculated the median $\frac{\ion{He}{I}}{\ion{H}{$\alpha$}}$ of the individual galaxies participating in each \sfr{}-\stellarmass{}-\distance{} bin. This resulted in 189 out of 239 bins with available measurements for $\frac{\ion{He}{I}}{\ion{H}{$\alpha$}}$. The remaining 50 bins did not include any galaxies with \ion{He}{I} and \ion{H}{$\alpha$} measurements so we did not consider them in our analysis. In Fig. \ref{Fig:M14_L16_SFR_M_D_stacks_color_code_with_HeIHa} we show the distribution of the \sfr{}-\stellarmass{}-\distance{} stacks in the \lx{}-\sfr{} and the \lx{}/\sfr{}-sSFR plane, respectively. The stacks are color-coded with the median $\frac{\ion{He}{I}}{\ion{H}{$\alpha$}}$. The open blue squares and the blue down-arrows indicate the stacks for which \ion{He}{I} and \ion{H}{$\alpha$} emission line measurements were not available. For comparison, we overplot the scaling relations from M14 and L16.

As we see there is a clear correlation of the elevated X-ray luminosity with the median $\frac{\ion{He}{I}}{\ion{H}{$\alpha$}}$ of the stellar populations in each \sfr{}-\stellarmass{}-\distance{} bin. For a given \sfr{} the ratio increases towards higher \lx{} (including the bins with uncertain luminosities which show systematically lower $\frac{\ion{He}{I}}{\ion{H}{$\alpha$}}$ ratios). In particular, in the lower \sfr{} regime (${<}1\,\rm{M_{\odot}\,yr^{-1}}$) where the excess from the M14 scaling relation (Fig. \ref{Fig:M14_excess_SFR_M_D_stacks}) and the scatter are larger, the gradient of the $\frac{\ion{He}{I}}{\ion{H}{$\alpha$}}$ ratio for the same \sfr{} is stronger. This indicates that the stacks with high \lx{} include galaxies hosting younger stellar populations (${\lesssim}\,8\,\rm{Myr}$) compared to stacks with lower \lx{} which tend to include older stellar populations (${<} 30\,\rm{Myrs}$). For higher SFRs the effect of age is not so strong, and the gradient in the $\frac{\ion{He}{I}}{\ion{H}{$\alpha$}}$ ratio is weaker. This is expected since at these high \sfr{}s the stellar populations of these galaxies are unlikely to be dominated by short-duration very recent (${\lesssim}\,10\,\rm{Myr}$) star-formation bursts. Instead, they tend to be dominated by longer star-forming episodes, reducing the effect of strong age or metallicity variations in their X-ray output and subsequently the observed scatter and deviation from the average scaling relations. 

These results are also confirmed by the trends in the \lx{}/\sfr{}-sSFR plane. For a given sSFR the ratio of the age tracers increases towards higher \lx{}/\sfr{} indicating that the average population of galaxies participating in those \sfr{}-\stellarmass{}-\distance{} bins is dominated by younger stellar populations. On the other hand, stacks with lower sSFRs show systematically lower $\frac{\ion{He}{I}}{\ion{H}{$\alpha$}}$ ratio. This is normal since for the galaxies with less active star-formation we expect that the HMXB populations will be correlated with slightly older stellar populations which will fall in the age window traced by the \ion{H}{$\alpha$} and their emission in the \ion{He}{I} will be weak. 

\begin{figure}
        \includegraphics[width=\columnwidth]{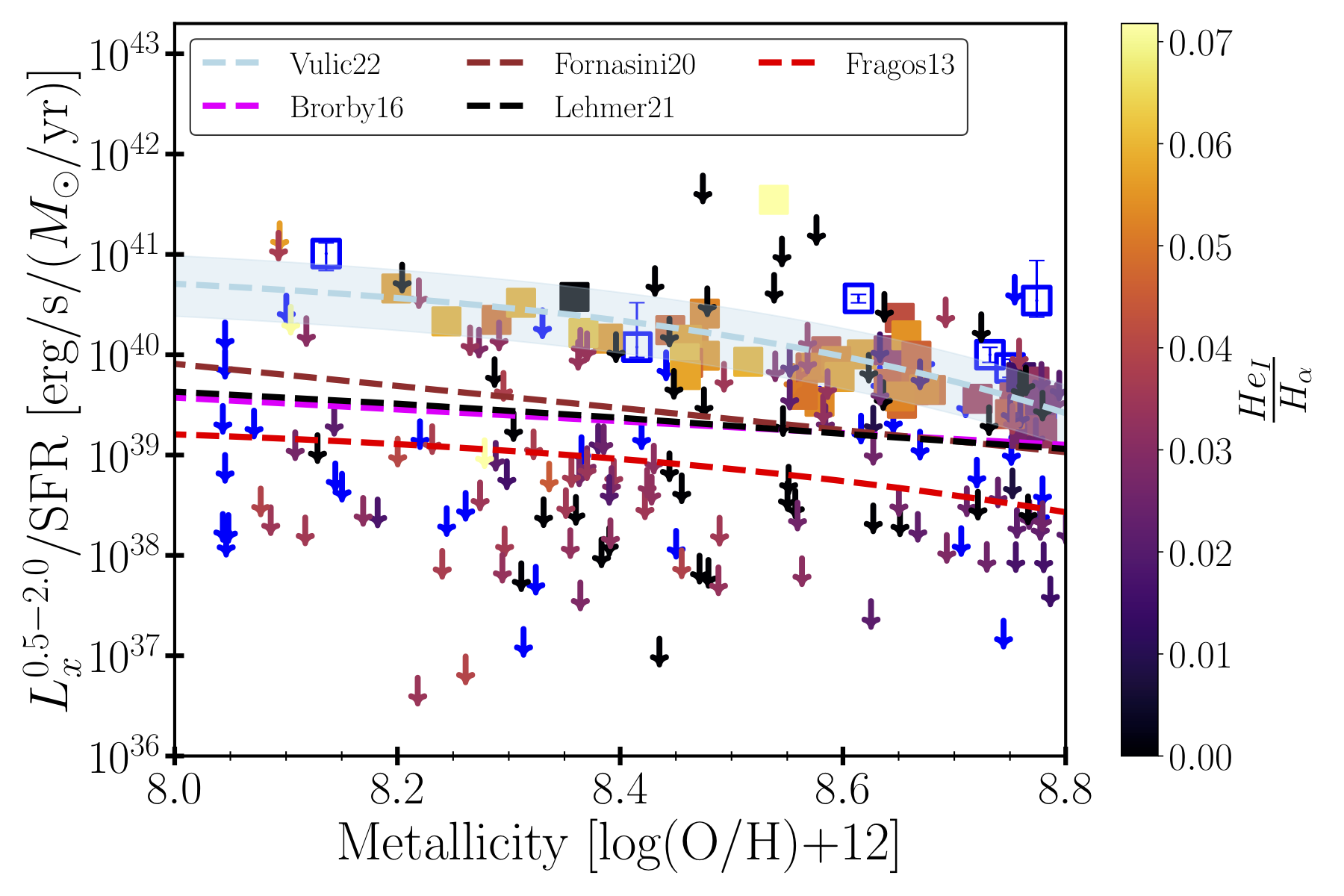}
        \caption{\lx{}$^{0.5-2}$/\sfr{} as a function of galaxy gas-phase metallicity for the \sfr{}-\stellarmass{}-\distance{} stacks in the color coded with the \ion{He}{I}/\ion{H}{$\alpha$} ratio. For comparison we overplot the empirical relations from \citet{brorby16} (dashed magenta line), \citet{fornasini20} (dashed brown line), \citet{lehmer21} (dashed black line), and \citet{vulic22} (dashed cyan line). The shaded region indicates the $1\sigma$ scatter around the latter fit. We also overlay the theoretical prediction from the XRB population synthesis best model from \citet{fragos13} (red dashed line).}
\label{Fig:Lx_05_2_Metallicity_SFR_M_D_stacks_color_code_with_HeIHa}
\end{figure}

In order to adress the combined effect of age and metallicity, in Fig. \ref{Fig:Lx_05_2_Metallicity_SFR_M_D_stacks_color_code_with_HeIHa} we present the X-ray luminosity in the 0.5--2 \energyunits{} band normalized by the \sfr{} as a function of the gas-phase metallicity, color coded with the median $\frac{\ion{He}{I}}{\ion{H}{$\alpha$}}$ ratio. The metallicity and the \sfr{} values of the stacked data correspond to the median and the mean values of the galaxy population included in each \sfr{}-\stellarmass{}-\distance{} bin, respectively.  We see that the increased \lx{}/\sfr{} towards subsolar metallicities is also correlated with an increase in the median $\frac{\ion{He}{I}}{\ion{H}{$\alpha$}}$ ratio. This is generally expected since lower metallicity star-forming galaxies tend to be dwarf galaxies dominated by recent star-formation episodes. However, we also see that for a given metallicity the gradient in the $\frac{\ion{He}{I}}{\ion{H}{$\alpha$}}$ increases towards higher \lx{}/\sfr{} implying that the age of the stellar populations plays a significant role in the observed \lx{} excess and scatter. This gradient is significantly decreased (for solar or super-solar metallicity) and the \lx{}/\sfr{} we measured shows smaller deviations from the scaling relations. Furthermore, the scatter is also decreased.

The results of Fig. \ref{Fig:M14_L16_SFR_M_D_stacks_color_code_with_HeIHa} and Fig. \ref{Fig:Lx_05_2_Metallicity_SFR_M_D_stacks_color_code_with_HeIHa} indicate that both metallicity and age of the stellar populations of the average galaxy population in each \sfr{}-\stellarmass{}-\distance{} bin are highly correlated with the X-ray luminosity excess from the standard scaling relations of M14 and L16. However, based on this qualitative analysis we cannot conclude which of them shows a stronger correlation. 

In order to assess the relative importance of the metallicity and stellar population age in driving the X-ray luminosity excess, we used the generalization of the partial Kendal-$\tau$ coefficient for censored data \citep{akritas96}. The presence of a significant number of stacks with upper limits (especially in the low luminosity bins; Fig.~\ref{Fig:M14_L16_SFR_M_D_stacks_color_code_with_HeIHa}) requires the use of survival analysis methods in order to obtain unbiased results. We calculated the partial Kendal-$\tau$ coefficient for the measured excess of the \lx{}/\sfr{} with respect to the L16 and the M14 scaling relations, against the metallicity and the $\frac{\ion{He}{I}}{\ion{H}{$\alpha$}}$ ratio of each \sfr{}-\stellarmass{}-\distance{} bin. We find that for both reference scaling relations, the null hypothesis for no partial correlation when the independent variable is the metallicity and the test variable is the $\frac{\ion{He}{I}}{\ion{H}{$\alpha$}}$ ratio cannot be rejected. On the other hand, when the independent variable is the $\frac{\ion{He}{I}}{\ion{H}{$\alpha$}}$ ratio (and the test variable is the metallicity) the zero partial correlation null hypothesis is rejected at the 5\% confidence level.  This indicates that the primary driver for the excess is the age of the stellar populations rather than the metallicity, in agreement with the qualitative picture seen in Fig.~\ref{Fig:Lx_05_2_Metallicity_SFR_M_D_stacks_color_code_with_HeIHa}. 
This is also in agreement with the theoretical models of \citet{fragos13} which show that at the youngest ages ($\lesssim10-20$\,Myr) there is stronger evolution of the X-ray output of a stellar population as a function of the stellar population age than the metallicity. 

\subsection{Stochasticity effect}\label{stochasticity effet}
In Sect.~\ref{sec:Excess from scaling relations} we showed that the average HEC-eR1 star-forming galaxy population shows a significant (up to $\sim$2 dex) excess above the standard scaling relations from M14 and L16. As we discussed, this excess is the result of extremely luminous individual galaxies possibly due to lower metallicities (Sect.~\ref{Sec:Metallicity dependence}) and/or very young stellar populations (Sect.~\ref{Sec:Stellar age dependence}) hosted by these galaxies. However, a reasonable question that arises is whether this excess is the result of stochastic sampling of individual XRBs associated with the galaxies participating in each bin, and has no physical origin \citep[e.g.][]{gilfanov04}.
 
To answer this question, we performed a simulation study by calculating the expected X-ray luminosity due to stochastic sampling of the HMXBs XLF, for a fiducial galaxy with \sfr{},\stellarmass{} and metallicity representative for each \sfr{}-\stellarmass{} bin. We then compared the resulting luminosity distribution with the observed X-ray luminosity of the galaxies in each bin. 
In particular, by using the mean \sfr{}, and the median gas-phase metallicity per \sfr{}-\stellarmass{} bin (see Table \ref{tab:SFR-M-D bins}) we first calculated the expected number of HMXBs per \sfr{}-\stellarmass{} bin using the metallicity-depended XLF from \citet{lehmer21} (L21):
\begin{equation}
N_{exp}(SFR,12+log(O/H)) = \int^{L_{max}}_{L_{min}}\frac{dN_{HMXB}}{dL}\cdot dL ,
\end{equation}
assuming their best-fit parameters which are a function of metallicity (from their Table 2), limiting luminosity $L_{min} = 10^{36}\,\rm{erg}\,\rm{s^{-1}}$, and maximum luminosity $L_{max} = 5\times10^{41}\,\rm{erg}\,\rm{s^{-1}}$. We note that changing the low integration limit had little effect in the final result. Then, by sampling from a Poisson distribution with a mean equal to the number of expected HMXBs calculated above we produced 20000 draws of the expected number of HMXBs 
\begin{equation}
N_{exp,i}^{inst} \sim Pois(N_{exp}(SFR,12+log(O/H))),\quad\quad\text{i=1,...20000}.
\end{equation}
Then we sampled the HMXB XLF, by drawing $N_{exp,i}^{inst}$ sources each time. The total X-ray luminosity of the overall XRB population for each of the 20000 draws is given by
\begin{equation}
L_{tot\,exp, i}^{inst} = \sum_{j=1}^{{N_{exp,i}^{inst}}}L_{X}^{j},\quad\quad\text{i=1,...20000}
\end{equation}
where $L_{X}^{j}$ is the luminosity for each source drawn from the XLF for each instance. This is akin to simulating the X-ray emission of 20000 galaxies with the typical properties of the galaxies in each \sfr{}-\stellarmass{} bin while accounting for fluctuations on the number of sources in each simulated galaxy, as well as, stochastic effects on sampling their XLF. For each of these distributions, we calculated the mean and mode and the $68\%$, $90\%$, $99\%$, and the $99.9\%$ upper and lower C.Is. Since the XLF from \citet{lehmer21} is given in the 0.5--8.0 \energyunits{} energy band we converted all the derived luminosities in our adopted band (0.5--2 \energyunits{}) by using the conversion factor c$_{7}$ from Table \ref{tab:conversion factors}. For comparison, we repeated the analysis above by integrating the HMXBs XLF from \citet{mineo12a} for each \sfr{}-\stellarmass{} bin, in the same integration limits [$L_{min}$,$L_{max}$]. We converted again the X-ray luminosities from their 0.5--8 \energyunits{} energy band to 0.5--2 \energyunits{} using the corresponding conversion factor c$_{8}$ from Table \ref{tab:conversion factors}.

\begin{figure}
        \includegraphics[width=\columnwidth]{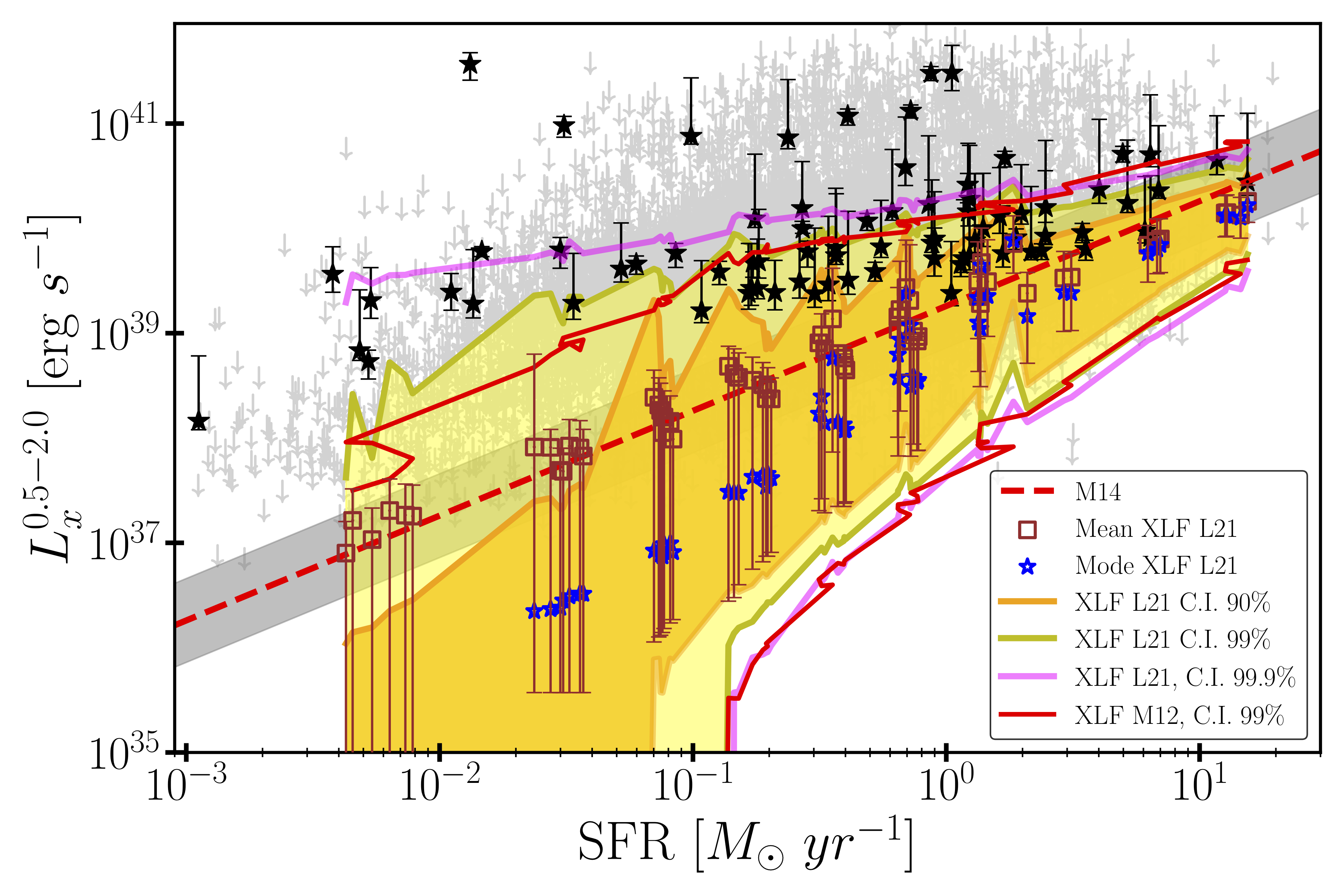}
        \caption{Total expected X-ray luminosity (0.5-2 \energyunits{}) due to the stochastic sampling of the HMXBs XLF from \citet{lehmer21} (L21), as a function of \sfr{}. Orange and yellow shaded regions indicate the 99$\%$ and the 99.9$\%$ C.I. of the expected \lx{}$^{0.5-2}$ distribution per \sfr{}-\stellarmass{} bin, due to the stochastic sampling of the XLF. The magenta line shows the 99.9$\%$ C.I. of the same distribution and the red line the 99$\%$ C.I from the \citet{mineo12a} (M12) XLF. Brown open squares and blue stars show the mean and mode values of the distributions, respectively. The uncertainties of the means correspond to the 68$\%$ C.I. Black stars (reliable flux) and gray down-arrows (uncertain flux) indicate the individual HEC-eR1 star-forming galaxies. The red dashed line shows the \lx{}-\sfr{} scaling relation from \citet{mineo14} (M14) with its 1$\sigma$ scatter }
        \label{Fig:M12_L21_SFR_M_bins_stochasticity_effect}
\end{figure}

In Fig. \ref{Fig:M12_L21_SFR_M_bins_stochasticity_effect} we present the distribution of the total expected X-ray luminosity (0.5--2 \energyunits{}) due to the stochastic sampling of the HMXBs XLF from L21, as a function of the \sfr{}. Orange and yellow shaded regions indicate the 90$\%$, and the 99$\%$ C.I., respectively of the expected \lx{}$^{0.5-2}$ distribution for each \sfr{}-\stellarmass{} bin. The magenta line shows the 99.9$\%$ C.I. of the same distribution and the red line the 99$\%$ C.I. from \citet{mineo12a} (M12). Brown open squares and blue stars show the mean and mode of the L21 distributions, respectively. The errorbars around the mean correspond to the 68$\%$ C.I. Black stars (reliable flux) and gray down-arrows (uncertain flux) indicate the individual HEC-eR1 star-forming galaxies. The red dashed line shows the \lx{}-\sfr{} scaling relation from M14 with its 1$\sigma$ scatter.

We see that the stochastic sampling of the HMXB XLF is not adequate to explain the most X-ray luminous individual galaxies in our sample. In particular, we find that a sub-population of galaxies has X-ray luminosities ${\sim}1$ dex higher than the upper 99.9$\%$ C.I. of the expected X-ray luminosity due to stochastic sampling. This population is spread in a \sfr{} range of ${\sim}0.01-2\,\rm{M_{\odot}\,yr^{-1}}$ and the majority of them has X-ray luminosity higher than $10^{41}\,\rm{erg s^{-1}}$. These galaxies are strong outliers with respect to the average galaxy population in each \sfr{}-\stellarmass{} bin and the probability of stochastically observing them is lower than $0.05\%$. 

We find that both, the L21 and M12 expected X-ray luminosity distributions for each \sfr{}-\stellarmass{} bin are narrower for high \sfr{} values while they become broader going towards lower \sfr{}s. This is expected because for higher \sfr{}s
we expect larger population of HMXB and as per the central limit theorem the sampling of their integrated X-ray emission converges to the mean of their expected value. On the other hand, for lower \sfr{}s the sampling of the XLF tends to be more stochastic, and the expected total X-ray luminosity per \sfr{} can be dominated by a few HMXBs drawn from the upper end of the XLF. In addition, we see that the mean values of the distributions are in excellent agreement with the \lx{}-\sfr{} scaling relation from \citet{mineo14} within the full \sfr{} range. On the other hand, the mode values are in much lower luminosities for lower \sfr{} values. This result implies that the previous studies tend to observe the most dominant X-ray luminous galaxies in the lower \sfr{} regimes, lacking the bulk of the population of those galaxies. This underscores the need for blind galaxy surveys such as eRASS1 which can provide a less biased census of X-ray galaxies in the local Universe. 

By comparing the upper 99$\%$ C.I.s of the expected \lx{}$^{0.5-2}$ based on sampling from the M12 and L21 HMXBs XLFs, we find that for the latter we systematically sample more luminous HMXBs in the lower \sfr{} regimes. This is due to the lower metallicity of the galaxies in those \sfr{} (see Sect.~\ref{Sec:Metallicity dependence}) bins. The M12 XLF, does not take into account the effect of metallicity. However, the L21 XLF which takes into account the metallicity show a flattening at the high luminosity end of the XLF for lower metallicities. 
In their work \citet{lehmer21} showed also that for super-solar metallicities the high-end of the XLF has a much steeper slope compared to the XLF for lower metallicities. This metallicity effect is clearly seen for $\sfr{}{>}1\,\rm{M_{\odot}\,\rm{yr^{-1}}}$ where the mode and the mean values of the sampled distributions are systematically in slightly lower X-ray luminosities than those suggested from the M14 scaling relation. The galaxy sample from \citet{mineo14} includes mostly solar and/or sub-solar metallicity galaxies. On the other hand, our sample spans a range of solar and supersolar metallicities at these SFR regimes since it is more representative of the average galaxy population, resulting in steeper actual XLFs yielding HMBXs with lower X-ray luminosities. That explains why the mean and the mode values are below the M14 scaling relation for $\sfr{}{>}1\,\rm{M_{\odot}\,\rm{yr^{-1}}}$. 

\begin{figure}        
         \includegraphics[width=\columnwidth]{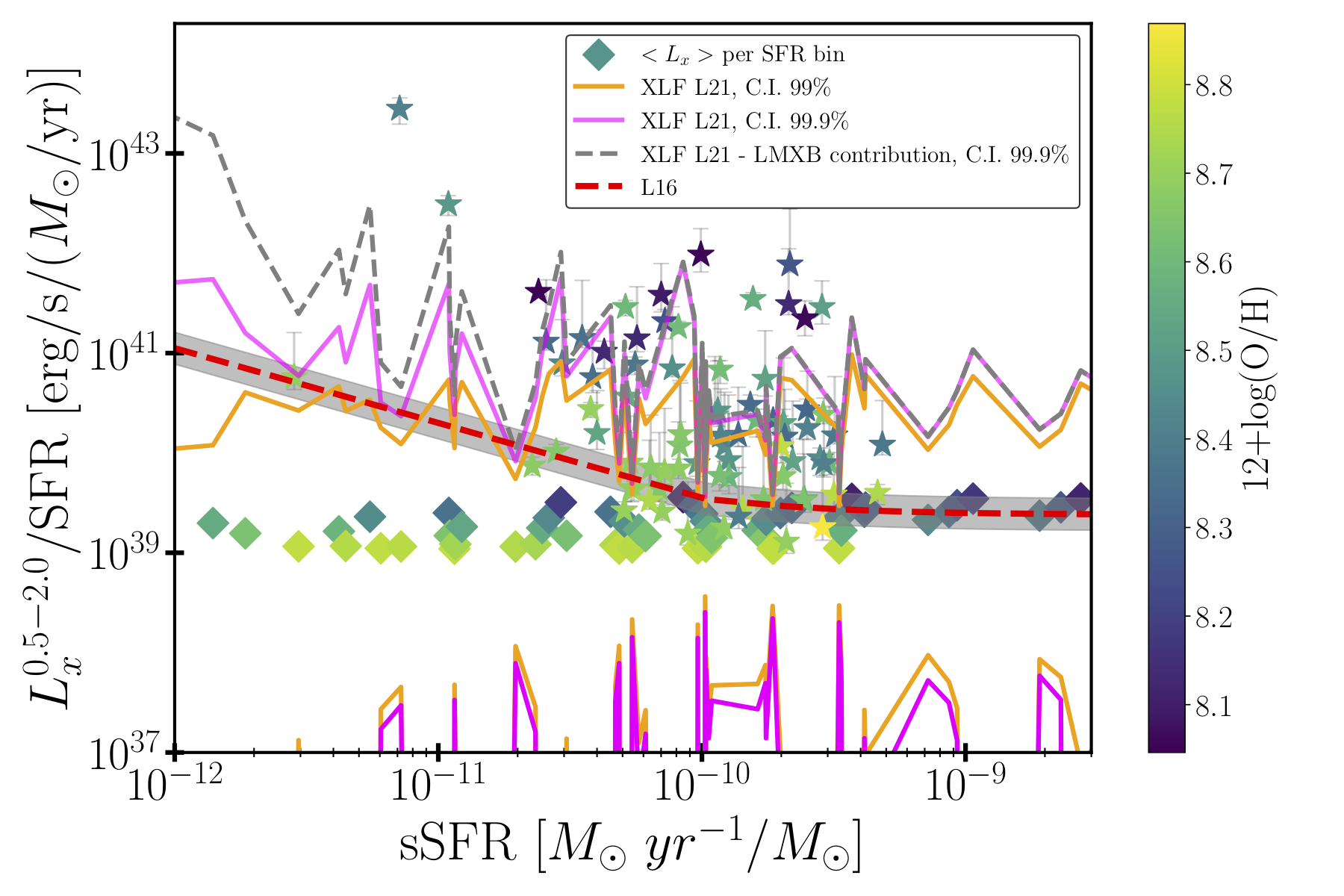}
    \caption{Total expected \lx{}/\sfr{} distribution due to stochastic sampling of HMXBs XLF as a function of the sSFR. The rhombuses indicate the mean \lx{}/\sfr{} per \sfr{}-\stellarmass{} bin, color-coded with the gas phase metallicity. The orange and magenta lines correspond to the upper and lower $99\%$, and $99.9\%$ C.Is. of the HMXBs expected \lx{}/\sfr{}, respectively. Gray dashed line depicts the upper $99.9\%$ C.I of the total expected luminosity accounting for the stochasticity of both, LMXBs and HMXBs. We also overplot with a red line the \lx{}/\sfr{} - sSFR L16 scaling relation, and its corresponding 1$\sigma$ scatter}
    \label{Fig:Lx_SFR_sSFR_LMXBs_stochasticity_effect}
\end{figure}

The analysis so far has assumed that the contribution of LMXBs is negligible. In order to address the role of LMXBs in the population of galaxies with significantly stronger X-ray emission with respect to the standard \lx{}-\sfr{} scaling relation 
in Fig. \ref{Fig:Lx_SFR_sSFR_LMXBs_stochasticity_effect} we present the total expected \lx{}/\sfr{} distribution due to stochastic sampling of HMXBs XLF as a function of the sSFR. The diamonds indicate the mean \lx{}/\sfr{} per \sfr{}-\stellarmass{} bin, color-coded with the gas phase metallicity. The sSFR and the metallicity were calculated using the mean \sfr{}, \stellarmass{}, and the median gas-phase metallicity of the galaxies participated in each \sfr{}-\stellarmass{} bin. The orange and magenta lines correspond to the upper and lower $99\%$, and $99.9\%$ C.Is. of the HMXBs expected \lx{}/\sfr{}, respectively. We also overplot with a red dashed line the \lx{}/\sfr{} - sSFR L16 scaling relation, and its corresponding 1$\sigma$ scatter. 

We see that the mean \lx{}/\sfr{} values from the stochastic sampling of the L21 XLF are in excellent agreement with the L16 scaling relation for sSFR${>}10^{-10}\,\rm{M_{\odot}}\,\rm{yr^{-1}}/\rm{M_{\odot}}$ while for lower sSFRs the L16 scaling relation suggests an elevation of the \lx{}/\sfr{} because of the contribution of the LMXBs in these sSFR regimes. 
The \lx{}/\sfr{} from the L21 relation remains constant because their XLF accounts only for the HMXBs contribution.
By calculating the difference between the expected mean \lx{}/\sfr{} per sSFR from the previous analysis and the \lx{}/\sfr{} suggested by the L16 we can estimate the contribution of LMXBs per sSFR. By adding this LMXB contribution to the upper $99.9\%$ C.I of the X-ray luminosity due to the stochastic sampling of the HMXBs XLF, we can estimate the upper $99.9\%$ C.I. of the total luminosity including both, LMXBs and HMXBs. This result is depicted by the gray dashed line. As we can see, for high sSFR values where the contribution of LMXBs is negligible the gray and the magenta lines are identical. On the other hand, for lower sSFRs the gray line is slightly higher than the magenta because the LMXBs have stronger contribution to the X-ray output of the galaxies. Interestingly enough, we find that even accounting for the contribution of LMXBs on the stochastic sampling of LMXBs and HMXBs, the sub-population of galaxies (color-coded stars) is still well above the upper $99.9\%$ C.I. of the expected X-ray luminosity. This result indicates that their observed X-ray luminosities have physical origin and they are not due to stochasticity effect.

\subsection{Low-Luminosity AGNs or Tidal Disruption events ?}\label{Sec:low-AGN or TDE}
Although our sample is carefully selected to exclude any potential contamination from AGN, the enhanced X-ray emission that we observe can be still due to the presence of a low-luminosity AGN in the central black hole of the host galaxy that is not traced by the optical and/or IR diagnostics we used. 

To test this scenario we performed the following exercise. Assuming that the observed X-ray emission originates from accretion onto the supermassive black hole of the galaxy, we first calculated the black hole masses ($\rm{M_{BH}}$) of the HEC-eR1 galaxy sample with reliable flux measurements using the $\rm{M_{BH}}$-\stellarmass{} scaling relation of \citet{green20}. This calculation resulted in masses spanning a range of $10^{3}-10^{8}\, \solarmass$ which includes the range of the Intermediate Mass Black Holes (IMBH) \citep{green20}. Based on this black-hole mass, we calculated the bolometric Eddington luminosity 
\begin{equation}
    L_{Edd}^{bol} = 1.26\times10^{38} \, (\rm{M_{BH}/M_{\odot}}) \,\,\, erg\,s^{-1} 
\end{equation}
and the corresponding Eddington fraction 
\begin{equation}
    f_{Edd}^{bol} = \frac{BC^{X}\times L_{x}^{0.5-2}}{L_{Edd}^{bol}}
\end{equation}
where $\rm{BC^{X}}=\frac{L_{x}^{0.5-2}}{L_{bol}}=0.042$ is the bolometric correction from the observed X-ray luminosity ($\rm{L_{x}^{0.5-2}}$) based on the SED of \citet{risaliti04}.
In addition, using the  same SED and the conversion factor to the optical band $\frac{L_{x}^{0.5-2}}{L_{opt}}=0.35$ we also calculated the optical luminosity ($L_{opt}$) that we would observe from the accretion disk of the central black hole. We find that this is higher than $3\times10^{39}\rm{erg\,s^{-1}}$ for all the galaxies. 
\begin{figure}        
         \includegraphics[width=\columnwidth]{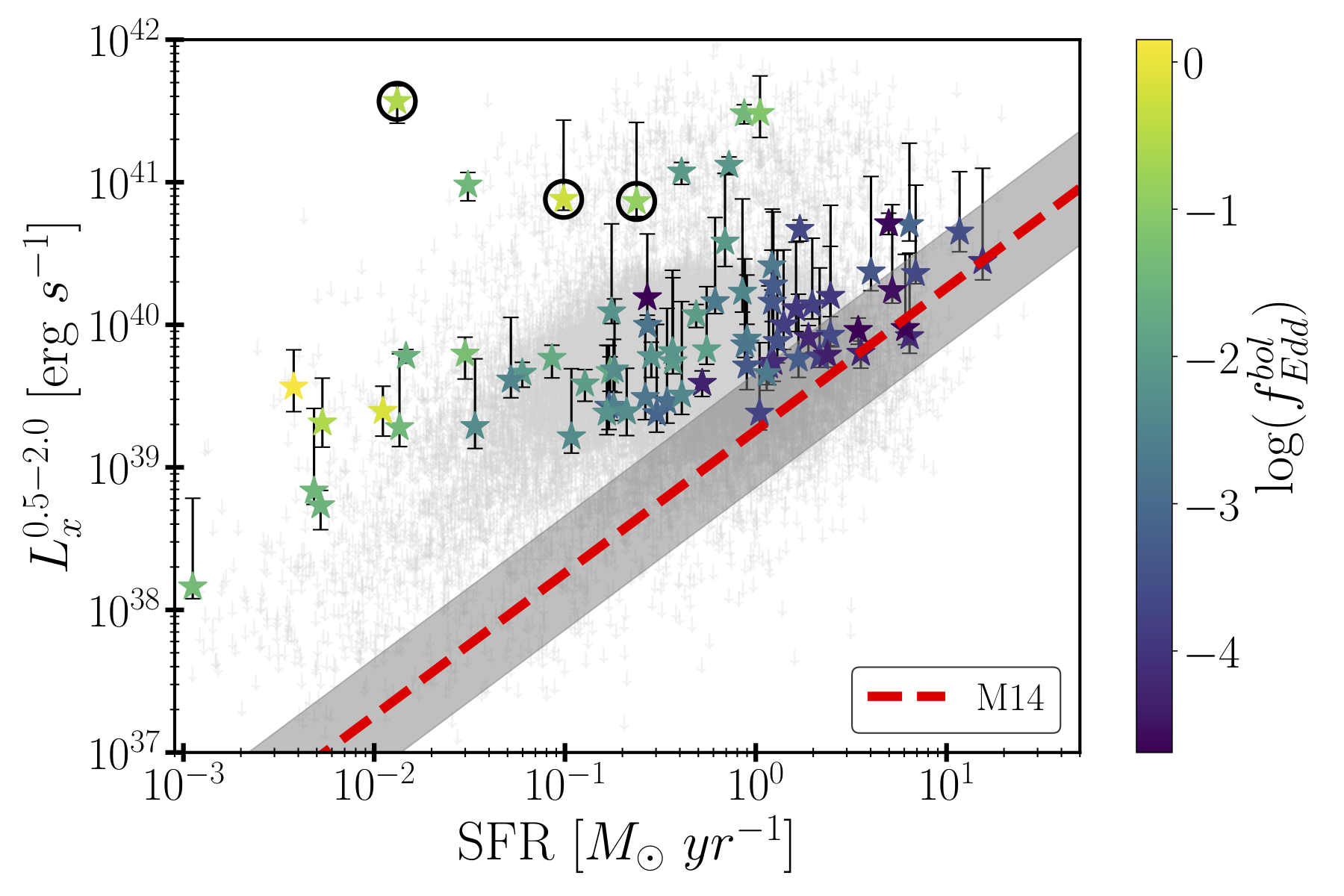}
    \caption{\lx{}-\sfr{} distribution of the HEC-eR1 star-forming galaxies with reliable flux measurement color coded with the log($f_{Edd}^{bol}$), computed under the assumption that the observed X-ray luminosity is produced by accretion onto a nuclear SMBH. Galaxies in the black circles belong to the most extreme X-ray luminous galaxies of our sample for which we have available \textit{SDSS} spectra. }
    \label{Fig:Lx_SFR_fedd}
\end{figure}

In Fig. \ref{Fig:Lx_SFR_fedd} we present the distribution of our galaxy sample in the \lx{}-\sfr{} plane color-coded with the Eddington fraction. Interestingly enough, we find that the most X-ray luminous galaxies ($L_{x}^{0.5-2}> 4{\times}10^{40} \rm{erg\,s^{-1}}$), as well as, the galaxies in the lower \sfr{} regime which are responsible for the observed excess, have systematically higher Eddington fraction which are very close to values typical for Seyfert 1 galaxies \citep{ho08}. However, if the observed X-ray luminosities are from so powerful accretion on the central black hole, and given that the $\rm{L_{opt}}$ for the majority of the galaxies is higher than $3\times10^{39}\rm{erg\,s^{-1}}$, we would expect that their optical spectra would exhibit AGN signatures (e.g. broad and/or high-excitation spectral lines). Furthermore, given the large Eddington fractions and the small black-hole masses of these sources, we would expect that if they are powered by AGN they would exhibit strong variability. However, none of these galaxies is in the sample of X-ray emitting AGN selected on the basis of optical/UV/IR variability \citep{arcodia23}.

\begin{figure}        
         \includegraphics[width=\columnwidth]{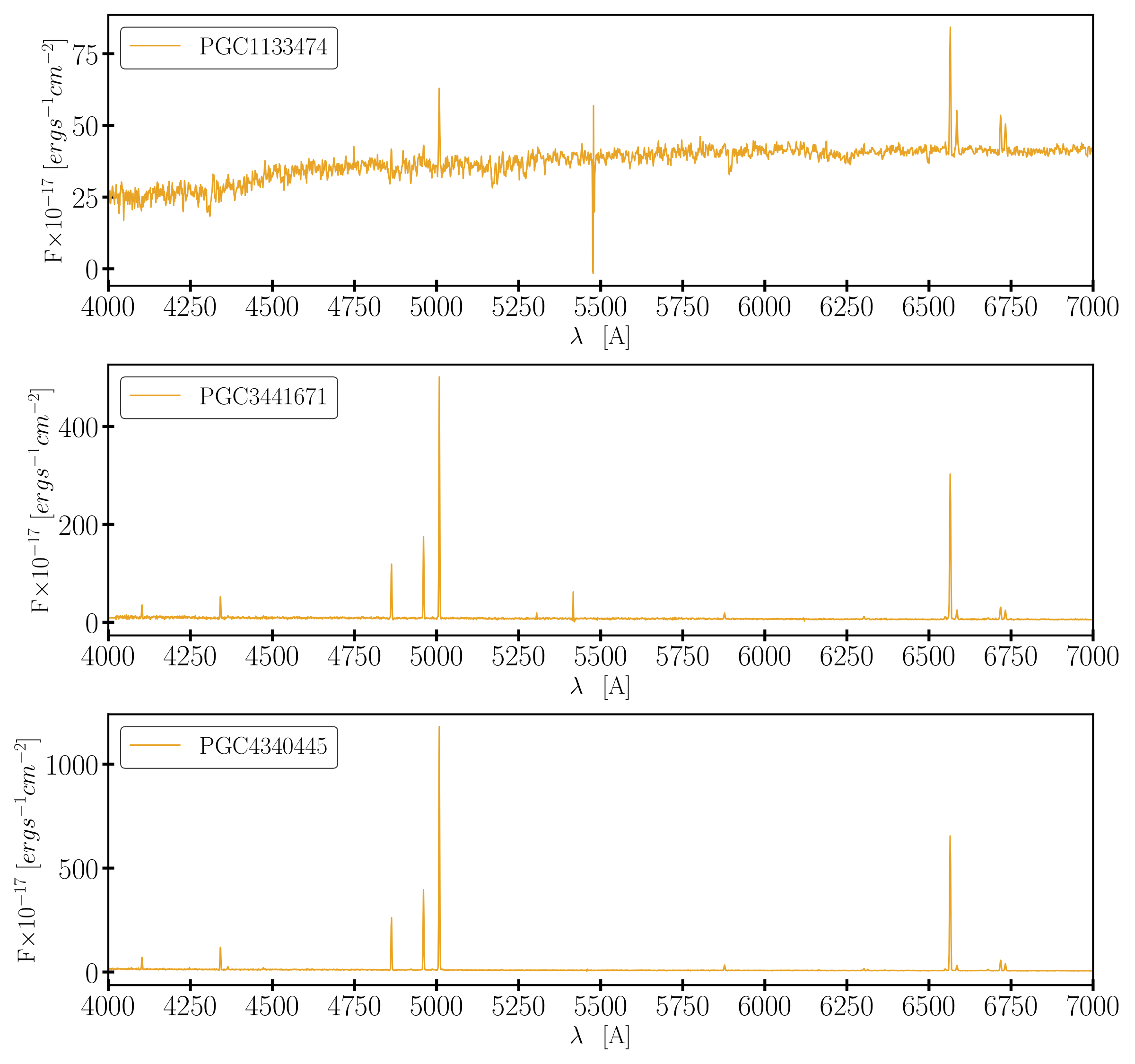}
    \caption{Optical spectra of the most X-ray luminous ($L_{x}^{0.5-2}{>}4{\times}10^{40} \rm{erg\,s^{-1}}$) HEC-eR1 star-forming galaxies. All three spectra are typical of star-forming galaxies, not showing any evidence for AGN activity. }
    \label{Fig:optical spectra}
\end{figure}

Focusing on the most extreme X-ray luminous galaxies ($L_{x}^{0.5-2}{>}4{\times}10^{40} \rm{erg\,s^{-1}}$) of our sample, three of them have optical SDSS spectra (PGC1133474, PGC3441671, and PGC4340445; open black circles in Fig. \ref{Fig:Lx_SFR_fedd}). These (Fig. \ref{Fig:optical spectra}) are typical of star-forming galaxies without any AGN signature such as high excitation forbidden lines, or broad Balmer lines.  
In contrast, they are representative of the most typical optical spectra of star-forming galaxies, and in the case of PGC3441671, and PGC4340445 indicative of very young stellar populations. These results imply that although our galaxies have very high X-ray luminosities, which could be potentially produced by low-luminosity AGN, there is not any observable signature in their optical spectra that indicate the presence of an AGN. 

Another possible source of contamination may be the X-ray emission produced by Tidal Disruption Events (TED). If a star that is close to a supermassive black hole approaches the event horizon it can be disrupted, resulting in observed X-ray flares from tidal debris that falls back onto the black hole \citep[for a review see;][]{gezari21}.
Recent studies have shown that TDEs are predominantly observed in post-starburst galaxies or galaxies that fall in the green valley in the color-mass diagram \citep{lawsmith17,vanvelzen21,yao23}. However, given that our galaxy sample is composed of carefully selected star-forming galaxies, their vast majority falls well within the region of the main-sequence of star-forming galaxies \citep[e.g.][]{peng10}.
As a result, given the HEC-eR1 host galaxy properties, we would not expect strong contamination by TDEs. On top of that, based on the most up-to-date optical TDE rate for  star-forming galaxies \citep[$1.5{\times}10^{-5}\,\rm{galaxy^{-1}}\,\rm{yr^{-1}}$; ][]{yao23}, and the size of the HEC-eR1 sample (18,790 galaxies) we estimate that we would expect to observe ${\sim}0.28$ optical TDEs within one year. As a result, given that the eRASS1 data are obtained from a 6-month scan, as well as, that the TDE rate in the X-rays is lower than in the optical we can conclude that the probability for our X-ray measurements to be contaminated by a TDE is very low, and that TDEs cannot explain the observed X-ray luminosity excess.

\subsection{A population of extreme X-ray luminous starbursts}
In Sect.~\ref{stochasticity effet}, we showed that the stochastic sampling of the HMXBs XLF (accounting for the contribution of the LMXBs) is not adequate to explain the sub-population of the individual star-forming galaxies within the HEC-eR1 sample with extreme X-ray luminosities.

In order to evaluate the significance of this population of luminous galaxies we estimated their expected number based on the parent population of star-forming galaxies and the previously discussed stochasticity simulation for each \sfr{}-\stellarmass{} bin. The parent population in each bin consists of the HEC-eR1 star-forming galaxies which based on the sensitivity of the eRASS1 survey and their distance would be detectable beyond the 99.9$\%$ C.I of the expected X-ray luminosity distribution due to stochasticity (\lx{}$^{99.9\%}$) for each corresponding bin. This is because only these galaxies would be potentially detected as outliers in our sample. This way we find that the number of outliers from this sample that would be detected above 99.9$\%$ C.I. by chance would be 0.05$\%$ of this parent sample.

\begin{figure}        
         \includegraphics[width=\columnwidth]{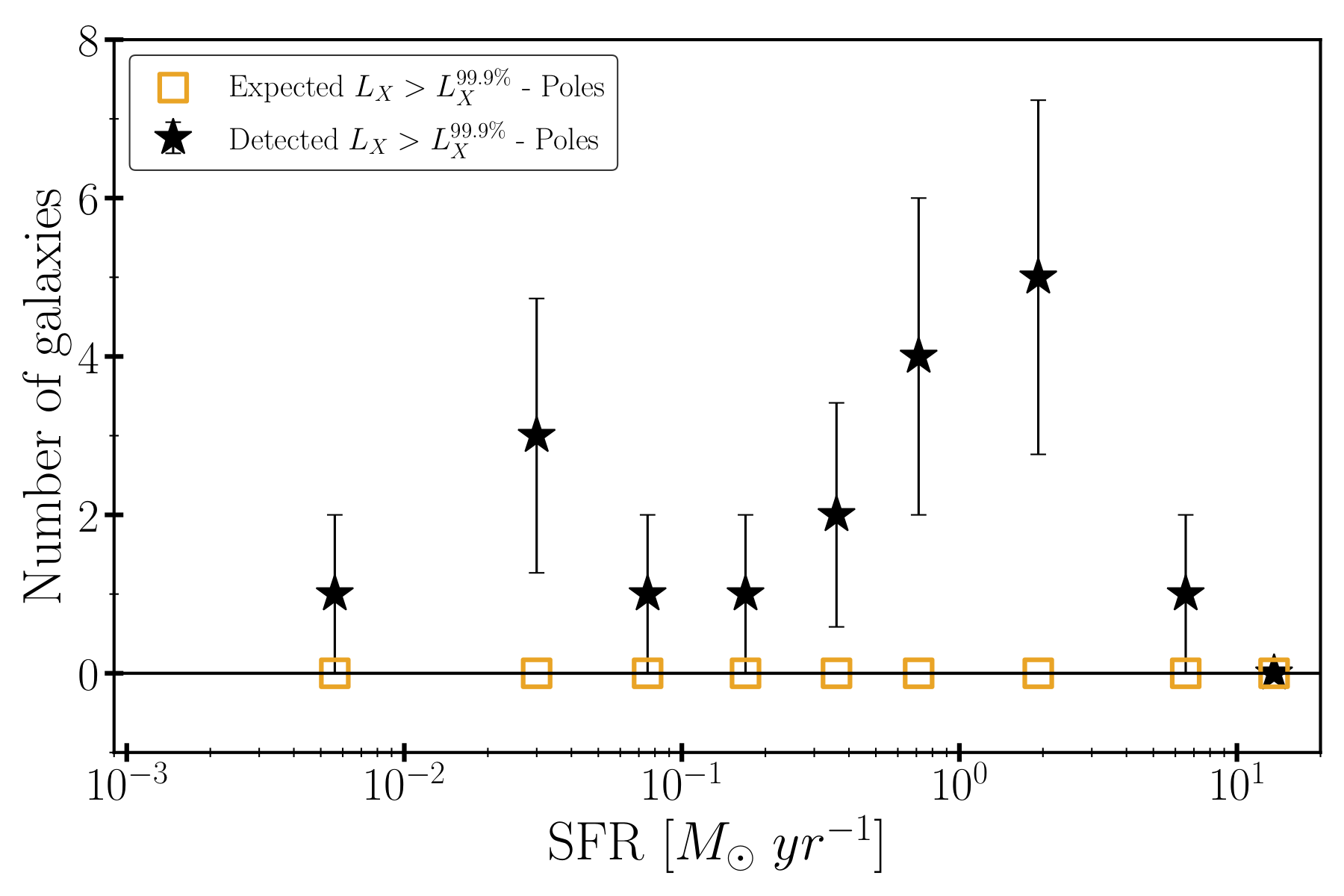}
    \caption{Number of detected galaxies with $\lx{}>\lx{}^{99.9\%}$ per SFR bin (black stars). Number of expected galaxies due to stochastic sampling of the XLF, with $\lx{}>\lx{}^{99.9\%}$ (open orange squares).}
    \label{Fig:Extreme_population_of_galaxies}
\end{figure}

In Fig. \ref{Fig:Extreme_population_of_galaxies} we present the result of this analysis. Open orange squares and black stars indicate the number of the expected and the detected galaxies, respectively, with \lx{}$>\lx{}^{99.9\%}$ per \sfr{} bin. The error bars correspond to the $\sqrt{N}$, where $N$ is the number of galaxies. We show only the results using the eRASS1 sensitivity at the poles as a conservative limit given the longer integration time resulting in higher sensitivity and larger number of galaxies.
Surprisingly enough, we find that even in this deeper case the number of expected galaxies is zero. 
The fact that the expected number of galaxies per \sfr{} bin with \lx{}$>\lx{}^{99.9\%}$ is zero, strengthens the argument that the sub-population of extreme X-ray luminous galaxies we find, is real and has an astrophysical origin. Although the significance of the detected galaxies compared to the expected is not very high for all the \sfr{} bins, for some of them it is higher than 1$\sigma$. This marginal significance is expected given the shallowness of the eRASS1 survey. However, the significance of our findings will be increased substantially with the forthcoming deeper eRASS:4 and eRASS:8 surveys, which will increase the population of the detected galaxies.

Given that our findings suggest that this sub-population of HEC-eR1 star-forming galaxies has a physical origin, it is worth examining its stellar population properties. In Fig. \ref{Fig:Lx_SFR_sSFR_LMXBs_stochasticity_effect} we see that it is mainly comprised of galaxies with higher sSFRs and lower metallicities. In particular, the vast majority of these galaxies span a sSFR range of ${\sim}5\times10^{-10}-2\times10^{-9}\,\rm{M_{\odot}}\,\rm{yr^{-1}}/\rm{M_{\odot}}$ and subsolar gas-metallicity in the range [log(O/H)+12]${\sim}8.1-8.6$. In addition, these galaxies host systematically younger stellar populations (see Sec.\ref{Sec:Stellar age dependence}). Given their higher sSFRs, their lower metallicities, and the age of their stellar populations, we can conclude that most likely these galaxies are experiencing a recent star formation episode, which produces a larger number of luminous HMXBs resulting in an elevated \lx{}/\sfr{} comparing to the standard scaling relations. This result is also consistent with the work of \citet{vulic22} who detected in the eFEDS field a population of metal-poor dwarf starburst galaxies with similar characteristics as our population of galaxies. 

This analysis reinforces the argument that large blind surveys can provide a more complete picture of the X-ray emitting galaxy population and their diversity. These intriguing results have important implications for understanding the population of X-ray binaries contributing in the most X-ray luminous galaxies. This is particularly important for understanding the cosmological evolution of the galactic X-ray emission (e.g. as inferred from deep and wide area surveys \citep{lehmer16,aird17,fornasini20} and their role in the preheating and reionization in the early universe. Stochasticity is particularly important for the latter since it will influence the shape of the cosmological 21 cm power spectrum \citep[c.f.][]{21cm_stochasticity_Kaur}. The next eRASS surveys will allow us to obtain a more clear picture of this intriguing population of luminous galaxies by obtaining more accurate measurements of their luminosity and setting better constrains on their overall scaling relations and scatter. 

\section{Summary and Conclusions}\label{sec:conclusion}
In this work, we present the results of the first unbiased all-sky survey of star-forming galaxies with the \erosita{} X-ray telescope. By combining the HECATE v2.0 value-added galaxy catalogue with the X-ray data obtained from the first complete eROSITA all-sky scan (eRASS1), and by applying stringent optical and mid-infrared activity classification criteria we constructed a sample of 18790 bona-fide star-forming galaxies (HEC-eR1 galaxy sample) with measurements of their integrated X-ray luminosity. 
This resulted in 77 star-forming galaxies with $3\sigma$ C.I. reliable X-ray fluxes, and 18713 galaxies with uncertain flux measurements based on their posterior net count distribution. This sample probes star-forming galaxies at luminosities as low as $10^{39}\, \rm{erg s^{-1}}$ out to distances of ${\sim}70$ Mpc, and luminosities of $10^{40}\, \rm{erg s^{-1}}$ out to distances of ${\sim}200$ Mpc.
In this way, we studied the relation between the X-ray emission of normal galaxies with their stellar population parameters (i.e. \sfr{},\stellarmass{}, Metallicity, stellar population age) for the largest galaxy sample so far.

Our main results are summarized as follows:
   \begin{enumerate}
      \item By stacking the X-ray spectra of all the galaxies in the HEC-eR1 sample, we derived the average X-ray spectrum of star-forming galaxies in the local universe which is well-fitted by an XRB and a hot-gas component ($\Gamma = 1.75^{+0.12}_{-0.07}$, $kT = 0.70^{0.06}_{-0.07} \, \energyunits{}$). Our best-fit results indicate that the average diffuse X-ray emission of the HEC-eR1 star-forming galaxies is $12.86\%$. This implies that only a small fraction of the total average integrated flux is due to hot gas indicating that the measured X-ray emission is dominated by a power-law component associated with XRBs.
      \item We find that the integrated X-ray luminosity of the individual HEC-eR1 star-forming galaxies is significantly elevated (up to 
      $10^{42}\,\rm{erg\,s^{-1}}$) with respect to the expected X-ray luminosity from the standard scaling relations M14 and L16.
      This excess is stronger in the low \sfr{} regime (${\lesssim}1 \rm{M_{\odot}\,yr^{-1}}$) and it is decreasing towards higher SFRs. Furthermore, the observed scatter  (including reliable and uncertain flux measurements) is significantly larger than the 1$\sigma$ scatter from the best-fit lines of M14 and L16.
      \item By stacking the X-ray data in \sfr{}-\stellarmass{}-\distance{} bins we studied the correlation between the average X-ray luminosity and the average stellar population parameters. Although the distribution of the stacked data is closer to the scaling relations than the individual galaxies, we find that there is still an excess (reaching ${\sim}2$ dex  at  \sfr{}$\sim10^{-1}$ $M_{\odot}\,yr^{-1}$)  above the standard scaling relations that anticorrelates with the \sfr{}.  
    Bootstrap analysis showed that the observed excess is real and not due to strong statistical fluctuations of a sub-population of X-ray luminous individual galaxies that dominate the stacked signal in each bin.
      \item This excess is not due to background AGN observed within the aperture of the sample galaxies or a population of LMXBs. The median AGN contamination fraction in our sample is 17$\%$ and the LMXB contribution fraction is ${\lesssim}10\%$ for the low \sfr{}s and it is slightly larger for some of the higher \sfr{} bins.
It is also unlikely that low-luminosity AGN, or TDEs contribute significantly in the measured emission of the star-forming galaxies. 
      \item 
     Fits of the \lx{}-\sfr{} scaling relation, accounting for the intrinsic scatter, suggest that our relation is higher than almost all other works, with a slightly shallower slope, reflecting the population of X-ray luminous galaxies, especially in the lower SFR regimes. Furthermore, we find that the \textit{intrinsic} scatter in the X-ray emission of the average population of star-forming galaxies anti-correlates with the \sfr{}.       
     \item We find that the \lx{}-\sfr{}-Metallicity scaling relation (spanning $\sim$1\,dex in metallicity) confirms the anticorrelation of \lx{}/\sfr{} with the gas-phase metallicity also reported in previous works. However, our best-fit relation results in ${\sim}0.5$dex  higher X-ray luminosities than the current scaling relations, due to 
     the effect of a previously unrecognized population of X-ray luminous galaxies, particularly in lower metallicities.
     \item We find that the stacks with higher X-ray luminosity tend to correlate more strongly with the $\frac{\ion{He}{I}}{\ion{H}{$\alpha$}}$ ratio of the galaxies they contain than their metallicity. This indicates that the primary driver for the excess is the age of the stellar populations rather than the metallicity and it is in agreement with theoretical model of binary population synthesis models.
     \item Our results suggest that the stochastic sampling of the HMXBs XLF (accounting for the metallicity and the contribution of the LMXBs) is not adequate to explain the sub-population of  individual star-forming galaxies with very high X-ray luminosities (up to ${\sim}10^{41}\,\rm{erg\,s^{-1}}$). This sub-population consists of star-forming galaxies with higher sSFRs, lower metallicities, and younger stellar populations. The latter is also confirmed by the optical spectra of some of the most X-ray luminous of them, which are typical of star-forming galaxies, indicating that they experience a recent star formation episode, resulting in a larger number of luminous HMXBs and  elevated \lx{}/\sfr{} ratio.  
   \end{enumerate}

These results demonstrate the power of large-area blind surveys in revealing rare populations of objects and recovering unbiased underlying correlations. The forthcoming eRASS surveys will provide a deeper census of the local galaxy populations which will help to better constrain the correlation between X-ray luminosity, \sfr{}--Metallicity and age in star-forming galaxies and their intrinsic scatter 

\begin{acknowledgements}
This work is based on data from eROSITA, the soft X-ray instrument aboard \textit{SRG}, a joint Russian-German science mission supported by
the Russian Space Agency (Roskosmos), in the interests of the Russian Academy of Sciences represented by its Space Research Institute (IKI),
and the Deutsches Zentrum f{\"u}r Luft- und Raumfahrt (DLR). The \textit{SRG} spacecraft was built by Lavochkin Association (NPOL) and its
subcontractors, and is operated by NPOL with support from the Max Planck Institute for Extraterrestrial Physics (MPE).
The development and construction of the eROSITA X-ray instrument was led by MPE, with contributions from the Dr. Karl Remeis Observatory
Bamberg \& ECAP (FAU Erlangen-N{\"u}rnberg), the University of Hamburg Observatory, the Leibniz Institute for Astrophysics Potsdam (AIP),
and the Institute for Astronomy and Astrophysics of the University of T{\"u}bingen, with the support of DLR and the Max Planck Society.
The Argelander Institute for Astronomy of the University of Bonn and the Ludwig Maximilians Universit{\"a}t Munich also participated in the
science preparation for eROSITA. The eROSITA data used here were processed with the \texttt{eSASS/NRTA} software system developed by the
German eROSITA consortium. EK acknowledges support from the Public Investments Program through a Matching Funds grant to
the IA-FORTH. The research leading to these
results has received funding from the European Union’s Horizon 2020 research and innovation programme under the Marie Skłodowska-Curie RISE action, Grant
Agreement n. 873089 (ASTROSTAT-II). AB acknowledges support by NASA under award number 80GSFC21M0002.
\end{acknowledgements}

%
%
\bibliographystyle{aa}
\bibliography{references}

\begin{thebibliography}{100}
\expandafter\ifx\csname natexlab\endcsname\relax\def\natexlab#1{#1}\fi

\bibitem[{{Abell}(1995)}]{abell74}
{Abell}, G.~O. 1995, VizieR Online Data Catalog, VII/4A

\bibitem[{{Aird} {et~al.}(2017){Aird}, {Coil}, \& {Georgakakis}}]{aird17}
{Aird}, J., {Coil}, A.~L., \& {Georgakakis}, A. 2017, \mnras, 465, 3390

\bibitem[{{Akritas} \& {Siebert}(1996)}]{akritas96}
{Akritas}, M.~G. \& {Siebert}, J. 1996, \mnras, 278, 919

\bibitem[{{Antoniou} \& {Zezas}(2016)}]{antoniou16}
{Antoniou}, V. \& {Zezas}, A. 2016, \mnras, 459, 528

\bibitem[{{Antoniou} {et~al.}(2019){Antoniou}, {Zezas}, {Drake}, {Badenes}, {Haberl}, {Wright}, {Hong}, {Di Stefano}, {Gaetz}, {Long}, {Plucinsky}, {Sasaki}, {Williams}, {Winkler}, \& {SMC XVP Collaboration}}]{antoniou19}
{Antoniou}, V., {Zezas}, A., {Drake}, J.~J., {et~al.} 2019, \apj, 887, 20

\bibitem[{{Arcodia} {et~al.}(2024){Arcodia}, {Merloni}, {Comparat}, {Dwelly}, {Seppi}, {Zhang}, {Buchner}, {Georgakakis}, {Haberl}, {Igo}, {Kyritsis}, {Liu}, {Nandra}, {Ni}, {Ponti}, {Salvato}, {Ward}, {Wolf}, \& {Zezas}}]{arcodia23}
{Arcodia}, R., {Merloni}, A., {Comparat}, J., {et~al.} 2024, \aap, 681, A97

\bibitem[{{Asplund} {et~al.}(2009){Asplund}, {Grevesse}, {Sauval}, \& {Scott}}]{asplund09}
{Asplund}, M., {Grevesse}, N., {Sauval}, A.~J., \& {Scott}, P. 2009, \araa, 47, 481

\bibitem[{{Ballo} {et~al.}(2004){Ballo}, {Braito}, {Della Ceca}, {Maraschi}, {Tavecchio}, \& {Dadina}}]{ballo04}
{Ballo}, L., {Braito}, V., {Della Ceca}, R., {et~al.} 2004, \apj, 600, 634

\bibitem[{{Basu-Zych} {et~al.}(2020){Basu-Zych}, {Hornschemeier}, {Haberl}, {Vulic}, {Wilms}, {Zezas}, {Kovlakas}, {Ptak}, \& {Dauser}}]{basuzych20}
{Basu-Zych}, A.~R., {Hornschemeier}, A.~E., {Haberl}, F., {et~al.} 2020, \mnras, 498, 1651

\bibitem[{{Basu-Zych} {et~al.}(2013){Basu-Zych}, {Lehmer}, {Hornschemeier}, {Gon{\c{c}}alves}, {Fragos}, {Heckman}, {Overzier}, {Ptak}, \& {Schiminovich}}]{basuzych13}
{Basu-Zych}, A.~R., {Lehmer}, B.~D., {Hornschemeier}, A.~E., {et~al.} 2013, \apj, 774, 152

\bibitem[{{Brinchmann} {et~al.}(2004){Brinchmann}, {Charlot}, {White}, {Tremonti}, {Kauffmann}, {Heckman}, \& {Brinkmann}}]{sdss2}
{Brinchmann}, J., {Charlot}, S., {White}, S.~D.~M., {et~al.} 2004, \mnras, 351, 1151

\bibitem[{{Brorby} {et~al.}(2016){Brorby}, {Kaaret}, {Prestwich}, \& {Mirabel}}]{brorby16}
{Brorby}, M., {Kaaret}, P., {Prestwich}, A., \& {Mirabel}, I.~F. 2016, \mnras, 457, 4081

\bibitem[{{Brunner} {et~al.}(2022){Brunner}, {Liu}, {Lamer}, {Georgakakis}, {Merloni}, {Brusa}, {Bulbul}, {Dennerl}, {Friedrich}, {Liu}, {Maitra}, {Nandra}, {Ramos-Ceja}, {Sanders}, {Stewart}, {Boller}, {Buchner}, {Clerc}, {Comparat}, {Dwelly}, {Eckert}, {Finoguenov}, {Freyberg}, {Ghirardini}, {Gueguen}, {Haberl}, {Kreykenbohm}, {Krumpe}, {Osterhage}, {Pacaud}, {Predehl}, {Reiprich}, {Robrade}, {Salvato}, {Santangelo}, {Schrabback}, {Schwope}, \& {Wilms}}]{brunner22_eSSAS}
{Brunner}, H., {Liu}, T., {Lamer}, G., {et~al.} 2022, \aap, 661, A1

\bibitem[{Burke {et~al.}(2023)Burke, Laurino, wmclaugh, Marie-Terrell, dtnguyen2, Günther, Siemiginowska, Budynkiewicz, Cheer, Aldcroft, Deil, Sipőcz, Buchner, Donath, Laginja, Leinweber, nplee, \& Todd}]{sherpa3}
Burke, D., Laurino, O., wmclaugh, {et~al.} 2023, sherpa/sherpa: Sherpa 4.15.1

\bibitem[{{Cluver} {et~al.}(2017){Cluver}, {Jarrett}, {Dale}, {Smith}, {August}, \& {Brown}}]{cluver17}
{Cluver}, M.~E., {Jarrett}, T.~H., {Dale}, D.~A., {et~al.} 2017, \apj, 850, 68

\bibitem[{{Daoutis} {et~al.}(2023){Daoutis}, {Kyritsis}, {Kouroumpatzakis}, \& {Zezas}}]{daoutis23}
{Daoutis}, C., {Kyritsis}, E., {Kouroumpatzakis}, K., \& {Zezas}, A. 2023, \aap, 679, A76

\bibitem[{{Das} {et~al.}(2017){Das}, {Mesinger}, {Pallottini}, {Ferrara}, \& {Wise}}]{das17}
{Das}, A., {Mesinger}, A., {Pallottini}, A., {Ferrara}, A., \& {Wise}, J.~H. 2017, \mnras, 469, 1166

\bibitem[{{Dickey} \& {Lockman}(1990)}]{dickey90}
{Dickey}, J.~M. \& {Lockman}, F.~J. 1990, \araa, 28, 215

\bibitem[{{Doe} {et~al.}(2007){Doe}, {Nguyen}, {Stawarz}, {Refsdal}, {Siemiginowska}, {Burke}, {Evans}, {Evans}, {McDowell}, {Houck}, \& {Nowak}}]{sherpa2}
{Doe}, S., {Nguyen}, D., {Stawarz}, C., {et~al.} 2007, in Astronomical Society of the Pacific Conference Series, Vol. 376, Astronomical Data Analysis Software and Systems XVI, ed. R.~A. {Shaw}, F.~{Hill}, \& D.~J. {Bell}, 543

\bibitem[{{Fabbiano}(2019)}]{fabbiano19}
{Fabbiano}, G. 2019, in The Chandra X-ray Observatory, ed. B.~{Wilkes} \& W.~{Tucker}, 7--1

\bibitem[{{Fabricant} {et~al.}(1998){Fabricant}, {Cheimets}, {Caldwell}, \& {Geary}}]{fabricant98}
{Fabricant}, D., {Cheimets}, P., {Caldwell}, N., \& {Geary}, J. 1998, \pasp, 110, 79

\bibitem[{{Fender} {et~al.}(2005){Fender}, {Maccarone}, \& {van Kesteren}}]{fender05}
{Fender}, R.~P., {Maccarone}, T.~J., \& {van Kesteren}, Z. 2005, \mnras, 360, 1085

\bibitem[{{Flewelling} {et~al.}(2020){Flewelling}, {Magnier}, {Chambers}, {Heasley}, {Holmberg}, {Huber}, {Sweeney}, {Waters}, {Calamida}, {Casertano}, {Chen}, {Farrow}, {Hasinger}, {Henderson}, {Long}, {Metcalfe}, {Narayan}, {Nieto-Santisteban}, {Norberg}, {Rest}, {Saglia}, {Szalay}, {Thakar}, {Tonry}, {Valenti}, {Werner}, {White}, {Denneau}, {Draper}, {Hodapp}, {Jedicke}, {Kaiser}, {Kudritzki}, {Price}, {Wainscoat}, {Chastel}, {McLean}, {Postman}, \& {Shiao}}]{panstarrs}
{Flewelling}, H.~A., {Magnier}, E.~A., {Chambers}, K.~C., {et~al.} 2020, \apjs, 251, 7

\bibitem[{{Fornasini} {et~al.}(2020){Fornasini}, {Civano}, \& {Suh}}]{fornasini20}
{Fornasini}, F.~M., {Civano}, F., \& {Suh}, H. 2020, \mnras, 495, 771

\bibitem[{{Fornasini} {et~al.}(2019){Fornasini}, {Kriek}, {Sanders}, {Shivaei}, {Civano}, {Reddy}, {Shapley}, {Coil}, {Mobasher}, {Siana}, {Aird}, {Azadi}, {Freeman}, {Leung}, {Price}, {Fetherolf}, {Zick}, \& {Barro}}]{fornasini19}
{Fornasini}, F.~M., {Kriek}, M., {Sanders}, R.~L., {et~al.} 2019, \apj, 885, 65

\bibitem[{{Fragos} {et~al.}(2013){Fragos}, {Lehmer}, {Naoz}, {Zezas}, \& {Basu-Zych}}]{fragos13}
{Fragos}, T., {Lehmer}, B.~D., {Naoz}, S., {Zezas}, A., \& {Basu-Zych}, A. 2013, \apjl, 776, L31

\bibitem[{{Freeman} {et~al.}(2001){Freeman}, {Doe}, \& {Siemiginowska}}]{sherpa1}
{Freeman}, P., {Doe}, S., \& {Siemiginowska}, A. 2001, in Society of Photo-Optical Instrumentation Engineers (SPIE) Conference Series, Vol. 4477, Astronomical Data Analysis, ed. J.-L. {Starck} \& F.~D. {Murtagh}, 76--87

\bibitem[{{Fruscione} {et~al.}(2006){Fruscione}, {McDowell}, {Allen}, {Brickhouse}, {Burke}, {Davis}, {Durham}, {Elvis}, {Galle}, {Harris}, {Huenemoerder}, {Houck}, {Ishibashi}, {Karovska}, {Nicastro}, {Noble}, {Nowak}, {Primini}, {Siemiginowska}, {Smith}, \& {Wise}}]{ciao06}
{Fruscione}, A., {McDowell}, J.~C., {Allen}, G.~E., {et~al.} 2006, in Society of Photo-Optical Instrumentation Engineers (SPIE) Conference Series, Vol. 6270, Society of Photo-Optical Instrumentation Engineers (SPIE) Conference Series, ed. D.~R. {Silva} \& R.~E. {Doxsey}, 62701V

\bibitem[{{Garofali} {et~al.}(2023){Garofali}, {Basu-Zych}, {Johnson}, {Tzanavaris}, {Jaskot}, {Richardson}, {Lehmer}, {Yukita}, {Hodges-Kluck}, {Hornschemeier}, {Ptak}, \& {Vulic}}]{garofali23}
{Garofali}, K., {Basu-Zych}, A.~R., {Johnson}, B.~D., {et~al.} 2023, arXiv e-prints, arXiv:2307.00050

\bibitem[{{Gezari}(2021)}]{gezari21}
{Gezari}, S. 2021, \araa, 59, 21

\bibitem[{{Gilbertson} {et~al.}(2022){Gilbertson}, {Lehmer}, {Doore}, {Eufrasio}, {Basu-Zych}, {Brandt}, {Fragos}, {Garofali}, {Kovlakas}, {Luo}, {Tozzi}, {Vito}, {Williams}, \& {Xue}}]{gilbertson22}
{Gilbertson}, W., {Lehmer}, B.~D., {Doore}, K., {et~al.} 2022, \apj, 926, 28

\bibitem[{{Gilfanov} {et~al.}(2022){Gilfanov}, {Fabbiano}, {Lehmer}, \& {Zezas}}]{gilfanov22}
{Gilfanov}, M., {Fabbiano}, G., {Lehmer}, B., \& {Zezas}, A. 2022, in Handbook of X-ray and Gamma-ray Astrophysics, 105

\bibitem[{{Gilfanov} {et~al.}(2004){Gilfanov}, {Grimm}, \& {Sunyaev}}]{gilfanov04}
{Gilfanov}, M., {Grimm}, H.~J., \& {Sunyaev}, R. 2004, \mnras, 351, 1365

\bibitem[{{G{\"o}tberg} {et~al.}(2020){G{\"o}tberg}, {de Mink}, {McQuinn}, {Zapartas}, {Groh}, \& {Norman}}]{gotberg20}
{G{\"o}tberg}, Y., {de Mink}, S.~E., {McQuinn}, M., {et~al.} 2020, \aap, 634, A134

\bibitem[{{Greene} {et~al.}(2020){Greene}, {Strader}, \& {Ho}}]{green20}
{Greene}, J.~E., {Strader}, J., \& {Ho}, L.~C. 2020, \araa, 58, 257

\bibitem[{{Helou} {et~al.}(1991){Helou}, {Madore}, {Schmitz}, {Bicay}, {Wu}, \& {Bennett}}]{ned}
{Helou}, G., {Madore}, B.~F., {Schmitz}, M., {et~al.} 1991, in Astrophysics and Space Science Library, Vol. 171, Databases and On-line Data in Astronomy, ed. M.~A. {Albrecht} \& D.~{Egret}, 89--106

\bibitem[{{HERA Collaboration} {et~al.}(2023){HERA Collaboration}, {Abdurashidova}, {Adams}, {Aguirre}, {Alexander}, {Ali}, {Baartman}, {Balfour}, {Barkana}, {Beardsley}, {Bernardi}, {Billings}, {Bowman}, {Bradley}, {Breitman}, {Bull}, {Burba}, {Carey}, {Carilli}, {Cheng}, {Choudhuri}, {DeBoer}, {de Lera Acedo}, {Dexter}, {Dillon}, {Ely}, {Ewall-Wice}, {Fagnoni}, {Fialkov}, {Fritz}, {Furlanetto}, {Gale-Sides}, {Garsden}, {Glendenning}, {Gorce}, {Gorthi}, {Greig}, {Grobbelaar}, {Halday}, {Hazelton}, {Heimersheim}, {Hewitt}, {Hickish}, {Jacobs}, {Julius}, {Kern}, {Kerrigan}, {Kittiwisit}, {Kohn}, {Kolopanis}, {Lanman}, {La Plante}, {Lewis}, {Liu}, {Loots}, {Ma}, {MacMahon}, {Malan}, {Malgas}, {Malgas}, {Maree}, {Marero}, {Martinot}, {McBride}, {Mesinger}, {Mirocha}, {Molewa}, {Morales}, {Mosiane}, {Mu{\~n}oz}, {Murray}, {Nagpal}, {Neben}, {Nikolic}, {Nunhokee}, {Nuwegeld}, {Parsons}, {Pascua}, {Patra}, {Pieterse}, {Qin}, {Razavi-Ghods}, {Robnett}, {Rosie}, {Santos}, {Sims}, {Singh}, {Smith}, {Swarts}, {Tan},
  {Thyagarajan}, {Wilensky}, {Williams}, {van Wyngaarden}, \& {Zheng}}]{abdurashidova23}
{HERA Collaboration}, {Abdurashidova}, Z., {Adams}, T., {et~al.} 2023, \apj, 945, 124

\bibitem[{{Hickson}(1982)}]{hickson82}
{Hickson}, P. 1982, \apj, 255, 382

\bibitem[{{Ho}(2008)}]{ho08}
{Ho}, L.~C. 2008, \araa, 46, 475

\bibitem[{{Hopp} {et~al.}(2000){Hopp}, {Engels}, {Green}, {Ugryumov}, {Izotov}, {Hagen}, {Kniazev}, {Lipovetsky}, {Pustilnik}, {Brosch}, {Masegosa}, {Martin}, \& {M{\'a}rquez}}]{hopp00}
{Hopp}, U., {Engels}, D., {Green}, R.~F., {et~al.} 2000, \aaps, 142, 417

\bibitem[{{Jarrett} {et~al.}(2000){Jarrett}, {Chester}, {Cutri}, {Schneider}, {Skrutskie}, \& {Huchra}}]{2mass}
{Jarrett}, T.~H., {Chester}, T., {Cutri}, R., {et~al.} 2000, \aj, 119, 2498

\bibitem[{{Jones} {et~al.}(2009){Jones}, {Read}, {Saunders}, {Colless}, {Jarrett}, {Parker}, {Fairall}, {Mauch}, {Sadler}, {Watson}, {Burton}, {Campbell}, {Cass}, {Croom}, {Dawe}, {Fiegert}, {Frankcombe}, {Hartley}, {Huchra}, {James}, {Kirby}, {Lahav}, {Lucey}, {Mamon}, {Moore}, {Peterson}, {Prior}, {Proust}, {Russell}, {Safouris}, {Wakamatsu}, {Westra}, \& {Williams}}]{jones09}
{Jones}, D.~H., {Read}, M.~A., {Saunders}, W., {et~al.} 2009, \mnras, 399, 683

\bibitem[{{Jones} {et~al.}(2004){Jones}, {Saunders}, {Colless}, {Read}, {Parker}, {Watson}, {Campbell}, {Burkey}, {Mauch}, {Moore}, {Hartley}, {Cass}, {James}, {Russell}, {Fiegert}, {Dawe}, {Huchra}, {Jarrett}, {Lahav}, {Lucey}, {Mamon}, {Proust}, {Sadler}, \& {Wakamatsu}}]{jones04}
{Jones}, D.~H., {Saunders}, W., {Colless}, M., {et~al.} 2004, \mnras, 355, 747

\bibitem[{{Kauffmann} {et~al.}(2003{\natexlab{a}}){Kauffmann}, {Heckman}, {Tremonti}, {Brinchmann}, {Charlot}, {White}, {Ridgway}, {Brinkmann}, {Fukugita}, {Hall}, {Ivezi{\'c}}, {Richards}, \& {Schneider}}]{sdss1}
{Kauffmann}, G., {Heckman}, T.~M., {Tremonti}, C., {et~al.} 2003{\natexlab{a}}, \mnras, 346, 1055

\bibitem[{{Kauffmann} {et~al.}(2003{\natexlab{b}}){Kauffmann}, {Heckman}, {Tremonti}, {Brinchmann}, {Charlot}, {White}, {Ridgway}, {Brinkmann}, {Fukugita}, {Hall}, {Ivezi{\'c}}, {Richards}, \& {Schneider}}]{kauffmann03}
{Kauffmann}, G., {Heckman}, T.~M., {Tremonti}, C., {et~al.} 2003{\natexlab{b}}, \mnras, 346, 1055

\bibitem[{{Kaur} {et~al.}(2022){Kaur}, {Qin}, {Mesinger}, {Pallottini}, {Fragos}, \& {Basu-Zych}}]{21cm_stochasticity_Kaur}
{Kaur}, H.~D., {Qin}, Y., {Mesinger}, A., {et~al.} 2022, \mnras, 513, 5097

\bibitem[{{Kennicutt} \& {Evans}(2012)}]{kennicutt12}
{Kennicutt}, R.~C. \& {Evans}, N.~J. 2012, \araa, 50, 531

\bibitem[{{Kewley} {et~al.}(2001){Kewley}, {Dopita}, {Sutherland}, {Heisler}, \& {Trevena}}]{kewley01}
{Kewley}, L.~J., {Dopita}, M.~A., {Sutherland}, R.~S., {Heisler}, C.~A., \& {Trevena}, J. 2001, \apj, 556, 121

\bibitem[{{Kewley} \& {Ellison}(2008)}]{kewley_elison08}
{Kewley}, L.~J. \& {Ellison}, S.~L. 2008, \apj, 681, 1183

\bibitem[{{Kim} {et~al.}(2007){Kim}, {Wilkes}, {Kim}, {Green}, {Barkhouse}, {Lee}, {Silverman}, \& {Tananbaum}}]{kim07}
{Kim}, M., {Wilkes}, B.~J., {Kim}, D.-W., {et~al.} 2007, \apj, 659, 29

\bibitem[{{Kouroumpatzakis} {et~al.}(2023){Kouroumpatzakis}, {Zezas}, {Kyritsis}, {Salim}, \& {Svoboda}}]{sfr_stellar_mass_kouroumpatzakis}
{Kouroumpatzakis}, K., {Zezas}, A., {Kyritsis}, E., {Salim}, S., \& {Svoboda}, J. 2023, \aap, 673, A16

\bibitem[{{Kouroumpatzakis} {et~al.}(2020){Kouroumpatzakis}, {Zezas}, {Sell}, {Kovlakas}, {Bonfini}, {Willner}, {Ashby}, {Maragkoudakis}, \& {Jarrett}}]{kouroumpatzakis20}
{Kouroumpatzakis}, K., {Zezas}, A., {Sell}, P., {et~al.} 2020, \mnras, 494, 5967

\bibitem[{{Kouroumpatzakis} {et~al.}(2021){Kouroumpatzakis}, {Zezas}, {Wolter}, {Fruscione}, {Anastasopoulou}, \& {Prestwich}}]{kouroumpatzakis21}
{Kouroumpatzakis}, K., {Zezas}, A., {Wolter}, A., {et~al.} 2021, \mnras, 500, 962

\bibitem[{{Kovlakas} {et~al.}(2021){Kovlakas}, {Zezas}, {Andrews}, {Basu-Zych}, {Fragos}, {Hornschemeier}, {Kouroumpatzakis}, {Lehmer}, \& {Ptak}}]{hecate}
{Kovlakas}, K., {Zezas}, A., {Andrews}, J.~J., {et~al.} 2021, \mnras, 506, 1896

\bibitem[{{Law-Smith} {et~al.}(2017){Law-Smith}, {Ramirez-Ruiz}, {Ellison}, \& {Foley}}]{lawsmith17}
{Law-Smith}, J., {Ramirez-Ruiz}, E., {Ellison}, S.~L., \& {Foley}, R.~J. 2017, \apj, 850, 22

\bibitem[{{Lehmer} {et~al.}(2016){Lehmer}, {Basu-Zych}, {Mineo}, {Brandt}, {Eufrasio}, {Fragos}, {Hornschemeier}, {Luo}, {Xue}, {Bauer}, {Gilfanov}, {Ranalli}, {Schneider}, {Shemmer}, {Tozzi}, {Trump}, {Vignali}, {Wang}, {Yukita}, \& {Zezas}}]{lehmer16}
{Lehmer}, B.~D., {Basu-Zych}, A.~R., {Mineo}, S., {et~al.} 2016, \apj, 825, 7

\bibitem[{{Lehmer} {et~al.}(2021){Lehmer}, {Eufrasio}, {Basu-Zych}, {Doore}, {Fragos}, {Garofali}, {Kovlakas}, {Williams}, {Zezas}, \& {Santana-Silva}}]{lehmer21}
{Lehmer}, B.~D., {Eufrasio}, R.~T., {Basu-Zych}, A., {et~al.} 2021, \apj, 907, 17

\bibitem[{{Lehmer} {et~al.}(2022){Lehmer}, {Eufrasio}, {Basu-Zych}, {Garofali}, {Gilbertson}, {Mesinger}, \& {Yukita}}]{lehmer22}
{Lehmer}, B.~D., {Eufrasio}, R.~T., {Basu-Zych}, A., {et~al.} 2022, \apj, 930, 135

\bibitem[{{Lehmer} {et~al.}(2017){Lehmer}, {Eufrasio}, {Markwardt}, {Zezas}, {Basu-Zych}, {Fragos}, {Hornschemeier}, {Ptak}, {Tzanavaris}, \& {Yukita}}]{lehmer17}
{Lehmer}, B.~D., {Eufrasio}, R.~T., {Markwardt}, L., {et~al.} 2017, \apj, 851, 11

\bibitem[{{Lehmer} {et~al.}(2019){Lehmer}, {Eufrasio}, {Tzanavaris}, {Basu-Zych}, {Fragos}, {Prestwich}, {Yukita}, {Zezas}, {Hornschemeier}, \& {Ptak}}]{lehmer19}
{Lehmer}, B.~D., {Eufrasio}, R.~T., {Tzanavaris}, P., {et~al.} 2019, \apjs, 243, 3

\bibitem[{{Leroy} {et~al.}(2019){Leroy}, {Sandstrom}, {Lang}, {Lewis}, {Salim}, {Behrens}, {Chastenet}, {Chiang}, {Gallagher}, {Kessler}, \& {Utomo}}]{leroy19}
{Leroy}, A.~K., {Sandstrom}, K.~M., {Lang}, D., {et~al.} 2019, \apjs, 244, 24

\bibitem[{{Linden} {et~al.}(2010){Linden}, {Kalogera}, {Sepinsky}, {Prestwich}, {Zezas}, \& {Gallagher}}]{linden10}
{Linden}, T., {Kalogera}, V., {Sepinsky}, J.~F., {et~al.} 2010, \apj, 725, 1984

\bibitem[{{Liu} {et~al.}(2019){Liu}, {Liu}, {Dong}, {Zhou}, {Wang}, {Lu}, \& {Yuan}}]{liu19}
{Liu}, H.-Y., {Liu}, W.-J., {Dong}, X.-B., {et~al.} 2019, \apjs, 243, 21

\bibitem[{{Louppe}(2014)}]{louppe04}
{Louppe}, G. 2014, arXiv e-prints, arXiv:1407.7502

\bibitem[{{Madau} \& {Fragos}(2017)}]{madau17}
{Madau}, P. \& {Fragos}, T. 2017, \apj, 840, 39

\bibitem[{{Makarov} {et~al.}(2014){Makarov}, {Prugniel}, {Terekhova}, {Courtois}, \& {Vauglin}}]{hyperleda}
{Makarov}, D., {Prugniel}, P., {Terekhova}, N., {Courtois}, H., \& {Vauglin}, I. 2014, \aap, 570, A13

\bibitem[{{Marchant} {et~al.}(2016){Marchant}, {Langer}, {Podsiadlowski}, {Tauris}, \& {Moriya}}]{marchant16}
{Marchant}, P., {Langer}, N., {Podsiadlowski}, P., {Tauris}, T.~M., \& {Moriya}, T.~J. 2016, \aap, 588, A50

\bibitem[{{Merloni} {et~al.}(2014){Merloni}, {Bongiorno}, {Brusa}, {Iwasawa}, {Mainieri}, {Magnelli}, {Salvato}, {Berta}, {Cappelluti}, {Comastri}, {Fiore}, {Gilli}, {Koekemoer}, {Le Floc'h}, {Lusso}, {Lutz}, {Miyaji}, {Pozzi}, {Riguccini}, {Rosario}, {Silverman}, {Symeonidis}, {Treister}, {Vignali}, \& {Zamorani}}]{merloni14}
{Merloni}, A., {Bongiorno}, A., {Brusa}, M., {et~al.} 2014, \mnras, 437, 3550

\bibitem[{{Merloni} {et~al.}(2024){Merloni}, {Lamer}, {Liu}, {Ramos-Ceja}, {Brunner}, {Bulbul}, {Dennerl}, {Doroshenko}, {Freyberg}, {Friedrich}, {Gatuzz}, {Georgakakis}, {Haberl}, {Igo}, {Kreykenbohm}, {Liu}, {Maitra}, {Malyali}, {Mayer}, {Nandra}, {Predehl}, {Robrade}, {Salvato}, {Sanders}, {Stewart}, {Tub{\'\i}n-Arenas}, {Weber}, {Wilms}, {Arcodia}, {Artis}, {Aschersleben}, {Avakyan}, {Aydar}, {Bahar}, {Balzer}, {Becker}, {Berger}, {Boller}, {Bornemann}, {Br{\"u}ggen}, {Brusa}, {Buchner}, {Burwitz}, {Camilloni}, {Clerc}, {Comparat}, {Coutinho}, {Czesla}, {Dannhauer}, {Dauner}, {Dauser}, {Dietl}, {Dolag}, {Dwelly}, {Egg}, {Ehl}, {Freund}, {Friedrich}, {Gaida}, {Garrel}, {Ghirardini}, {Gokus}, {Gr{\"u}nwald}, {Grandis}, {Grotova}, {Gruen}, {Gueguen}, {H{\"a}mmerich}, {Hamaus}, {Hasinger}, {Haubner}, {Homan}, {Ider Chitham}, {Joseph}, {Joyce}, {K{\"o}nig}, {Kaltenbrunner}, {Khokhriakova}, {Kink}, {Kirsch}, {Kluge}, {Knies}, {Krippendorf}, {Krumpe}, {Kurpas}, {Li}, {Liu}, {Locatelli}, {Lorenz}, {M{\"u}ller},
  {Magaudda}, {Mannes}, {McCall}, {Meidinger}, {Michailidis}, {Migkas}, {Mu{\~n}oz-Giraldo}, {Musiimenta}, {Nguyen-Dang}, {Ni}, {Olechowska}, {Ota}, {Pacaud}, {Pasini}, {Perinati}, {Pires}, {Pommranz}, {Ponti}, {Poppenhaeger}, {P{\"u}hlhofer}, {Rau}, {Reh}, {Reiprich}, {Roster}, {Saeedi}, {Santangelo}, {Sasaki}, {Schmitt}, {Schneider}, {Schrabback}, {Schuster}, {Schwope}, {Seppi}, {Serim}, {Shreeram}, {Sokolova-Lapa}, {Starck}, {Stelzer}, {Stierhof}, {Suleimanov}, {Tenzer}, {Traulsen}, {Tr{\"u}mper}, {Tsuge}, {Urrutia}, {Veronica}, {Waddell}, {Willer}, {Wolf}, {Yeung}, {Zainab}, {Zangrandi}, {Zhang}, {Zhang}, \& {Zheng}}]{merloni24}
{Merloni}, A., {Lamer}, G., {Liu}, T., {et~al.} 2024, \aap, 682, A34

\bibitem[{{Mineo} {et~al.}(2014){Mineo}, {Gilfanov}, {Lehmer}, {Morrison}, \& {Sunyaev}}]{mineo14}
{Mineo}, S., {Gilfanov}, M., {Lehmer}, B.~D., {Morrison}, G.~E., \& {Sunyaev}, R. 2014, \mnras, 437, 1698

\bibitem[{{Mineo} {et~al.}(2012{\natexlab{a}}){Mineo}, {Gilfanov}, \& {Sunyaev}}]{mineo12a}
{Mineo}, S., {Gilfanov}, M., \& {Sunyaev}, R. 2012{\natexlab{a}}, \mnras, 419, 2095

\bibitem[{{Mineo} {et~al.}(2012{\natexlab{b}}){Mineo}, {Gilfanov}, \& {Sunyaev}}]{mineo12b}
{Mineo}, S., {Gilfanov}, M., \& {Sunyaev}, R. 2012{\natexlab{b}}, \mnras, 426, 1870

\bibitem[{{Owen} \& {Warwick}(2009)}]{owen09}
{Owen}, R.~A. \& {Warwick}, R.~S. 2009, \mnras, 394, 1741

\bibitem[{{Park} {et~al.}(2006){Park}, {Kashyap}, {Siemiginowska}, {van Dyk}, {Zezas}, {Heinke}, \& {Wargelin}}]{park06}
{Park}, T., {Kashyap}, V.~L., {Siemiginowska}, A., {et~al.} 2006, \apj, 652, 610

\bibitem[{{Peng} {et~al.}(2010){Peng}, {Lilly}, {Kova{\v{c}}}, {Bolzonella}, {Pozzetti}, {Renzini}, {Zamorani}, {Ilbert}, {Knobel}, {Iovino}, {Maier}, {Cucciati}, {Tasca}, {Carollo}, {Silverman}, {Kampczyk}, {de Ravel}, {Sanders}, {Scoville}, {Contini}, {Mainieri}, {Scodeggio}, {Kneib}, {Le F{\`e}vre}, {Bardelli}, {Bongiorno}, {Caputi}, {Coppa}, {de la Torre}, {Franzetti}, {Garilli}, {Lamareille}, {Le Borgne}, {Le Brun}, {Mignoli}, {Perez Montero}, {Pello}, {Ricciardelli}, {Tanaka}, {Tresse}, {Vergani}, {Welikala}, {Zucca}, {Oesch}, {Abbas}, {Barnes}, {Bordoloi}, {Bottini}, {Cappi}, {Cassata}, {Cimatti}, {Fumana}, {Hasinger}, {Koekemoer}, {Leauthaud}, {Maccagni}, {Marinoni}, {McCracken}, {Memeo}, {Meneux}, {Nair}, {Porciani}, {Presotto}, \& {Scaramella}}]{peng10}
{Peng}, Y.-j., {Lilly}, S.~J., {Kova{\v{c}}}, K., {et~al.} 2010, \apj, 721, 193

\bibitem[{{Predehl} {et~al.}(2021){Predehl}, {Andritschke}, {Arefiev}, {Babyshkin}, {Batanov}, {Becker}, {B{\"o}hringer}, {Bogomolov}, {Boller}, {Borm}, {Bornemann}, {Br{\"a}uninger}, {Br{\"u}ggen}, {Brunner}, {Brusa}, {Bulbul}, {Buntov}, {Burwitz}, {Burkert}, {Clerc}, {Churazov}, {Coutinho}, {Dauser}, {Dennerl}, {Doroshenko}, {Eder}, {Emberger}, {Eraerds}, {Finoguenov}, {Freyberg}, {Friedrich}, {Friedrich}, {F{\"u}rmetz}, {Georgakakis}, {Gilfanov}, {Granato}, {Grossberger}, {Gueguen}, {Gureev}, {Haberl}, {H{\"a}lker}, {Hartner}, {Hasinger}, {Huber}, {Ji}, {Kienlin}, {Kink}, {Korotkov}, {Kreykenbohm}, {Lamer}, {Lomakin}, {Lapshov}, {Liu}, {Maitra}, {Meidinger}, {Menz}, {Merloni}, {Mernik}, {Mican}, {Mohr}, {M{\"u}ller}, {Nandra}, {Nazarov}, {Pacaud}, {Pavlinsky}, {Perinati}, {Pfeffermann}, {Pietschner}, {Ramos-Ceja}, {Rau}, {Reiffers}, {Reiprich}, {Robrade}, {Salvato}, {Sanders}, {Santangelo}, {Sasaki}, {Scheuerle}, {Schmid}, {Schmitt}, {Schwope}, {Shirshakov}, {Steinmetz}, {Stewart}, {Str{\"u}der},
  {Sunyaev}, {Tenzer}, {Tiedemann}, {Tr{\"u}mper}, {Voron}, {Weber}, {Wilms}, \& {Yaroshenko}}]{predehl}
{Predehl}, P., {Andritschke}, R., {Arefiev}, V., {et~al.} 2021, \aap, 647, A1

\bibitem[{{Renzini} \& {Peng}(2015)}]{renzini15}
{Renzini}, A. \& {Peng}, Y.-j. 2015, \apjl, 801, L29

\bibitem[{{Ricci} {et~al.}(2015){Ricci}, {Ueda}, {Koss}, {Trakhtenbrot}, {Bauer}, \& {Gandhi}}]{ricci15}
{Ricci}, C., {Ueda}, Y., {Koss}, M.~J., {et~al.} 2015, \apjl, 815, L13

\bibitem[{{Riccio} {et~al.}(2023){Riccio}, {Yang}, {Ma{\l}ek}, {Boquien}, {Junais}, {Pistis}, {Hamed}, {Grespan}, {Paolillo}, \& {Torbaniuk}}]{riccio23}
{Riccio}, G., {Yang}, G., {Ma{\l}ek}, K., {et~al.} 2023, \aap, 678, A164

\bibitem[{{Risaliti} \& {Elvis}(2004)}]{risaliti04}
{Risaliti}, G. \& {Elvis}, M. 2004, in Astrophysics and Space Science Library, Vol. 308, Supermassive Black Holes in the Distant Universe, ed. A.~J. {Barger}, 187

\bibitem[{{Salim} {et~al.}(2014){Salim}, {Lee}, {Ly}, {Brinchmann}, {Dav{\'e}}, {Dickinson}, {Salzer}, \& {Charlot}}]{salim14}
{Salim}, S., {Lee}, J.~C., {Ly}, C., {et~al.} 2014, \apj, 797, 126

\bibitem[{{Schaerer} {et~al.}(2019){Schaerer}, {Fragos}, \& {Izotov}}]{schaerer19}
{Schaerer}, D., {Fragos}, T., \& {Izotov}, Y.~I. 2019, \aap, 622, L10

\bibitem[{{Silva} {et~al.}(2005){Silva}, {Massey}, {DeGioia-Eastwood}, \& {Henning}}]{silva05}
{Silva}, D.~R., {Massey}, P., {DeGioia-Eastwood}, K., \& {Henning}, P.~A. 2005, \apj, 623, 148

\bibitem[{{Smith} {et~al.}(2001){Smith}, {Brickhouse}, {Liedahl}, \& {Raymond}}]{smith01}
{Smith}, R.~K., {Brickhouse}, N.~S., {Liedahl}, D.~A., \& {Raymond}, J.~C. 2001, \apjl, 556, L91

\bibitem[{{Stampoulis} {et~al.}(2019){Stampoulis}, {van Dyk}, {Kashyap}, \& {Zezas}}]{stampoulis19}
{Stampoulis}, V., {van Dyk}, D.~A., {Kashyap}, V.~L., \& {Zezas}, A. 2019, \mnras, 485, 1085

\bibitem[{{Sunyaev} {et~al.}(2021){Sunyaev}, {Arefiev}, {Babyshkin}, {Bogomolov}, {Borisov}, {Buntov}, {Brunner}, {Burenin}, {Churazov}, {Coutinho}, {Eder}, {Eismont}, {Freyberg}, {Gilfanov}, {Gureyev}, {Hasinger}, {Khabibullin}, {Kolmykov}, {Komovkin}, {Krivonos}, {Lapshov}, {Levin}, {Lomakin}, {Lutovinov}, {Medvedev}, {Merloni}, {Mernik}, {Mikhailov}, {Molodtsov}, {Mzhelsky}, {M{\"u}ller}, {Nandra}, {Nazarov}, {Pavlinsky}, {Poghodin}, {Predehl}, {Robrade}, {Sazonov}, {Scheuerle}, {Shirshakov}, {Tkachenko}, \& {Voron}}]{sunyav21}
{Sunyaev}, R., {Arefiev}, V., {Babyshkin}, V., {et~al.} 2021, \aap, 656, A132

\bibitem[{{Tanimoto} {et~al.}(2022){Tanimoto}, {Ueda}, {Odaka}, {Yamada}, \& {Ricci}}]{tanimoto22}
{Tanimoto}, A., {Ueda}, Y., {Odaka}, H., {Yamada}, S., \& {Ricci}, C. 2022, \apjs, 260, 30

\bibitem[{{Taylor}(2005)}]{taylor05}
{Taylor}, M.~B. 2005, in Astronomical Society of the Pacific Conference Series, Vol. 347, Astronomical Data Analysis Software and Systems XIV, ed. P.~{Shopbell}, M.~{Britton}, \& R.~{Ebert}, 29

\bibitem[{{Tremonti} {et~al.}(2004){Tremonti}, {Heckman}, {Kauffmann}, {Brinchmann}, {Charlot}, {White}, {Seibert}, {Peng}, {Schlegel}, {Uomoto}, {Fukugita}, \& {Brinkmann}}]{sdss3}
{Tremonti}, C.~A., {Heckman}, T.~M., {Kauffmann}, G., {et~al.} 2004, \apj, 613, 898

\bibitem[{{van Dyk} {et~al.}(2001){van Dyk}, {Connors}, {Kashyap}, \& {Siemiginowska}}]{vandyk01}
{van Dyk}, D.~A., {Connors}, A., {Kashyap}, V.~L., \& {Siemiginowska}, A. 2001, \apj, 548, 224

\bibitem[{{van Velzen} {et~al.}(2021){van Velzen}, {Gezari}, {Hammerstein}, {Roth}, {Frederick}, {Ward}, {Hung}, {Cenko}, {Stein}, {Perley}, {Taggart}, {Foley}, {Sollerman}, {Blagorodnova}, {Andreoni}, {Bellm}, {Brinnel}, {De}, {Dekany}, {Feeney}, {Fremling}, {Giomi}, {Golkhou}, {Graham}, {Ho}, {Kasliwal}, {Kilpatrick}, {Kulkarni}, {Kupfer}, {Laher}, {Mahabal}, {Masci}, {Miller}, {Nordin}, {Riddle}, {Rusholme}, {van Santen}, {Sharma}, {Shupe}, \& {Soumagnac}}]{vanvelzen21}
{van Velzen}, S., {Gezari}, S., {Hammerstein}, E., {et~al.} 2021, \apj, 908, 4

\bibitem[{{Vulic} {et~al.}(2022){Vulic}, {Hornschemeier}, {Haberl}, {Basu-Zych}, {Kyritsis}, {Zezas}, {Salvato}, {Ptak}, {Bogdan}, {Kovlakas}, {Wilms}, {Sasaki}, {Liu}, {Merloni}, {Dwelly}, {Brunner}, {Lamer}, {Maitra}, {Nandra}, \& {Santangelo}}]{vulic22}
{Vulic}, N., {Hornschemeier}, A.~E., {Haberl}, F., {et~al.} 2022, \aap, 661, A16

\bibitem[{{Wang} {et~al.}(2014){Wang}, {Rowan-Robinson}, {Norberg}, {Heinis}, \& {Han}}]{iras}
{Wang}, L., {Rowan-Robinson}, M., {Norberg}, P., {Heinis}, S., \& {Han}, J. 2014, \mnras, 442, 2739

\bibitem[{{Wen} {et~al.}(2013){Wen}, {Wu}, {Zhu}, {Lam}, {Wu}, {Wicker}, \& {Zhao}}]{wen13}
{Wen}, X.-Q., {Wu}, H., {Zhu}, Y.-N., {et~al.} 2013, \mnras, 433, 2946

\bibitem[{{Wilms} {et~al.}(2000){Wilms}, {Allen}, \& {McCray}}]{wilms00}
{Wilms}, J., {Allen}, A., \& {McCray}, R. 2000, \apj, 542, 914

\bibitem[{{Wright} {et~al.}(2010){Wright}, {Eisenhardt}, {Mainzer}, {Ressler}, {Cutri}, {Jarrett}, {Kirkpatrick}, {Padgett}, {McMillan}, {Skrutskie}, {Stanford}, {Cohen}, {Walker}, {Mather}, {Leisawitz}, {Gautier}, {McLean}, {Benford}, {Lonsdale}, {Blain}, {Mendez}, {Irace}, {Duval}, {Liu}, {Royer}, {Heinrichsen}, {Howard}, {Shannon}, {Kendall}, {Walsh}, {Larsen}, {Cardon}, {Schick}, {Schwalm}, {Abid}, {Fabinsky}, {Naes}, \& {Tsai}}]{wise}
{Wright}, E.~L., {Eisenhardt}, P. R.~M., {Mainzer}, A.~K., {et~al.} 2010, \aj, 140, 1868

\bibitem[{{Yao} {et~al.}(2023){Yao}, {Ravi}, {Gezari}, {van Velzen}, {Lu}, {Schulze}, {Somalwar}, {Kulkarni}, {Hammerstein}, {Nicholl}, {Graham}, {Perley}, {Cenko}, {Stein}, {Ricarte}, {Chadayammuri}, {Quataert}, {Bellm}, {Bloom}, {Dekany}, {Drake}, {Groom}, {Mahabal}, {Prince}, {Riddle}, {Rusholme}, {Sharma}, {Sollerman}, \& {Yan}}]{yao23}
{Yao}, Y., {Ravi}, V., {Gezari}, S., {et~al.} 2023, \apjl, 955, L6

\bibitem[{{Zaw} {et~al.}(2019){Zaw}, {Chen}, \& {Farrar}}]{zaw19}
{Zaw}, I., {Chen}, Y.-P., \& {Farrar}, G.~R. 2019, \apj, 872, 134

\bibitem[{{Zezas}(2021)}]{zezas21}
{Zezas}, A. 2021, High-Energy Star-Formation Rate Indicators, ed. A.~Zezas \& V.~E. Buat, Cambridge Astrophysics (Cambridge University Press), 243

\bibitem[{{Zezas} {et~al.}(2006){Zezas}, {Fabbiano}, {Baldi}, {Schweizer}, {King}, {Ponman}, \& {Rots}}]{zezas06}
{Zezas}, A., {Fabbiano}, G., {Baldi}, A., {et~al.} 2006, \apjs, 166, 211

\end{thebibliography}


\clearpage
\begin{landscape}
\onecolumn
\captionof{table}{Properties of HEC-eR1 star-forming galaxies with reliable X-ray flux measurements.}\label{Tab:Fluxes}
\begin{longtable}{ccccccccccccc}
\hline\hline
\toprule
    PGC & RA &  DEC &  D & SFR &  M$_{\star}$ &log(O/H)+12 &  \multicolumn{3}{c}{F$_{X}^{0.6-2.3}$}& \multicolumn{3}{c}{L$_{X}^{0.5-2}$}\\
    & J2000 &  J2000 &   &  &  & & Mode & Upper & Low & Mode & Upper & Low  \\
    & deg. &  deg. &  Mpc & M$_{\odot}$\,yr$^{-1}$ &  $10^{9}$M$_{\odot}$ & & \multicolumn{3}{c}{10$^{-14}$erg\,s$^{-1}$\,cm$^{-1}$}& \multicolumn{3}{c}{10$^{40}$erg\,s$^{-1}$} \\
    (1)& (2) &  (3) & (4)  & (5) & (6) & (7)& (8) & (9) & (10) & (11) & (12) & (13)  \\
\midrule
\endfirsthead
\toprule
    PGC & RA &  DEC &  D & SFR &  M$_{\star}$ &log(O/H)+12 &  \multicolumn{3}{c}{F$_{X}^{0.6-2.3}$}& \multicolumn{3}{c}{L$_{X}^{0.5-2}$} \\
    & J2000 &  J2000 &   &  &  & & Mode & Upper & Low & Mode & Upper & Low  \\
\midrule
\endhead
\midrule
\multicolumn{13}{r}{{Continued on next page}} \\
\midrule
\endfoot
\bottomrule
\endlastfoot
   9279 &  36.604500 & -15.225847 &  50.46 &  0.264 &         2.16 &        8.503 &               1.05 &                 2.57 &                 0.32 &             0.31 &               0.76 &               0.09 \\
4078746 &  42.137688 & -29.579645 &  26.98 &  0.004 &         0.04 &        8.055 &               4.35 &                 3.56 &                 1.43 &             0.37 &               0.30 &               0.12 \\
  10966 &  43.639803 & -10.028284 &  17.95 &  0.127 &         0.83 &        8.371 &              10.34 &                 2.66 &                 2.54 &             0.39 &               0.10 &               0.09 \\
  12607 &  50.468278 & -19.481766 &  55.19 &  0.343 &         2.72 &        8.533 &               0.82 &                 2.87 &                 0.24 &             0.29 &               1.01 &               0.09 \\
3662384 &  53.892646 & -14.499583 &  64.41 &  0.368 &         1.15 &        8.416 &               1.15 &                 3.30 &                 0.20 &             0.55 &               1.59 &               0.10 \\
  13317 &  54.111312 &  -4.701264 &  85.48 &  0.870 &         5.57 &        8.622 &              35.76 &                 5.53 &                 5.41 &            30.27 &               4.68 &               4.58 \\
  14448 &  61.376541 &   4.411750 &  71.68 &  1.399 &        21.87 &        8.747 &               1.68 &                 3.95 &                 0.55 &             1.00 &               2.35 &               0.33 \\
  14475 &  61.707176 & -21.172653 &   8.14 &  0.005 &         0.12 &        8.136 &               7.03 &                 1.96 &                 2.27 &             0.05 &               0.01 &               0.02 \\
  15051 &  66.160848 &  -0.759896 &  63.07 &  0.877 &         3.97 &        8.582 &               1.57 &                 4.72 &                 0.29 &             0.72 &               2.17 &               0.13 \\
 146315 &  66.912420 & -11.725722 & 135.15 &  4.018 &        19.78 &        8.741 &               1.11 &                 4.07 &                 0.29 &             2.35 &               8.62 &               0.62 \\
  15533 &  68.475042 &  16.912028 &  58.08 &  1.235 &        22.91 &        8.750 &               1.40 &                 3.65 &                 0.37 &             0.55 &               1.43 &               0.15 \\
  15950 &  71.474083 & -17.278931 &  38.18 &  0.170 &         4.25 &        8.590 &               1.59 &                 2.64 &                 0.50 &             0.27 &               0.45 &               0.08 \\
  16018 &  71.909937 & -17.433986 &  53.23 &  0.410 &         1.41 &        8.444 &               0.98 &                 3.45 &                 0.26 &             0.32 &               1.13 &               0.09 \\
  16988 &  78.839152 & -26.471559 &  50.36 &  0.366 &         1.46 &        8.448 &               2.21 &                 6.00 &                 0.52 &             0.65 &               1.76 &               0.15 \\
2816473 &  80.340125 & -15.542583 &  44.33 &  0.302 &         3.16 &        8.552 &               1.05 &                 3.37 &                 0.27 &             0.24 &               0.77 &               0.06 \\
  18749 &  94.091796 & -21.372709 &  36.14 &  4.974 &       177.63 &        8.732 &              33.87 &                 6.22 &                 5.48 &             5.12 &               0.94 &               0.83 \\
  22894 & 122.350163 &   0.609478 &  24.21 &  0.034 &         0.88 &        8.378 &               2.85 &                 5.65 &                 0.86 &             0.19 &               0.38 &               0.06 \\
  22962 & 122.846190 &   3.633118 &  56.43 &  6.080 &        68.76 &        8.778 &               2.56 &                 5.90 &                 0.59 &             0.94 &               2.18 &               0.22 \\
  23218 & 124.250390 & -25.370182 &  19.15 &  0.005 &         0.08 &        8.095 &               4.83 &                 5.11 &                 1.57 &             0.21 &               0.22 &               0.07 \\
  23303 & 124.690000 & -25.499500 &  17.78 &  1.887 &        34.22 &        8.769 &              22.26 &                 4.71 &                 5.08 &             0.82 &               0.17 &               0.19 \\
  24071 & 128.639086 &  -2.546869 &  66.68 &  0.270 &        95.11 &        8.770 &               3.01 &                 5.41 &                 0.75 &             1.55 &               2.78 &               0.38 \\
  24178 & 129.064125 &  28.060194 &  49.67 &  0.898 &         7.09 &        8.650 &               1.80 &                 6.01 &                 0.58 &             0.52 &               1.72 &               0.16 \\
  24469 & 130.700826 &  14.265250 &  29.99 &  0.182 &         1.58 &        8.527 &               4.63 &                 9.93 &                 1.41 &             0.48 &               1.04 &               0.15 \\
  25231 & 134.735486 &  -4.901998 &  63.90 &  1.626 &        30.43 &        8.764 &               2.75 &                 5.29 &                 0.85 &             1.30 &               2.50 &               0.40 \\
  25515 & 136.380136 & -19.042773 &  31.92 &  0.052 &         1.77 &        8.475 &               3.48 &                 6.07 &                 0.87 &             0.41 &               0.72 &               0.10 \\
  27077 & 143.042052 &  21.501566 &   9.43 &  2.376 &        33.68 &        8.768 &              59.54 &                 9.57 &                10.62 &             0.61 &               0.10 &               0.11 \\
  27622 & 145.180089 &   3.959247 &  22.70 &  0.014 &         0.24 &        8.138 &               3.19 &                 7.31 &                 0.85 &             0.19 &               0.44 &               0.05 \\
  28010 & 146.515055 &   1.668362 &  28.32 &  1.047 &         7.60 &        8.431 &               2.59 &                 5.47 &                 0.62 &             0.24 &               0.51 &               0.06 \\
  28310 & 147.547083 &  16.286580 &  83.30 &  1.237 &        14.88 &        8.734 &               2.37 &                 5.33 &                 0.49 &             1.91 &               4.28 &               0.39 \\
4080350 & 149.547729 & -29.576808 &  30.60 &  0.166 &         0.80 &        8.366 &               2.22 &                 4.40 &                 0.66 &             0.24 &               0.48 &               0.07 \\
  28805 & 149.587874 &  32.369890 &  20.33 &  0.211 &         1.05 &        8.867 &               5.12 &                 5.21 &                 1.64 &             0.25 &               0.25 &               0.08 \\
  29366 & 151.638789 & -29.935274 &  14.98 &  0.172 &         0.69 &        8.345 &              18.08 &                 4.84 &                 4.95 &             0.47 &               0.13 &               0.13 \\
  31359 & 158.846577 & -24.754115 &   7.48 &  0.015 &         0.61 &        8.044 &              92.82 &                10.47 &                 9.77 &             0.60 &               0.07 &               0.06 \\
  31435 & 159.088822 &  13.711433 &  47.86 &  1.171 &        16.98 &        8.730 &               1.95 &                 5.12 &                 0.51 &             0.52 &               1.36 &               0.14 \\
  31691 & 159.851568 &  -0.389254 &  76.36 &  0.690 &         3.97 &        8.582 &               5.64 &                11.44 &                 1.85 &             3.81 &               7.72 &               1.25 \\
  32424 & 162.458791 &  32.982833 &  19.64 &  0.030 &         0.41 &        8.276 &              13.98 &                 4.34 &                 4.66 &             0.62 &               0.19 &               0.21 \\
  32434 & 162.479479 &  32.990823 &  21.78 &  0.900 &         3.21 &        8.490 &              14.45 &                 3.89 &                 4.72 &             0.79 &               0.21 &               0.26 \\
  32517 & 162.743466 &  -2.150347 &  62.63 &  6.401 &        30.62 &        8.765 &               1.82 &                 5.18 &                 0.43 &             0.83 &               2.35 &               0.20 \\
  32954 & 164.445880 &  36.260757 &  12.11 &  0.011 &         0.05 &        8.062 &              14.56 &                 7.24 &                 4.82 &             0.25 &               0.12 &               0.08 \\
 170135 & 170.712874 &  -7.588486 &  95.84 &  2.467 &        34.18 &        8.769 &               1.49 &                 4.99 &                 0.42 &             1.59 &               5.31 &               0.45 \\
  35294 & 172.046015 &  25.661080 &  45.85 &  0.553 &         1.14 &        8.415 &               2.75 &                 4.86 &                 0.58 &             0.67 &               1.18 &               0.14 \\
  35942 & 174.226392 &  19.971530 &  98.21 &  6.889 &        28.42 &        8.710 &               2.05 &                 6.48 &                 0.32 &             2.29 &               7.24 &               0.35 \\
  36445 & 175.874399 & -16.796435 &  50.82 &  1.674 &         6.84 &        8.646 &               1.93 &                 3.55 &                 0.50 &             0.58 &               1.06 &               0.15 \\
  36658 & 176.396880 &   3.231860 &  86.03 &  5.191 &        82.09 &        8.831 &               2.03 &                 6.09 &                 0.38 &             1.74 &               5.22 &               0.32 \\
  37285 & 178.419261 &  -3.996432 &  23.01 &  0.108 &         0.79 &        8.363 &               2.68 &                 5.35 &                 0.63 &             0.16 &               0.33 &               0.04 \\
  38040 & 180.709269 &  18.015579 &  68.17 &  2.460 &        14.35 &        8.717 &               1.57 &                 5.02 &                 0.27 &             0.85 &               2.70 &               0.15 \\
  38093 & 180.863897 &  16.485705 &  58.53 &  2.162 &        15.14 &        8.838 &               1.50 &                 4.81 &                 0.24 &             0.60 &               1.91 &               0.09 \\
  38285 & 181.346511 &  17.885981 &  72.11 &  1.977 &        24.25 &        8.753 &               2.29 &                 4.43 &                 0.47 &             1.38 &               2.67 &               0.28 \\
3098529 & 181.354660 &  15.514361 &  72.96 &  0.612 &         5.46 &        8.620 &               2.36 &                 6.81 &                 0.42 &             1.46 &               4.20 &               0.26 \\
  39578 & 184.706731 &  14.416792 &  14.06 &  3.553 &        29.28 &        8.763 &              27.64 &                 5.96 &                 5.97 &             0.63 &               0.14 &               0.14 \\
  39904 & 185.297146 &  17.638618 &  16.98 &  0.005 &         0.14 &        8.384 &               2.05 &                 5.71 &                 0.40 &             0.07 &               0.19 &               0.01 \\
  40153 & 185.728667 &  15.822249 &  15.42 &  3.443 &        67.87 &        8.778 &              33.21 &                 7.28 &                 6.70 &             0.91 &               0.20 &               0.18 \\
  41789 & 188.535251 &   2.653794 &  13.82 &  1.141 &         2.46 &        8.827 &              20.56 &                 4.35 &                 4.89 &             0.46 &               0.10 &               0.11 \\
  41850 & 188.670396 &   3.323671 &  66.25 &  1.297 &        11.11 &        8.694 &               1.50 &                 5.03 &                 0.44 &             0.76 &               2.56 &               0.22 \\
  42002 & 188.990341 &  27.959684 &   8.78 &  0.271 &         5.17 &        8.614 &             111.05 &                10.54 &                10.70 &             0.99 &               0.09 &               0.10 \\
  42020 & 189.033922 &  19.322671 &  17.80 &  0.283 &         1.52 &        8.351 &              16.35 &                 4.22 &                 4.72 &             0.60 &               0.15 &               0.17 \\
  42089 & 189.208080 &  13.163395 &   9.86 &  0.524 &        23.05 &        8.750 &              34.59 &                 7.60 &                 6.70 &             0.39 &               0.09 &               0.08 \\
4340445 & 191.416708 &  22.052195 &  99.61 &  0.098 &         0.46 &        8.289 &               6.61 &                17.13 &                 1.07 &             7.59 &              19.68 &               1.23 \\
  43051 & 191.518446 &   8.476259 &   9.82 &  0.001 &         0.04 &        8.371 &               1.30 &                 4.14 &                 0.23 &             0.01 &               0.05 &               0.00 \\
  43238 & 192.056948 &  -3.332820 &  18.21 &  0.085 &         0.73 &        8.681 &              15.19 &                 3.53 &                 4.15 &             0.58 &               0.14 &               0.16 \\
  44112 & 194.042147 &  -8.151319 &  56.38 &  0.408 &         7.96 &        8.662 &              32.10 &                 5.27 &                 5.86 &            11.82 &               1.94 &               2.16 \\
  44186 & 194.194335 &   5.916345 & 132.99 &  6.388 &        20.96 &        8.661 &               2.47 &                 6.70 &                 0.58 &             5.05 &              13.73 &               1.18 \\
  44445 & 194.588127 &   4.885624 & 151.02 & 15.452 &        53.84 &        8.965 &               1.05 &                 3.69 &                 0.27 &             2.79 &               9.74 &               0.72 \\
1013976 & 194.777416 &  -7.727722 &  48.56 &  0.031 &         2.84 &        8.539 &              35.07 &                 7.79 &                 7.95 &             9.58 &               2.13 &               2.17 \\
  45608 & 197.322021 &  -5.272778 &  45.13 &  1.214 &         7.54 &        8.656 &              11.01 &                 3.15 &                 3.58 &             2.60 &               0.74 &               0.84 \\
1133474 & 197.982541 &  -0.836541 &  78.24 &  0.013 &         1.85 &        8.447 &              52.09 &                14.77 &                15.53 &            36.93 &              10.47 &              11.01 \\
3093942 & 200.870937 & -16.978208 & 100.53 &  0.852 &         4.18 &        8.588 &               1.45 &                 5.07 &                 0.40 &             1.70 &               5.94 &               0.47 \\
  88902 & 203.493001 & -28.274849 & 187.16 &  1.053 &         3.69 &        8.572 &               7.50 &                 6.22 &                 2.42 &            30.42 &              25.25 &               9.83 \\
  48130 & 204.383420 &   8.885176 &  13.38 &  0.060 &         1.07 &        8.405 &              22.15 &                 4.15 &                 4.55 &             0.46 &               0.09 &               0.09 \\
 170294 & 206.402250 &  -7.326361 & 104.78 &  1.219 &        14.80 &        8.720 &               1.14 &                 3.96 &                 0.31 &             1.44 &               5.03 &               0.39 \\
  49511 & 208.904238 &  15.970625 &  70.27 &  0.176 &         2.28 &        8.509 &               2.16 &                 6.77 &                 0.38 &             1.23 &               3.87 &               0.21 \\
  50083 & 210.853722 &  -6.069867 &  39.81 &  1.702 &        44.99 &        8.776 &              25.48 &                 4.11 &                 4.47 &             4.68 &               0.75 &               0.82 \\
  50345 & 211.729271 &  -5.460500 &  42.85 &  0.724 &         8.89 &        8.674 &              61.99 &                 8.70 &                 8.94 &            13.19 &               1.85 &               1.90 \\
  50666 & 212.869005 & -26.203938 &  36.17 &  0.180 &         1.53 &        8.455 &               1.77 &                 3.45 &                 0.36 &             0.27 &               0.52 &               0.05 \\
  51849 & 217.676401 &   1.881474 & 143.39 & 11.705 &        36.96 &        8.851 &               1.88 &                 3.10 &                 0.52 &             4.48 &               7.38 &               1.23 \\
3441671 & 217.930349 &  -2.211974 & 127.00 &  0.237 &         1.11 &        8.145 &               3.92 &                10.14 &                 0.85 &             7.33 &              18.95 &               1.58 \\
  52412 & 220.047953 &  -0.289004 &  22.23 &  0.487 &         1.69 &        8.725 &              20.60 &                 3.65 &                 3.88 &             1.18 &               0.21 &               0.22 \\
\end{longtable}
\begin{tablenotes}
    \small
        \item \textbf{Notes}: Columns (1)-(7) give the galaxy characteristics from HECATE v2.0. Columns (8)-(13) present the results of our X-ray analysis for the HEC-eR1 star-forming galaxies with reliable fluxes. Column (1): PGC number;  Column (2): Right ascension; Column (3): Declination; Column (4): Distance to the galaxy; Columns (5) and (6): star-formation rate (\sfr{}) and stellar mass (\stellarmass{}); Column (7): Gas-phase metallicity; Columns (8)-(10): Mode of the posterior flux distribution in the 0.6-2.3 \energyunits{} band along with the upper and lower 68$\%$ C.I.;
        Columns (11)-(13): Mode of the posterior X-ray luminosity distribution in the 0.5-2 \energyunits{} band along with the upper and lower 68$\%$ C.I.
\end{tablenotes}
\end{landscape}
\clearpage
\twocolumn

\end{document}